\pdfoutput=1
\documentclass[twocolumn]{aastex62}

\usepackage{amsmath}
\usepackage{graphicx}
\usepackage{url}
\usepackage{tabularx}
\usepackage{footmisc}

\graphicspath{{figures/}} 

\newcommand{\Kepler}{{\it Kepler}}
\newcommand{\Keplers}{{\it Kepler's}}

\submitjournal{AAS Journals}

\shorttitle{Occurrence Rates with DR25}
\shortauthors{Bryson et al.}

\begin{document}

\title{A Probabilistic Approach to {\it Kepler} Completeness and Reliability for Exoplanet Occurrence Rates}

\correspondingauthor{S. Bryson}
\email{steve.bryson@nasa.gov}

\author[0000-0003-0081-1797]{S. Bryson}
\affiliation{NASA Ames Research Center, Moffett Field, CA 94901}

\author[0000-0003-1634-9672]{J. Coughlin}
\affiliation{NASA Ames Research Center, Moffett Field, CA 94901}
\affiliation{SETI Institute, Mountain View, CA}

\author{N. M. Batalha}
\affiliation{University of California Santa Cruz, Santa Cruz, CA}

\author{T. Berger}
\affiliation{Institute for Astronomy, University of Hawai`i, 2680 Woodlawn Drive, Honolulu, HI 96822, USA}

\author{D. Huber}
\affiliation{Institute for Astronomy, University of Hawai`i, 2680 Woodlawn Drive, Honolulu, HI 96822, USA}

\author{C. Burke}
\affiliation{MIT Kavli Institute, Cambridge, MA}

\author{J. Dotson}
\affiliation{NASA Ames Research Center, Moffett Field, CA 94901}

\author[0000-0001-7106-4683]{S. E. Mullally}
\affiliation{Space Telescope Science Institute, 3700 San Martin Dr, Baltimore, MD 21218}



\begin{abstract} 
Exoplanet catalogs produced by surveys suffer from a lack of {\it completeness} (not every planet is detected) and less than perfect {\it reliability} (not every planet in the catalog is a true planet), particularly near the survey's detection limit.  Exoplanet occurrence rate studies based on such a catalog must be corrected for completeness and reliability.  The final \Kepler\ data release, DR25, features a uniformly vetted planet candidate catalog and data products that facilitate corrections.  We present a new probabilistic approach to the characterization of \Kepler\ completeness and reliability, making full use of the \Kepler\ DR25 products.  We illustrate the impact of completeness and reliability corrections with a Poisson-likelihood occurrence rate method, using a recent stellar properties catalog that incorporates Gaia stellar radii and essentially uniform treatment of the stellar population.  Correcting for reliability has a significant impact: the exoplanet occurrence rate for orbital period and radius within 20\% of Earth's around GK dwarf stars, corrected for reliability, is $0.015^{+0.011}_{-0.007}$, whereas not correcting results in $0.034^{+0.018}_{-0.012}$ --- correcting for reliability reduces this occurrence rate by more than a factor of two. We further show that using Gaia-based vs. DR25 stellar properties impacts the same occurrence rate by a factor of two. We critically examine the the DR25 catalog and the assumptions behind our occurrence rate method. We propose several ways in which confidence in both the \Kepler\ catalog and occurrence rate calculations can be improved.  This work provides an example of how the community can use the DR25 completeness and reliability products.
\end{abstract}

\keywords{\Kepler\ --- DR25 --- exoplanets --- exoplanet occurrence rates --- catalogs --- surveys}


\section{Introduction} \label{section:introduction}
The \Kepler\ space telescope \citep{Borucki2010,Koch2010} has delivered unique data that enables the characterization of exoplanet population statistics, from hot Jupiters in short-period orbits to terrestrial-size rocky planets in orbits with periods up to one year\footnote{\label{footnote:occurrenceRatePapaers}\url{https://exoplanetarchive.ipac.caltech.edu/docs/occurrence_rate_papers.html}}. 
By observing $>$150,000 stars nearly continuously for four years looking for transiting exoplanets, \Kepler\ detected several thousand planet candidates (PCs) \citep{Thompson2018}, leading to the confirmation or statistical validation of over 2,300 exoplanets.  This rich trove of exoplanet data has delivered many insights into exoplanet structure and formation, and promises deeper insights with further analysis. One of the most exciting insights to be gained from \Kepler\ data is the occurrence rate of temperate, terrestrial-size planets orbiting Sun-like stars (often referred to as $\eta_{\oplus}$).  This occurrence rate is also a critical input to the design of future space telescopes designed to discover and characterize habitable exoplanets, such as HabEx and LUVOIR.

Fully exploiting \Kepler\ data requires a thorough understanding of how well it reflects the underlying exoplanet population.  There are several ways in which the \Kepler\ planet candidate catalog does not directly measure the real planet population:

\begin{itemize}
    \item the catalog is {\it incomplete}, missing real planets
    \item it may be {\it unreliable}, with the planet candidate catalog being polluted with false positives
    \item it may be {\it inaccurate} due to observational errors leading to incorrect planet properties.
\end{itemize}
Lack of completeness and reliability are particularly acute at the \Kepler\ detection limit, which happens to coincide with the period and radius of Earth-Sun analog exoplanets.  We therefore focus our attention on a period and radius range spanning the \Kepler\ detection limit.

In this paper we address {\it vetting incompleteness}, a significant component of incompleteness caused by incorrectly classifying detected true planets as false positives, and {\it vetting reliability}, caused by incorrectly classifying detections as planets candidate when they are in fact not true planets.  We address accuracy by using new, uniformly determined stellar properties based in part on Gaia observations, described in \S\ref{section:stellarCatalog}.

We focus our analysis on the final \Kepler\ data release DR25 \citep{Thompson2018} and its associated planet candidate catalog\footnote{\label{footnote:exoplanetArchive}\url{https://exoplanetarchive.ipac.caltech.edu}}.  DR25 contains several products designed to support the characterization of the completeness and reliability of the DR25 planet candidate catalog.  The primary contribution of this paper is a new probabilistic approach to using the DR25 completeness and reliability products to characterize vetting completeness and reliability.  We illustrate the impact of completeness and reliability with standard occurrence rate computations, and examine the impact on occurrence rates due to changes in various assumptions.  This is the first occurrence rate computation that fully uses the DR25 completeness and reliability products to characterize vetting reliability.

\subsection{Previous Work} \label{section:previousWork}
\Keplers\ survey of the Cygnus field involved four years of data collection and another four years of pipeline development, data processing, and survey characterization, culminating in the final deliveries referred to as Data Release 25 (DR25). Incremental data deliveries enabled preliminary science investigations, and several occurrence rate studies were executed as the survey progressed.  The simplest, first-look estimates used Gaussian cumulative distribution functions (CDF) as proxies for pipeline completeness \citep{Borucki2011}, restricted samples where completeness was assumed to be near unity \citep{Howard2012}, and linear approximations to a Gaussian CDF \citep{Fressin2013}. Lacking a full characterization of the \Kepler\ pipeline, others employed independent detection pipelines, including injection and recovery experiments operating on flux light curves to quantify the detection completeness \citep{Petigura2013,ForemanMackey2014,Dressing2015}.  \citet{Hsu2018} performed an occurrence rate calculation using approximate Bayesian computation (ABC) and \citet{zink2019} computed a habitable zone occurrence rate taking into account the effect of planet multiplicity on completeness.

The performance of the Kepler pipeline was characterized incrementally as more data were collected using transit injection and recovery operating on raw pixel fluxes as described in Section~\ref{section:compRelProducts} \citep{Christiansen2013,Christiansen2015,Christiansen2016,Christiansen2017}.  These early studies provided positive feedback to the Kepler pipeline whereby deficiencies were identified and improved upon \citep{Twicken2016}.  Occurrence rate calculations using 16 out of 17 quarters of data and the associated pipeline completeness offered a benchmark computation for testing methodologies and comparing independent pipelines \citep{Burke2015}.  Systematic errors were explored and the tallest tent poles were identified.  Among these tall tent poles were two standouts:  stellar property uncertainties and catalog reliability. 

The first studies to include a treatment of catalog reliability focused on identifying astrophysical false positives either deterministically through follow-up observations \citep{Santerne2012} or probabilistically via population synthesis \citep{Morton2011,Morton2014,Morton2016}.  And while most treatments culled or weighted the planet population, others sought to model both the planet and astrophysical sources as part of the planet occurrence estimation \citep{Fressin2013}, with  \citet{Farr2014} applying a mixture model approach. These efforts used idealized models of the false positive and false alarm populations.  An extremely useful astrophysical false positive probability statistic was developed by \citet{Morton2016}.  

The most significant effort to characterize the reliability of the \Kepler\ planet candidate catalog to date is the final \Kepler\ DR25 catalog paper \citet{Thompson2018}.  The DR25 catalog includes a {\it Robovetter score} which estimates the confidence with which the Robovetter vetted a TCE.  \citet{Thompson2018} suggests that restricting occurrence rate studies to high-Robovetter-score planet candidates avoids the problem of low-reliability candidates (we test this approach in \S\ref{section:variationsScoreCut}).  \citet{Hsu2018} and \citet{Mulders2018} apply this high-score approach in their occurrence rate studies.  

\citet{burke2019} analyzes DR25 reliability using the inverted and scrambled data, approaching reliability characterization via kernel density estimation.  The focus of \citet{burke2019} is how reliability impacts statistical exoplanet validation.  

\subsection{This Paper}

The primary purpose of this paper is to present a probabilistic characterization of \Kepler\ vetting completeness (\S\ref{section:vettingCompleteness}) and reliability against false alarms (\S\ref{section:vetReliability}) and show the impact on standard occurrence rate computations.  This probabilistic analysis is robust against sparse data and resolves detailed structure of the dependence of vetting completeness and reliability on orbital period and transit signal strength.  We explore the impact of using this characterization to correct occurrence rates based on \Kepler\ data by performing a standard Poisson-likelihood-based occurrence rate, following \citet{Burke2015}.  Our characterization depends on the exoplanet population, which in turn depends on the parent stellar sample that is searched for planets.  We restrict our analysis to GK dwarf stars.

Our method of accounting for completeness and reliability proceeds by executing the following steps:
\begin{itemize}
    \item Select a subset of the target star population, which will be our parent population of stars that are searched for planets.  We apply various cuts intended to select well-behaved and well-observed stars, and we restrict our analysis to GK dwarfs, as described in \S\ref{section:stellarCatalog}.
    \item Use the injected data to characterize vetting completeness, described in \S\ref{section:vettingCompleteness}.
    \item Compute the summed detection completeness, incorporating vetting completeness, described in \S\ref{section:detectionCompleteness}.
    \item Use observed, inverted and scrambled data to characterize false alarm reliability, described in \S\ref{section:vetReliability}.
    \item Assemble the collection of planet candidates, including computing the reliability of each candidate from the false alarm reliability and False Positive Probability.
    \item Compute the desired occurrence rates, presented in \S\ref{section:occurrence}.
\end{itemize}

We perform our completeness and reliability analysis on the planet radius range $0.5 \leq \mathrm{radius} \leq 15$ $R_{\oplus}$.  We perform our completeness analysis on the period range $50 \leq \mathrm{period} \leq 500$ days and reliability analysis on the period range $50 \leq \mathrm{period} \leq 600$ days.  Our occurrence rates are focused on illustrating the impact of vetting completeness and reliability where both are low, so our occurrence rates are analyzed for $50 \leq \mathrm{period} \leq 400$ days and $0.75 \leq \mathrm{radius} \leq 2.5$ $R_{\oplus}$.

Following an introduction to the details of completeness, reliability and the data products that support their computation in \S\ref{section:vettingIntro}, this paper has four major parts: \S\ref{section:catalogs} assembles the stellar and planet catalogs we use in our analysis.  We choose a stellar catalog that incorporates Gaia stellar radii and features an essentially uniform treatment of the parent stellar sample. \S\ref{section:completeness} describes catalog completeness and describes our characterization of vetting completeness.  \S\ref{section:reliabilityModel} describes our characterization of catalog reliability.  In \S\ref{section:occurrence} we perform our baseline occurrence rate computations corrected for vetting completeness, emphasizing the difference between correcting and not correcting for reliability.  We then explore alternative occurrence rate calculations, including using an alternative stellar properties catalog, demonstrating the impact of not correcting for vetting completeness and restricting our analysis to planet candidates with Robovetter score $> 0.9$.

Throughout this paper we present results with confidence intervals that are the 14\textsuperscript{th} and 86\textsuperscript{th} percentiles of posterior distributions resulting from MCMC analysis using fixed inputs.  These confidence intervals do not account for uncertainties in the inputs.  We address the issue of uncertainties in the inputs in \S\ref{section:uncertainties}.

All results reported in this paper were produced with Python code, mostly in the form of Python Jupyter notebooks, found at the paper GitHub site\footnote{\label{footnote:github}\url{https://github.com/stevepur/DR25-occurrence-public}}.

\section{Completeness, Reliability and Occurrence Rates, } \label{section:vettingIntro}

As described above, completeness has two components.  {\it Detection completeness} is the fraction of true planets that are detected by the \Kepler\ pipeline.  {\it Vetting completeness} is the fraction of detected true planets that are correctly vetted as planet candidates.  {\it Vetting reliability} is the fraction of vetted planet candidates that are true planets.  

During catalog creation, the reliability of the planet candidate catalog is increased by detecting and removing false positives using a variety of tests.  When the transit signal is weak it can be difficult for these tests to distinguish false positives from true planets, so maximizing reliability via stringent tests can cause true planets to be classified as false alarms, reducing vetting completeness.  The DR25 catalog addressed this problem with uniform automated vetting via the {\it Robovetter}, using tests that were tuned to strike a balance between vetting completeness and reliability.  This uniform automated vetting made it possible to vet synthetic and modified data sets designed to statistically mimic true planets and false positives, described in \S\ref{section:compRelProducts}, in exactly the same way that the observed data was vetted.  In this way the completeness and reliability of the planet candidate catalog can be measured and corrected in occurrence rates.

\begin{figure*}
\centering
\includegraphics[width=\linewidth]{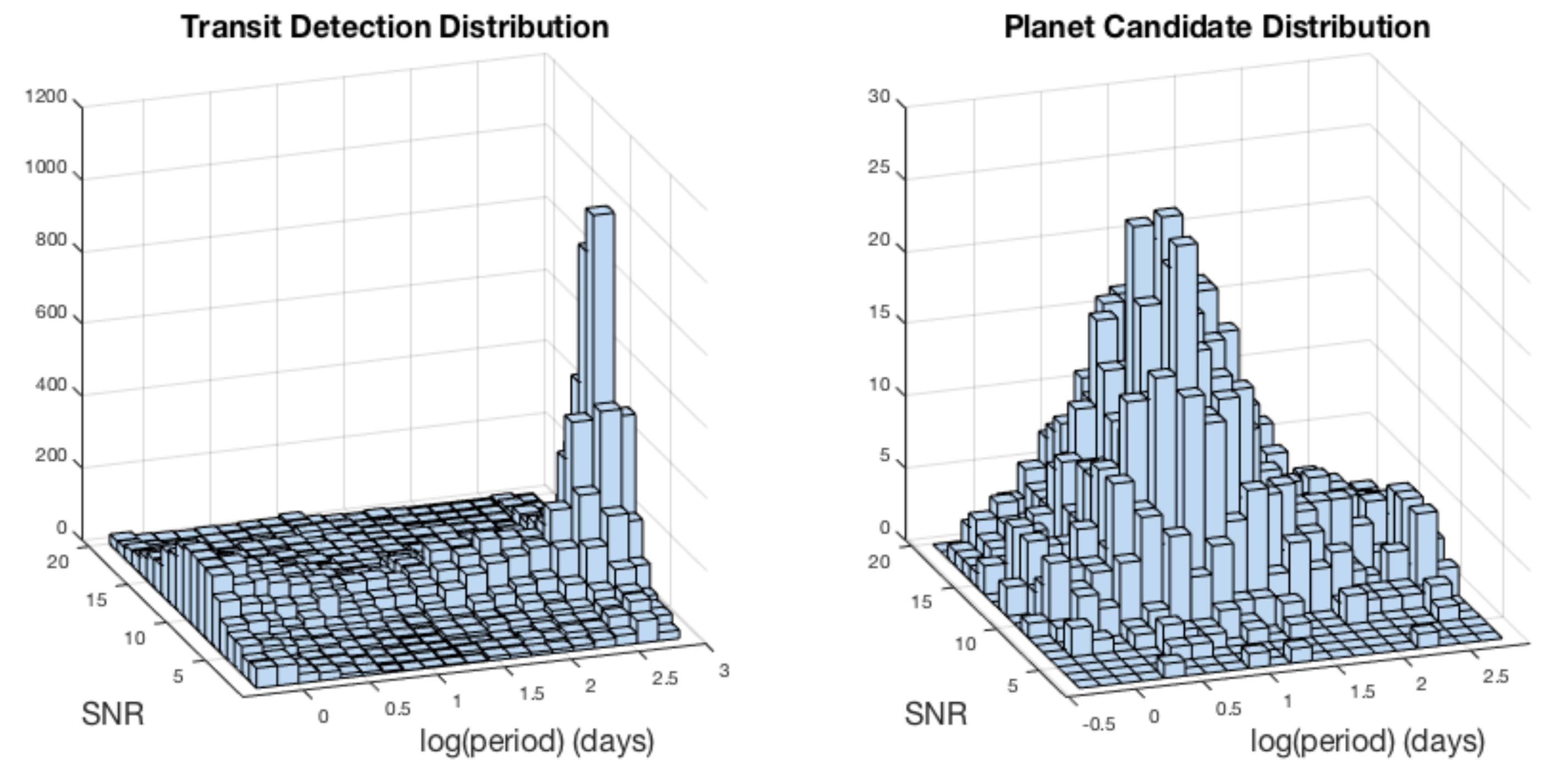} 
\caption{{\bf Left:} the distribution of transit signal detections in SNR-period space from the final \Kepler\ data release (DR25) \citet{Thompson2018}, showing a dramatic excess near the \Kepler\ orbital period of 372 days and SNR between 7 and 15.  {\bf Right}: the distribution of DR25 planet candidates (PCs) over the same space.  Note the scale change on the vertical axis. While the vast majority of detections have been identified as false alarms and removed from the planet candidate population, there remains a possible small excess of PCs near the \Kepler\ orbital period and with SNR between 7 and 15. } \label{figure:periodSnrDist}
\end{figure*}

We distinguish two broad classes of phenomena that pollute the planet candidate catalog:
\begin{itemize}
    \item {\bf Astrophysical False Positives}, such as grazing or eclipsing binaries, which produce a planet-transit-like signal with a regular ephemeris in observed light curves that are not due to planetary transits.  There has been extensive effort to identify and remove such false positives from \Kepler\ catalogs \citep[e.g.,][]{Morton2011,Bryson2013,Fressin2013,Coughlin2014,Morton2016,Thompson2018}, and for high SNR transits the resulting removal of astrophysical false positives is very effective.  For low SNR, however, it is more difficult to distinguish astrophysical false positives from true planetary transits.  In this paper we address astrophysical false positives via the probabilistic evaluation of \citet{Morton2016}.
    
    \item {\bf False Alarms}, which trigger a transit detection with a regular ephemeris, but are not due to regularly repeating astrophysical phenomena.  The dominant source of false alarms in \Kepler\ data is instrumental artifacts.  There are two important classes of instrumental artifacts that have been identified as responsible for the overwhelming majority of false alarms at long periods: {\it rolling bands} and {\it statistical and pixel fluctuations}.  
    
    \begin{itemize}
        \item {\bf Rolling bands} \citep{VanCleve2009,Caldwell2010} are thermally dependent quasi-sinusoidal electrical signals in the output of the \Kepler\ CCDs.  As the \Kepler\ telescope slowly changes its attitude relative to the Sun, different parts of the photometer are illuminated.  While the thermal insulation of the telescope and electronics is very good, it is not perfect and the resulting thermal variations cause the rolling bands to slowly move across the CCDs, often introducing signals that look very much like transits.  Because \Keplers\ attitude is determined by its 372-day orbit, rolling bands often induce transit-like signals that repeat with a nearly regular ephemeris with an approximately 372-day period.  This is the cause of the sharp peak of TCEs in the left panel of Figure~\ref{figure:periodSnrDist}.  Rolling bands are highly focal plane position dependent: some \Kepler\ CCD channels have much more severe rolling bands than others.
    
        \item {\bf Statistical and pixel fluctuations} are independent, unrelated dips in light curves due to cosmic ray hits, single transits, and statistical fluctuations that trigger transit detections when they accidentally fall into a regular ephemeris.  This class of false alarms becomes much more common for long-period ephemerides because they only require three or four events to fall on a regular ephemeris, and explains the broader ``shoulders'' of the tall peak in the left panel of Figure~\ref{figure:periodSnrDist}.
    \end{itemize}
\end{itemize}

In addition, stellar variability triggers false alarm transit detections on a regular ephemeris, typically at short periods.  False positives and false alarms are treated differently in our analysis, as described in \S\ref{section:reliabilityModel}.

In an ideal world, measuring planet occurrence rates from \Kepler\ data would be simple --- divide the number of detected planets by the number of observed stars, correcting for the geometrical probability of a planet transit.  To get an accurate occurrence rate, however, this approach must be corrected for completeness and reliability.  This correction can be large for Earth-Sun analog systems, which are at the \Kepler\ detection limit where both completeness and reliability are very low.  Specifically, instrumental false alarms are a significant source of false transit detections in the long-period, low signal-to-noise (SNR) region of most interest for habitable zone occurrence rate studies for G and K dwarf stars. In this regime, as shown in Figure~\ref{figure:periodSnrDist}, the number of instrumental false alarms is very large compared to the expected population of true exoplanet detections.  

\subsection{DR25 Vetting and Reliability Products} \label{section:compRelProducts}
The DR25 planet candidate catalog \citep{Thompson2018} contains 4034 identified planet candidates (PCs) out of 8054 \Kepler\ {\it Objects of Interest} (KOIs).  The KOIs were extracted from a catalog of 34,032 transit detections called {\it threshold crossing events} (TCEs), which are periodic transit-like events \citep[as identified by a matched filter;][]{Jenkins2002} that have a combined signal strength above a threshold (typically $7.1\sigma$). Identification of the PCs from the KOIs was performed by a fully automated {\it Robovetter}.  The Robovetter applies a variety of tests to each TCE, many of which are based on the synthetic test datasets described below, and planet candidates are TCEs that pass all tests while following a logic tree.  Such automated vetting (and transit detection) is critical for the production of a statistically uniform catalog that is amenable to statistical correction for completeness and reliability.  

The DR25 completeness products are based on {\it injected data} --- a ground-truth of transiting planets is obtained by injecting transit signals with specific characteristics on all observed stars at the pixel level \citep{Christiansen2017}.  This data is then analyzed by the \Kepler\ detection pipeline to produce a catalog of detections at the injected ephemerides called {\it injected and recovered TCEs}, which are then sent through the same Robovetter used to identify planet candidates.  The fraction of injected transits that are recovered as TCEs measures detection completeness, while the fraction of recovered TCEs that are vetted as planet candidates measures vetting completeness.  A large number of transits were also injected on a small number of target stars to measure the dependence of completeness on transit parameters and stellar properties.  This data is used to create high-resolution, per-target-star detection contours described in \S\ref{section:detectionCompleteness}, providing completeness for each target star as a function of planet orbital period and radius \citep{BurkeJCat2017}.

The rate of false alarms, measured for the first time in DR25, are characterized by manipulating observed data so that it contains no true astrophysical transiting exoplanet signals, creating a ground-truth in which any TCE or vetted planet candidate is an instrumental false alarm.  There are two basic manipulations that create the data used to characterize the rate of false alarms:
\begin{itemize}
    \itemsep0em 
    \item {\bf Data inversion} flips the light curves ``upside down'' so that true transiting signals increase in brightness and are therefore not identified as transits.  This is believed to preserve the quasi-sinusoidal rolling bands described above.  The distribution of TCEs detected in the inverted data reproduces well the sharp peak at a period of 372 days that is seen in the distribution of observed TCEs in the left panel of Figure~\ref{figure:periodSnrDist}.  
    \item {\bf Data scrambling} shuffles the \Kepler\ observational quarters in a way that destroys the regular ephemeris of astrophysical transit signals, preventing their detection by the \Kepler\ pipeline.  While this also prevents the detection of the same false alarms that are detected in the original observed data, it is believed to preserve the statistics of detections due to statistical and pixel fluctuations. The distribution of TCEs detected in the scrambled data is very similar to the broad shoulder near periods of one year in the distribution of observed TCEs seen in the left panel of Figure~\ref{figure:periodSnrDist}. Three different shuffles of the \Kepler\ data are available.
\end{itemize}
The DR25 Robovetter uses a number of metrics to identify instrumental false alarms, and the inverted and scrambled data sets were used to tune their pass/fail thresholds.  For an extensive discussion, see \citet{Thompson2018}.

For many Robovetter metrics, the distribution of values from the inverted/scrambled data overlaps the distribution of values in the observed data, making it difficult to distinguish false alarms from true planets, particularly at long period and low SNR.  This results in low reliability in some parameter spaces of the \Kepler\ planet candidate catalog, especially near the detection limit.  An effort to characterize this catalog reliability is described in Section 4 of \citet{Thompson2018}.  The work presented in this paper is an attempt to improve on this characterization.

\section{Input Catalogs} \label{section:catalogs}
\subsection{The Stellar Catalog} \label{section:stellarCatalog}

Our occurrence rate starts with a parent stellar population of GK dwarf stars that is searched for planetary transits.  The properties of each star determine the likelihood that a transiting planet of a given size will be observed. The radius of the planet is derived from the fitted radius ratio and stellar radius.  While the most accurate stellar properties for each star is desirable for understanding the properties of the transiting planet, a statistical occurrence rate requires the most {\it uniform} stellar properties possible.  This is an issue for \Kepler\ data because target stars with actual transit detections are much better characterized than most targets stars without transit detections.  This can potentially lead to unknown biases in the estimated detection completeness of \S\ref{section:vettingCompleteness}.

The stellar catalog associated with the DR25 exoplanet catalog, Q1-Q17 DR25 (with supplement)\footnote{\raggedright \label{footnote:dr25Stellar} \url{https://exoplanetarchive.ipac.caltech.edu/docs/Kepler\_stellar\_docs.html}}, is based on heterogeneous observations, with some stars having properties derived from asteroseismic data, others from spectral data, and most from photometric data, all fitted to Dartmouth isochrones \citep{Mathur2016,Mathur2017ApJS}.  \citet{berger18} combined the DR25 stellar catalog with Gaia parallaxes \citep{gaiaDR2Astrometric} to improve stellar radii, yielding an average radius precision of less than 10\% for most Kepler stars. However, the adopted effective temperatures were still heterogeneous, and no revisions of other stellar properties such as mass and surface gravities were performed.

For our parent stellar population we use the stellar catalog of \citet{berger20}, which extends \citet{berger18} by deriving a full set of stellar properties from isochrone fitting using broadband photometry, Gaia parallaxes and spectroscopic metallicities where available.  This catalog is based on the homogeneous derivation of temperatures and luminosities, which previously have been the dominant sources of systematic errors in stellar (and thus planet) radii.  These consistently fitted stellar properties provide more uniformly derived stellar radii over the entire parent population than the DR25 stellar properties. We recompute the 4-parameter stellar limb darkening model coefficients using these stellar properties via the table {\it tableeq5.dat} in \citet{Claret2011}, assuming a microturbulent velocity of 2 km s$^{-1}$. In \S\ref{section:variationsKic} we compare the resulting baseline occurrence rates with those using the DR25 stellar properties.  We address the issue of possible bias against small planets in the \citet{berger20} catalog in \S\ref{section:variationsKic} and Appendix~\ref{app:catalogComparison}.

\begin{figure}[ht]
   \centering
   \includegraphics[width=\linewidth]{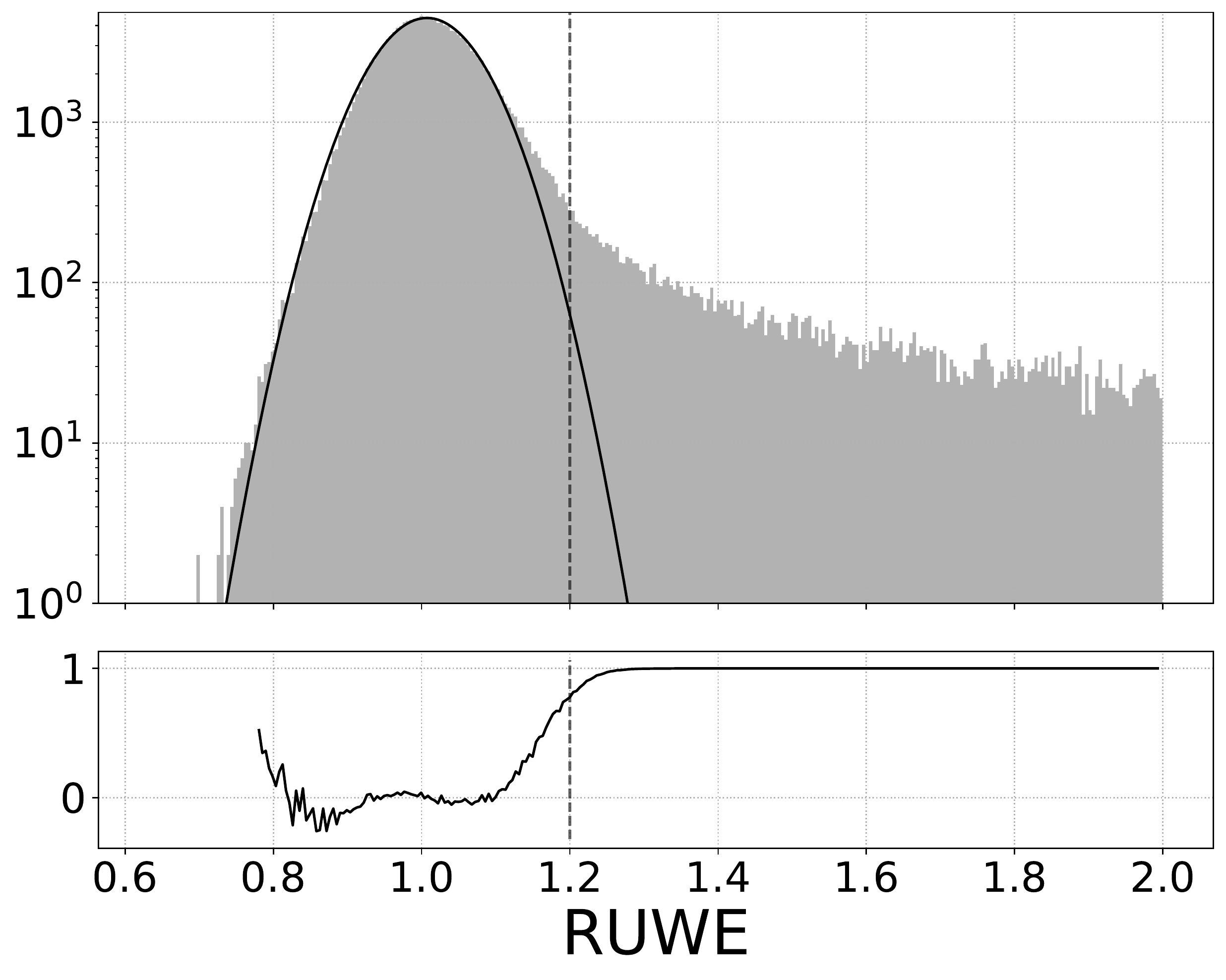}
   \caption{
   The distribution of Gaia Renormalized Unit Weight Error (RUWE) for the \citet{berger18} catalog.  {\bf Top panel}: the distribution in grey, with the black line showing the Gaussian fit to that distribution for stars with RUWE $< 1.15$.  The grey distribution above the black line indicates the number of stars with an excessively high RUWE, which can indicate stellar multiplicity.  {\bf Bottom panel}: the fraction of stars with excess RUWE.  The vertical dashed line indicates our chosen cutoff rejecting stars with RUWE $> 1.2$.
   } \label{figure:ruwe}
\end{figure}

Because we require information such as observational completeness from the DR25 catalog and the binary flag from \citet{berger18} for each target star, we merge \citet{berger20}, the DR25 stellar catalog (with supplement), and \citet{berger18}, keeping only the 177,798 stars that are in all three catalogs.  We remove possibly poorly characterized, binary and evolved stars using the following cuts:
\begin{itemize}
    \item Remove stars with \citet{berger20} goodness of fit ({\it iso\_gof}) $< 0.99$ and Gaia Renormalized Unit Weight Error (RUWE, provided by \citet{berger20}) \citep{gaiaRuwe2018} $> 1.2$, leaving 162,219 stars. {\it iso\_gof} measures the quality of the \citet{berger20} isochrone fitting, and RUWE combines several Gaia goodness-of-fit metrics.  RUWE is expected to have a Gaussian distribution \citep{gaiaRuwe2018} for single stars. Figure ~\ref{figure:ruwe} shows the distribution of RUWE for the \citet{berger20} catalog, with a Gaussian fit to those stars with RUWE $< 1.15$.  Above RUWE $> 1.15$ there is an apparent excess in RUWE, with that excess becoming dominant ($> 75\%$ of stars) at RUWE $\approx 1.2$. An excess of RUWE over a Gaussian distribution is believed to be a strong indicator of stellar multiplicity. For example, \citet{kraus19} finds that few \Kepler\ target stars with RUWE $> 1.2$ are single stars.   We find that the RUWE Gaussian distribution has a slight magnitude dependence: for $g \leq 13$ the fitted Gaussian has the mode at 0.98, while for $g>13$ we find the mode at 1.01.  We balance the loss of ``good'' stars against removing ``bad'' stars by choosing an RUWE cutoff of 1.2, in contrast to the cutoff of 1.4 discussed in the Gaia literature, {\it e.g.} \citet{gaiaRuwe2018}.
    
    \item Remove stars that, according to \citet{berger18}, are likely binaries (Bin flag = 1 or 3; we allow Bin = 2 because that indicates a nearby companion star found via high-resolution imaging, which was only performed on a subset of the target stars).  This leaves 160,633 stars.

    \item Remove stars that have evolved off the main sequence, recomputing the Evol flag described in \citet{berger18} using the \citet{berger20} stellar properties.  We use the evolstate package\footnote{\url{http://ascl.net/1905.003}} to determine the evolution state of each star using the isochrone-fitted $T_{\rm eff}$, radius and logg as inputs.   We remove those stars with Evol $> 0$, indicating that they are likely not main sequence dwarfs.  After removing these stars 105,118 stars remain.
\end{itemize}

We then remove stars whose observations were not well suited for long-period transit searches \citep{Burke2015, BurkeJCat2017}:
\begin{itemize}
    \item Remove noisy targets identified in the KeplerPorts package\footnote{\url{https://github.com/nasa/KeplerPORTs/blob/master/DR25_DEModel_NoisyTargetList.txt}}, leaving 103,626 stars.
    \item Remove stars with NaN limb darkening coefficients, leaving 103,371 stars.
    \item Remove stars with NaN observation duty cycle, leaving 102,909 stars.
    \item Remove stars with a decrease in observation duty cycle $> 30\%$ due to data removal from other transits detected on this star, leaving 98,672 stars.
    \item Remove stars with observation duty cycle $< 60\%$, leaving 95,335 stars.
    \item Remove stars with data span $< 1000$ days, leaving 87,765 stars.
    \item Remove stars with the DR25 stellar properties table {\it timeoutsumry} flag $\neq 1$, leaving 82,371 stars.  This flag = 1 indicates that the \Kepler\ pipeline completed its transit search on this star before timing out.
\end{itemize}

\begin{figure}[ht]
   \centering
   \includegraphics[width=\linewidth]{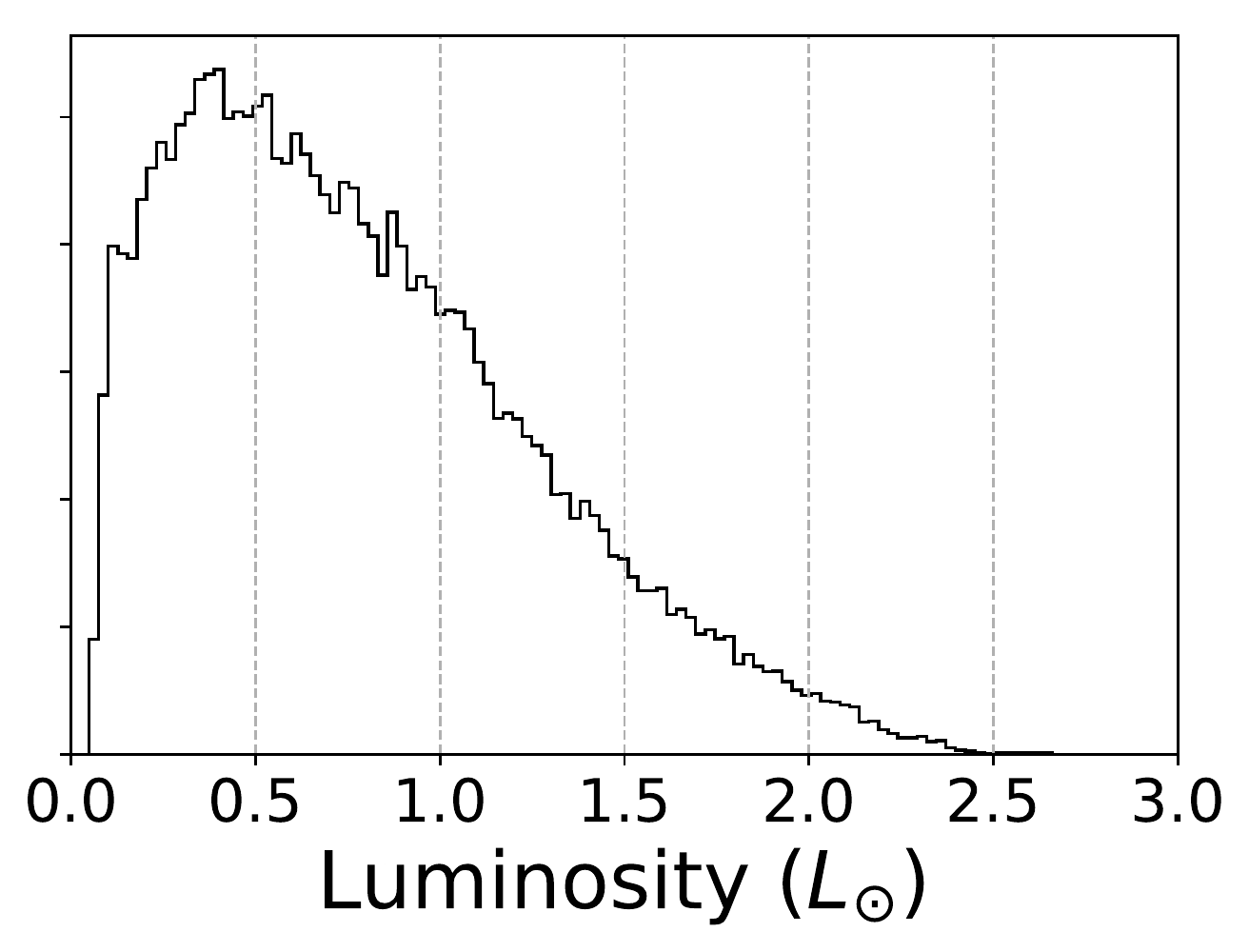}
   \caption{The distribution of stellar luminosities for our final GK parent stellar population.} \label{figure:stellarLum}
\end{figure}

Finally we select our GK population using the isochrone-fitted effective temperature as $3900\mathrm{K} \leq T_{\rm eff} < 6000\mathrm{K}$, using the temperature limits of \citet{Pecaut2013}, leaving 57,015 stars in our parent GK dwarf population.  The distribution of luminosities of these stars, computed as $R_*^2 T_{\rm eff}^4$ in Solar units, is shown in Figure~\ref{figure:stellarLum}.

\begin{figure}[ht]
   \centering
   \includegraphics[width=\linewidth]{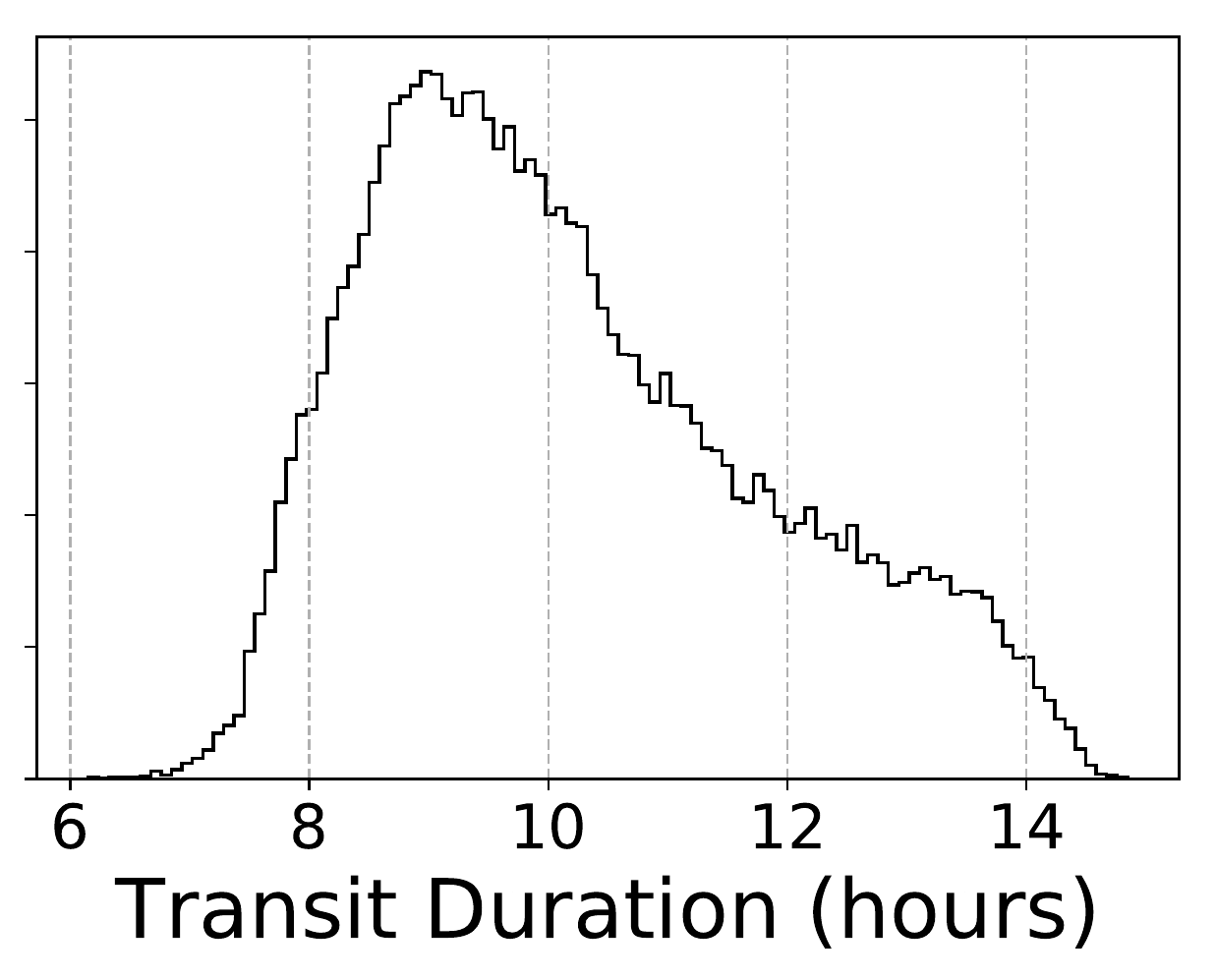}
   \caption{The distribution of transit durations for our baseline GK parent stellar population, assuming a 400-day orbit and zero eccentricity.} \label{figure:durationDist}
\end{figure}

The largest remaining GK star has a stellar radius of $1.536 R_{\odot}$.  \citet{BurkeJCat2017} states that the per-star detection completeness described in \S\ref{section:detectionCompleteness} is invalid for stars of radius $> 1.25 R_{\odot}$ because completeness is characterized only for transit duration below the maximum 15 hours searched by the \Kepler\ pipeline.  We extend this to $R_* > 1.35 R_{\odot}$ because we are restricting our orbital period to 400 days, which keeps transit duration under 15-hours for our GK stellar population. We do not impose the stellar radius criterion in our baseline, instead opting for the physically motivated selection based on the Evol flag (though we use the $R_* > 1.35 R_{\odot}$ radius cut when using the DR25 stellar properties catalog in \S\ref{section:variationsKic}).  Our baseline stellar population has 1,043 stars, or 1.83\%, with $R_* > 1.35 R_{\odot}$.  The maximum transit duration across our baseline population for a 400 day period and eccentricity = 0 is 14.85 hours.  The distribution of transit durations for our baseline stellar population assuming a 400-day period and eccentricity = 0 is shown in Figure~\ref{figure:durationDist}, which shows that our population gets close to, but does not exceed, the 15-hour duration limit.

\subsection{The Planetary Catalog} \label{section:planetCatalog}

Our planetary catalog is the Q1--Q17 DR25 Kepler Object of Interest (KOI) table at the exoplanet archive\footref{footnote:exoplanetArchive} \citep{Thompson2018}, restricted to planet candidates (KOIs with koi\_pdisposition = CANDIDATE) on stars in the catalog from \citet{berger20}.  We accept the CANDIDATE and FALSE POSITIVE dispositions resulting from the uniform robovetter run on the TCEs.  

For our baseline case, we recompute the planet radii $R_p$ (in Earth radii) from the stellar radii $R_*$ (in Solar radii) in \citet{berger20} and the ratio of the planet radius to the stellar radius, $A = R_p / R_*$ from the {\it koi\_ror} column of the KOI table, as $ R_p = A  R_*   R_{\odot} / R_{\oplus} $ where $R_{\odot}$ ($R_{\oplus}$) is the Solar (Earth) radius.  We compute the planet radius uncertainties $\sigma_{R_p}$ from the stellar radius uncertainties $\sigma_{R_*}$ and planet radius to the stellar radius ratio uncertainties $\sigma_{A}$ via standard propagation of uncertainties:
 $ \sigma_{R_p} = \sqrt{\sigma_{A}^2  R_*^2 + A^2  \sigma_{R_*}^2}  R_{\odot} / R_{\oplus}$,
where the upper and lower uncertainties are computed independently.

\section{The Completeness Model} \label{section:completeness}
As described in \S\ref{section:introduction}, the set of planet candidates in the DR25 KOI catalog is not expected to be complete: particularly near the \Kepler\ detection limit we expect that some transiting planets will be detected while others will be missed, and some of those detected will be mis-classified as false positives.  {\it Detection completeness} is a measure of the fraction of true transiting planets that are detected.  {\it Vetting completeness} is a measure of the fraction of detected true transiting planets that are correctly classified as planet candidates.  We expect detection and vetting completeness to be functions of the orbital period and the signal to noise ratio (SNR), which in the \Kepler\ data processing pipeline is measured as the Multiple Event Statistic (MES) \citep{Jenkins2002}.  MES measures the combined significance of all observed transits in the de-trended, whitened light curve.  

Detection and vetting completeness are both measured using the DR25 transit injection data products\footnote{\raggedright \label{footnote:simulatedData} \url{https://exoplanetarchive.ipac.caltech.edu/docs/KeplerSimulated.html}} \citep{Christiansen2017}.  \citet{Christiansen2013,Christiansen2015,Christiansen2016} used these injection products to produce average detection curves as a function of MES for various stellar populations, marginalized over period.  While these marginalized occurrence rate curves are convenient, the Poisson likelihood method we use for our occurrence rate, described in \S\ref{section:occurrence}, works best with completeness provided as a function of both MES and period.  We will use the star-by-star detection completeness model of \citet{BurkeJCat2017}, provided for each star as a two-dimensional function of MES and period that accounts for each star's detailed observational coverage.  These completeness models are derived from a comprehensive database of $1.2 \times 10^8$ transit injection and recovery trials, which we summarize in \S\ref{section:detectionCompleteness}.  In \S\ref{section:vettingCompleteness} we introduce a new probabilistic approach to modeling vetting completeness.  

\subsection{Combined Detection and Vetting Completeness} \label{section:detectionCompleteness}

We use the characterization of \Kepler\ detection completeness computed by a modified version of the  KeplerPorts code base\footref{footnote:github}. This software computes a completeness function $\eta_s \left( p, r \right)$ (not to be confused with $\eta_{\oplus}$) as a function of period $p$ and planet radius $r$ for each star $s$.  The completeness function is described in detail in \citet{BurkeJCat2017}.  We briefly summarize the main steps for calculating the completeness function and describe the augmentations that incorporate vetting completeness.

The detection completeness calculation begins with estimating the MES expected for a given planet period and radius based on the stellar properties of the host.
For each period and radius, a central crossing transit depth is estimated based on the stellar properties and limb darkening provided by the stellar catalog.  The central crossing transit depth is converted into an expected MES by interpolating in the tabulated values of the one-sigma depth functions for each target \citep{BurkeJCat2017}. The one-sigma depth function corresponds to the signal depth that results in a 1-sigma value for MES, and is a function of the planet period and the expected transit duration.  The resulting expected MES is mapped to detection completeness based on analysis of the injected data.  We treat this pipeline detection completeness estimate and the vetting completeness as independent.  Thus the detection completeness is multiplied by the vetting completeness function $\rho(p,\mathrm{expectedMES}, \boldsymbol{\theta})$ described in \S\ref{section:vettingCompleteness}.  

This produces the combined detection and vetting completeness for a central transit.  The impact of non-central transits are accounted for through {\it MES smearing}, which convolves completeness with a distribution derived from analysis of the injected data.  Completeness is then multiplied by the tabulated window function, which accounts for observational gaps for this star and the requirement of the Kepler pipeline of having at least three transit events for a detection, and the geometric transit probability assuming a uniform distribution of the cosine of orbital inclination angles.

The output is a collection of completeness functions $\eta_s \left( p, r \right)$, one for each star $s$ which includes detection completeness, vetting completeness and geometric transit probability.  We sum these functions to create $\eta \left( p, r \right)  = \sum_{s=1}^{N_*} \eta_s \left( p, r \right)$ where $N_*$ is the number of searched stars. The summed completeness $\eta \left( p, r \right)$ is used in the occurrence rate calculations in \S\ref{section:occurrence}. 
\subsection{Vetting Completeness}
\label{section:vettingCompleteness}

Vetting completeness is the fraction of detected TCEs that were correctly vetted as PCs.  This vetting is uniformly performed on both the observed TCEs and on the injected data TCEs, both resulting from the \Kepler\ data analysis pipeline, with the DR25 Robovetter \citep{Coughlin2017,Thompson2018} using the same thresholds in both cases.  Because in the injected data every TCE is by definition a PC, vetting completeness is simply the fraction of injected on-target TCEs that were identified as PC by the Robovetter.  We study the dependence of the injected vetting completeness on TCE period and expected MES by binning the detected injected TCEs on a regular grid.  Our approach treats vetting completeness as a statistical property of a stellar population, analyzed separately for each choice of stellar population or stellar properties or other choices that may change the stellar or planet population. We present our vetting completeness analysis of the baseline GK star population in detail.  Other cases described in \S\ref{section:variations} require independent analysis, which can be found in the {\it htmlArchive} folders on the paper GitHub site\footref{footnote:github}.

Previous treatments of vetting completeness, {\it e.g.} \citet{Thompson2018}, partitions the expected MES-orbital period plane into cells and divides the number of injected TCEs vetted as a PC by the total number of injected TCEs in each cell, which is an estimate of the vetting completeness in that cell.  \citet{Mulders2018} does the same on a radius-orbital period plane, and \citet{Coughlin2017} does the same with multiple parameter combinations (MES, period, planet radius, stellar radius, stellar temperature, and insolation flux).  Using this data, one can estimate the dependence of the vetting completeness as a function of expected MES or planet radius and orbital period based on the measured fraction in each cell via, for example, $\chi^2$ fitting to a parametric model assuming a Gaussian likelihood, as done in \citet{Mulders2018}.  When there are many TCEs and many detections, this method can be expected to work well.  Near the \Kepler\ detection limit, however, there will be few TCEs and fewer correct PC dispositions, leading to strong gridding dependence and the possibility of adjacent cells having very different values.  For example, if adjacent cells have only one TCE each and one is vetted as PC while the other is vetted as false positive, then these adjacent cells will have completeness 0 and 1. Cells with no TCEs require special treatment.  Addressing these problems by requiring cells large enough to contain many TCEs can result in large cells smoothing out details of the vetting completeness' functional dependence.

Rather than fitting a parametric model of a particular functional form using $\chi^2$ methodology, we take a probabilistic approach using a binomial likelihood that readily handles sparsely populated regions of parameter space.  Specifically, we treat the injected TCEs as a collection, with a rate $\rho$ of (correctly) vetted PCs and a rate $(1- \rho)$ of (incorrectly) vetted FPs, and vetting by the Robovetter as draws from this collection. This is a classic binomial problem, in which the probability distribution of correctly vetting a PC depends on $\rho$ and the number of TCEs in the underlying collection.  For example, if there is only one TCE in a cell with a $\rho = 50\%$ probability of being vetted as a PC, the probability distribution of (correctly) vetting that TCE as a PC is extremely broad, with equal probability of PC or FP.  Thus it is expected that such adjacent cells with single TCEs will have vetting completeness 0 and 1, with equal likelihood.  Cells with no TCEs are gracefully handled because they have zero probability of a vetted PC.  This allows the use of fine grids that can detect details of the functional dependence of vetting completeness.

By partitioning the expected MES -- period space with a grid and computing $\rho$ in each grid cell, we can measure the dependence of $\rho$ on expected MES and period, inferring the function $\rho(p, m)$.  This is what we do in the next section.

\subsubsection{Vetting Completeness for the GK Baseline}

\begin{figure*}[htp]
   \centering
   \includegraphics[width=\linewidth]{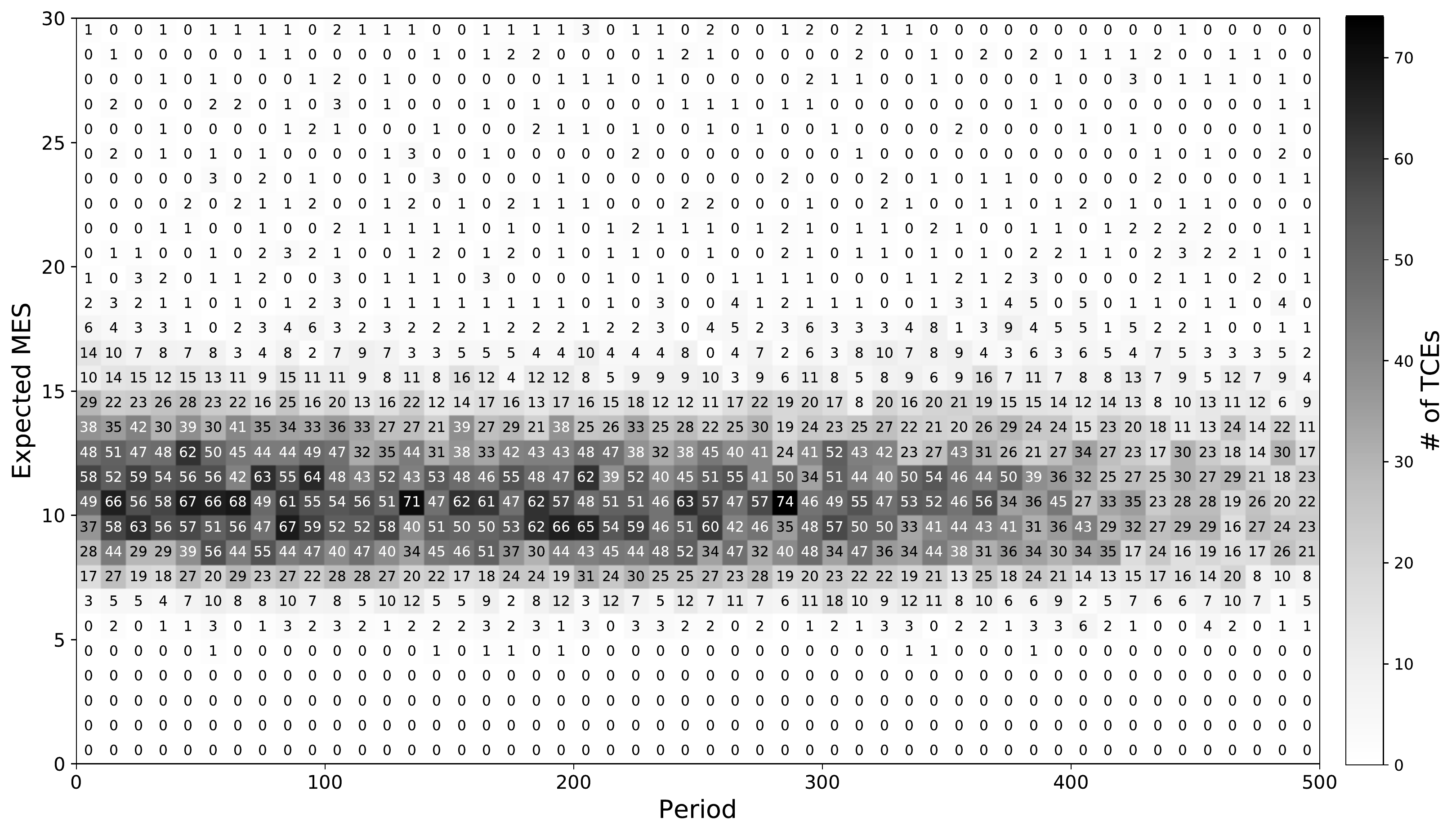} 
   \caption{The number of TCEs per cell found in the injected data.} \label{figure:nInjTCEs}
\end{figure*}

\begin{figure*}[htp]
   \centering
   \includegraphics[width=\linewidth]{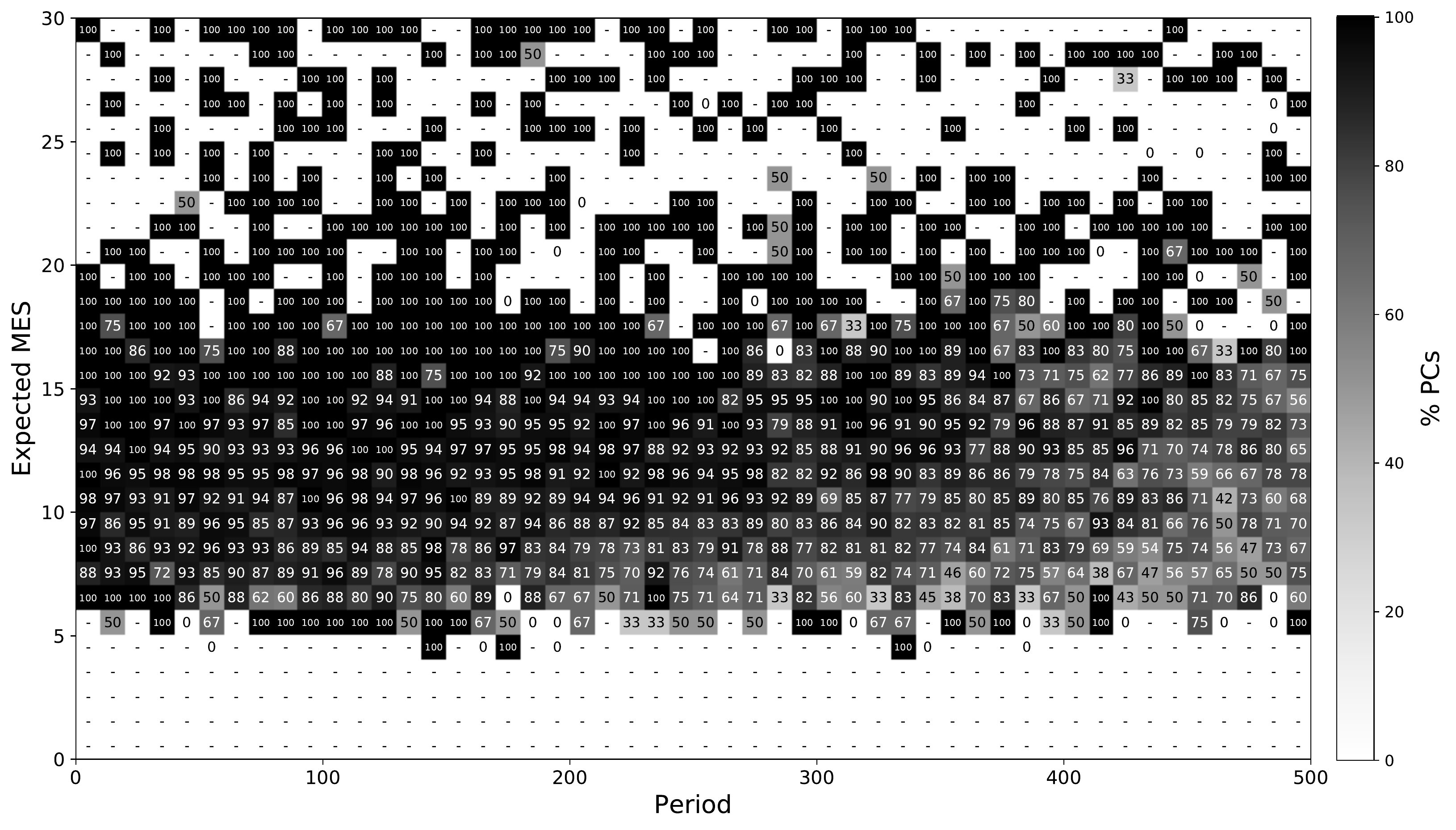} 
   \caption{The measured rate of correctly vetted injected PCs, measuring vetting completeness, displaying the functional dependence of the rate on period and expected MES.  Cells with no detected injected TCEs are marked with `-'.} \label{figure:injPCRate}
\end{figure*}

Figure~\ref{figure:nInjTCEs} shows the number of TCEs in each grid cell detected by the \Kepler\ pipeline in the injected data at the correct ephemeris in the GK baseline stellar population. The injections were with expected MES between about 8 and 15, and period less than 500 days \citep[see][for details]{Christiansen2017}. Figure~\ref{figure:injPCRate} shows the percentage of TCEs in each cell that were vetted as PC by the robovetter.  Perfect completeness is 100\%.  We see that for high expected MES and period $<$ 200 days the completeness in each cell is typically near 100\%, while for period $>$ 300 days and expected MES $<$ 15 the completeness drops off.  We will characterize this behavior using a function $\rho(p, m, \boldsymbol{\theta})$ of planet period $p$ and expected MES $m$, where the exact form of $\rho$ is specified below and $\boldsymbol{\theta}$ is the vector of function parameters. Given a specific form for $\rho$, we infer $\boldsymbol{\theta}$ from the number of PCs that are correctly vetted as PC in each cell.


\begin{figure*}[ht]
   \centering
   \includegraphics[width=\linewidth]{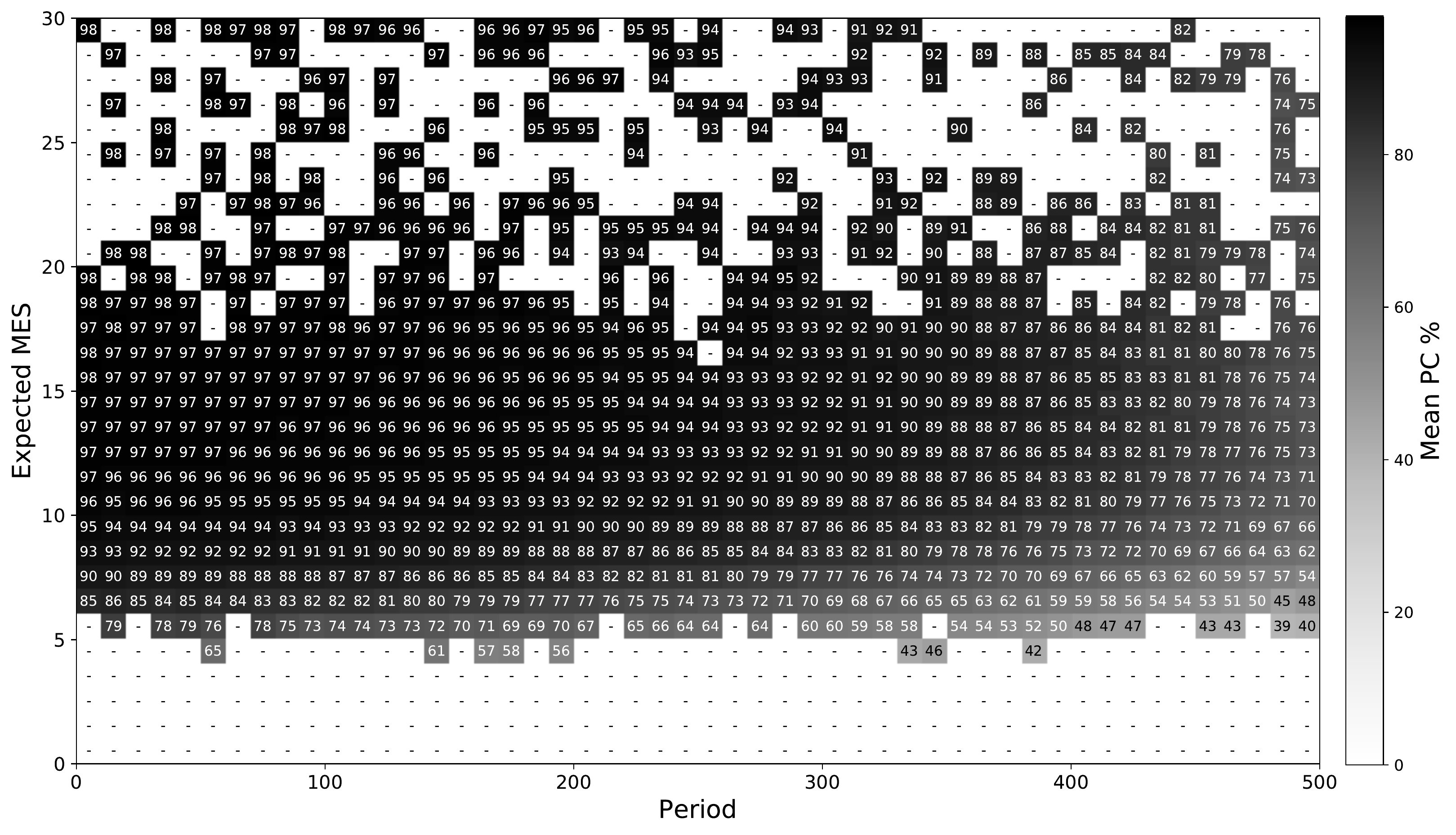} \\
   \includegraphics[scale=0.54]{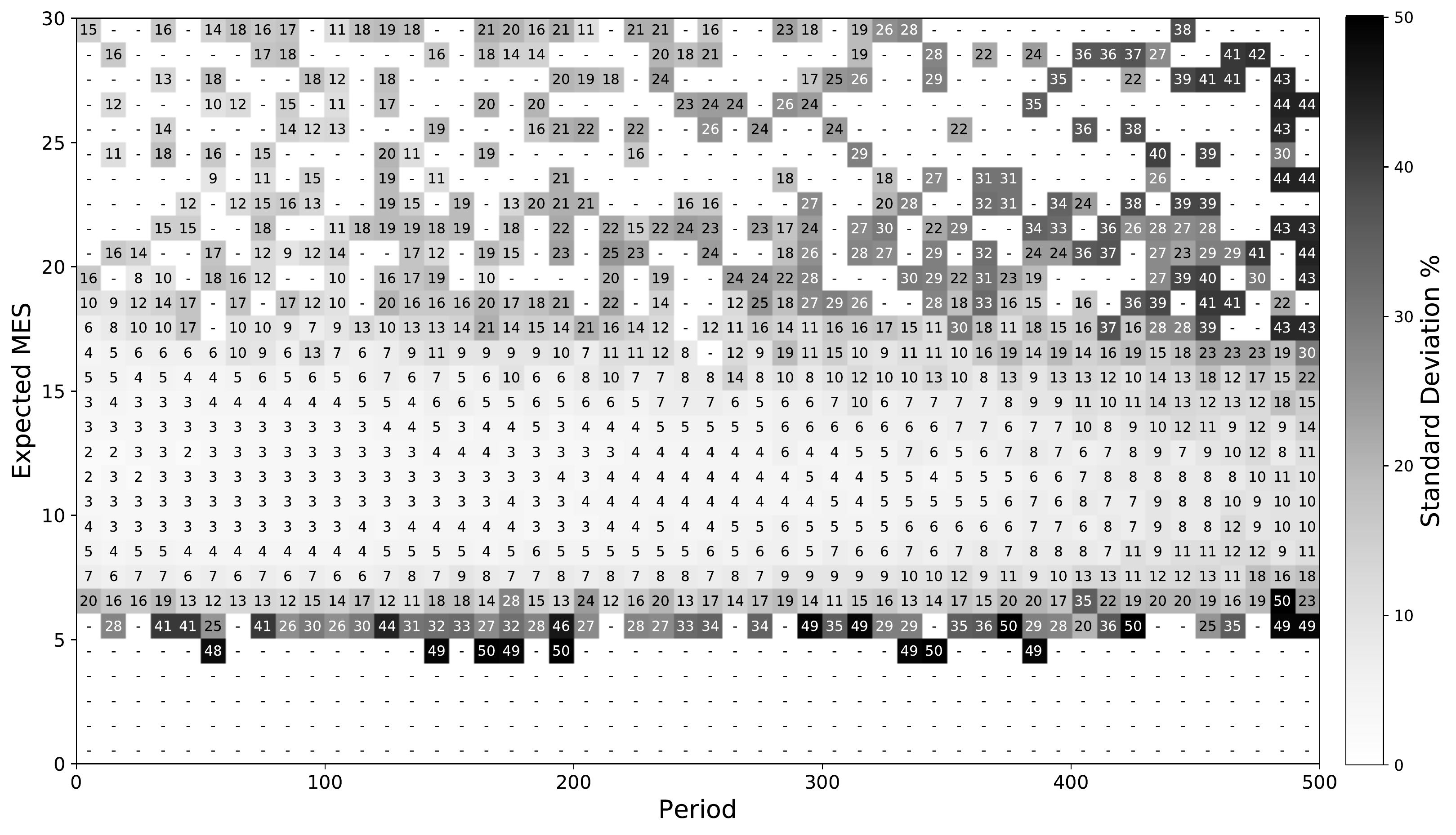} 
   \caption{
   The mean and standard deviation of 1000 realizations of the binomial completeness model $\rho(p_i,m_j,\boldsymbol{\theta})$ in Equation~\ref{eqn:completenessSolution}, where each realization $\boldsymbol{\theta}$ is drawn uniformly from the posterior distribution of $\rho(p_i,m_j,\boldsymbol{\theta})$.  We expect the observed completeness in Figure~\ref{figure:injPCRate} to be a realization of this model. Top: the mean completeness, showing an overall pattern similar to Figure~\ref{figure:injPCRate}.  Bottom: The standard deviation of the 1000 realizations, showing a similar variation to Figure~\ref{figure:injPCRate}.  The large cell-to-cell variation at low expected MES is due to the strong dependence of the binomial standard deviation on the number of TCEs in each cell ($n$ in Equation~\ref{eqn:binProb}), which is small at low expected MES (see Figure~\ref{figure:nInjTCEs}). 
   } \label{figure:compPostRealizations}
\end{figure*}

\begin{figure*}[ht]
   \centering
   \includegraphics[width=\linewidth]{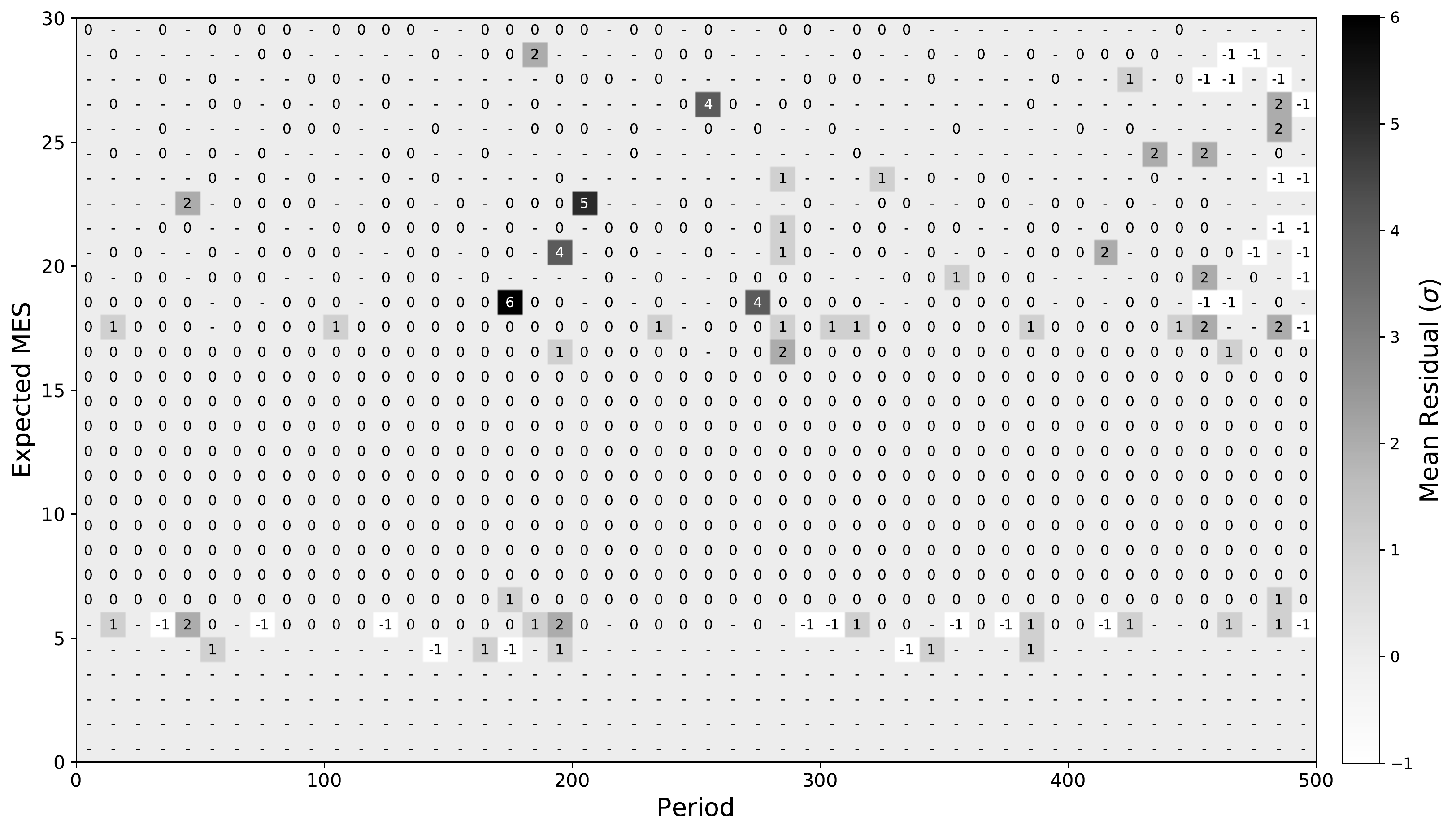} 
   \caption{The residuals of the measured completeness rate in Figure~\ref{figure:injPCRate} from the mean shown in Figure~\ref{figure:compPostRealizations}, normalized to the standard deviation in Figure~\ref{figure:compPostRealizations},  showing no significant bias.  The values are rounded for the nearest integer for clarity} \label{figure:compPostResiduals}
\end{figure*}

We conceptualize the determination of $\rho(p, m, \boldsymbol{\theta})$ as a binomial problem, thinking of each TCE as a draw from a population of PCs (= injected TCEs), which will be correctly vetted as PC by the robovetter with probability $\rho(p, m, \boldsymbol{\theta})$. If the number of TCEs in cell $(i,j)$ is $n_{i,j}$, then the probability of vetting $c_{i,j}$ TCEs correctly as PCs, given $\rho(p_{i,j}, m_{i,j}, \boldsymbol{\theta}) \equiv \rho_{i,j}(\boldsymbol{\theta})$, is given by the binomial distribution
\begin{equation} \label{eqn:binProb}
    P(c_{i,j}|n_{i,j},\boldsymbol{\theta}) = \binom{n_{i,j}}{c_{i,j}} \rho_{i,j}(\boldsymbol{\theta})^{c_{i,j}} \left(1-\rho_{i,j}(\boldsymbol{\theta}) \right)^{n_{i,j}-c_{i,j}}.
\end{equation}
For a particular choice of the functional form of $\rho$, equation~(\ref{eqn:binProb}) will be used to find the $\boldsymbol{\theta}$ that is most consistent with the number of PCs in each cell.

We will infer our rate function $\rho(p, m, \boldsymbol{\theta})$ via an MCMC Bayesian inference.  We treat each grid cell as independent identically distributed binomial realizations, which leads to the likelihood 
\begin{equation} \label{eqn:binLikelihood}
    L(\boldsymbol{c}, \boldsymbol{n}, \boldsymbol{\theta}) = \prod_{i,j} \binom{n_{i,j}}{c_{i,j}} \rho_{i,j}(\boldsymbol{\theta})^{c_{i,j}} \left(1-\rho_{i,j}(\boldsymbol{\theta}) \right)^{n_{i,j}-c_{i,j}}
\end{equation}
where $\boldsymbol{n} = \{n_{i,j}\}$ is the set of the number of injected TCEs in each cell, and $\boldsymbol{c} = \{c_{i,j}\}$ is the set of the number of injected TCEs vetted as PC in cell $(i,j)$.  

We perform the MCMC inference using the \emph{emcee} package\footnote{\url{https://emcee.readthedocs.io}}, which requires the log likelihood
\begin{equation} \label{eqn:binLogLikelihood}
\begin{split}
    \log(L) = &\sum_{i,j}
    \biggl[ 
    \log\binom{n_{i,j}}{c_{i,j}} + c_{i,j} \log\left(\rho_{i,j}(\boldsymbol{\theta}) \right)  \\
    &+  \left(n_{i,j}-c_{i,j} \right) \log \left(1-\rho_{i,j}(\boldsymbol{\theta}) \right) 
    \biggr].
\end{split}
\end{equation}

We considered several functional forms for $\rho(p, m, \boldsymbol{\theta})$, described in Appendix~\ref{app:modelSelect}.  Figure~\ref{figure:injPCRate} suggests a product of functions that are approximately, but not exactly, coordinate aligned. Qualitatively, Figure~\ref{figure:injPCRate} also suggests that a generalized logistic function 
\begin{equation} \label{eqn:logistic}
Y\left(x, x_0, k, \nu \right) = \left[1 + \exp( -k ( x - x_0 ) ) \right]^{-\frac{1}{\nu}}
\end{equation}
may be a good fit.  We construct many, though not all, of the functional forms considered in this paper from this generalized Logistic function.

\begin{figure}[ht]
   \centering
   \includegraphics[width=\linewidth]{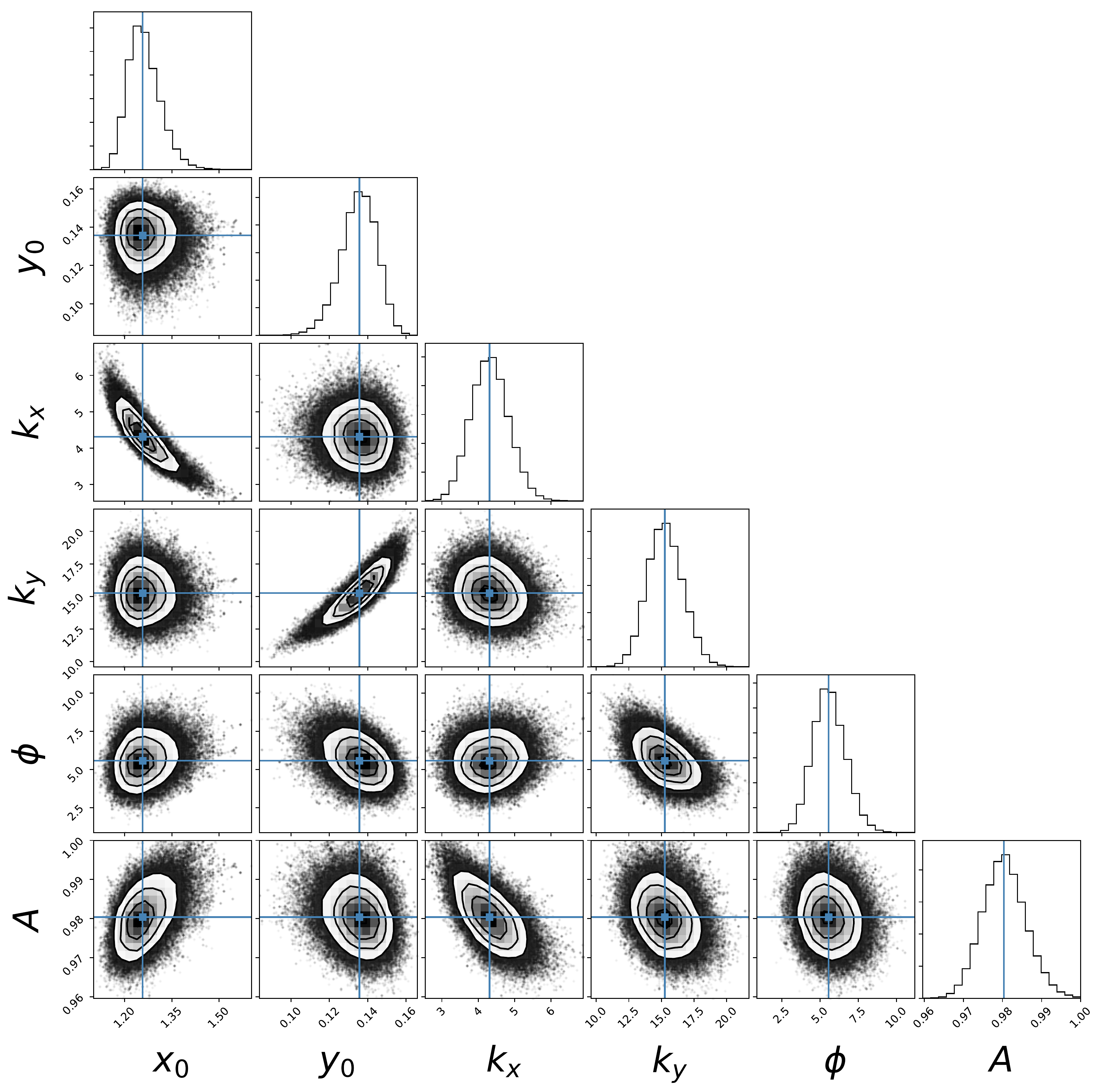}
   \caption{Posterior distributions for the components of the vetting completeness rate function parameters $\boldsymbol{\theta}$.  The straight lines indicate the median values.} \label{figure:compPost}
\end{figure}

Appendix~\ref{app:modelSelect} describes our use of the Akiake Information Criterion (AIC) and other considerations to select the form of $\rho$ that best fits the data.  In all cases, before applying the function we transform from (period, expected MES) coordinates to homogeneous coordinates on the unit square $[0,1] \times [0,1]$, which allows rotation.  Of the functions we considered, we find that a product of a non-rotated simplified logistic function in period $p$ times a rotated logistic in $p$ and expected MES $m$ to best fit the data: for $\boldsymbol{\theta} = [x_0, y_0, k_x, k_y, \phi, A]$, 
\begin{equation} \label{eqn:finalRho}
\begin{split}
    x =& \frac{\left( p - p_{\min}\right)}{\left( p_{\mathrm{max}} - p_{\min}\right)} \\
    y =& \frac{\left( m - m_{\min}\right)}{\left( m_{\mathrm{max}} - m_{\min}\right)} \\
    y_{\mathrm{rot}} =& (y - 0.5)*\cos(\phi) - (x - 0.5)*\sin(\phi) \\
    \rho =& A ~ Y\left(x, x_0, -k_x, 1 \right) \\
    &\times Y\left(y_{\mathrm{rot}} + 0.5, y_0, k_y, 1 \right). 
\end{split}
\end{equation}
We used the uniform priors $-1 \leq x_0, y_0 \leq 2$, $10^{-4} < k_x, k_y < 10^{4}$, $-180 < \phi < 180$, $0 < A < 1$, and initialized $\boldsymbol{\theta}$ by minimizing $-\log(L)$ using the Python \emph{optimize} package. Our MCMC computation used 100 walkers, and ran for 5000 steps after 5000 steps of burn-in. Figure~\ref{figure:compPost} shows the resulting posteriors, 
giving $\boldsymbol{\theta} = [x_0, y_0, k_x, k_y, \phi, A]$ as
\begin{equation} \label{eqn:completenessSolution}
\begin{split}
    x_0 &= 1.257^{+0.056}_{-0.044}, \qquad y_0 = 0.136^{+0.009}_{-0.010} \\
    k_x &= 4.311^{+0.523}_{-0.503}, \qquad k_y = 15.259^{+1.399}_{-1.305} \\
    \phi &= 5.566^{+1.106}_{-0.998}, \qquad A = 0.980^{+0.006}_{-0.005}
\end{split} \nonumber
\end{equation}
where the central values are the posterior median and the $+$ and $-$ errors are for the 84\textsuperscript{th} and 16\textsuperscript{th} percentiles. 
We denote the median parameter vector by $\boldsymbol{\Bar{\theta}}$.  The correlations between parameters apparent in Figure~\ref{figure:compPost} reflect the fact that the parameters of the logistic function are not fully independent.

\begin{figure*}[ht]
   \centering
   \includegraphics[width=\linewidth]{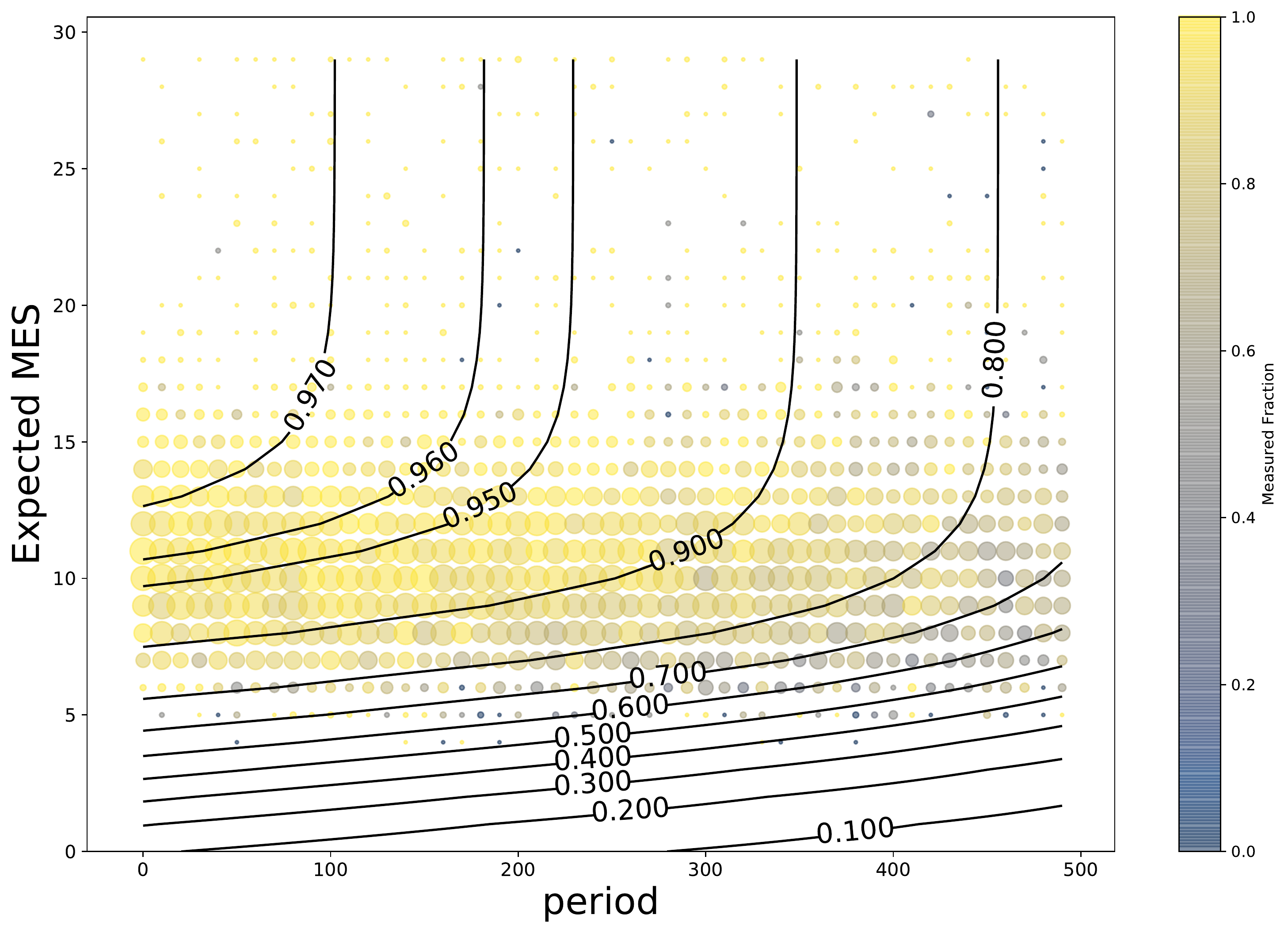}
   \caption{Contours of the vetting completeness rate function $\rho(p_i,m_j,\boldsymbol{\Bar{\theta}})$ for the median of the posteriors.  The colored shapes show the measured data in each grid cell, with the color indicating the measured rate, and the size indicating the number of TCEs in the cell.} \label{figure:compPostContours}
\end{figure*}

\begin{figure}[ht]
   \centering
   \includegraphics[width=\linewidth]{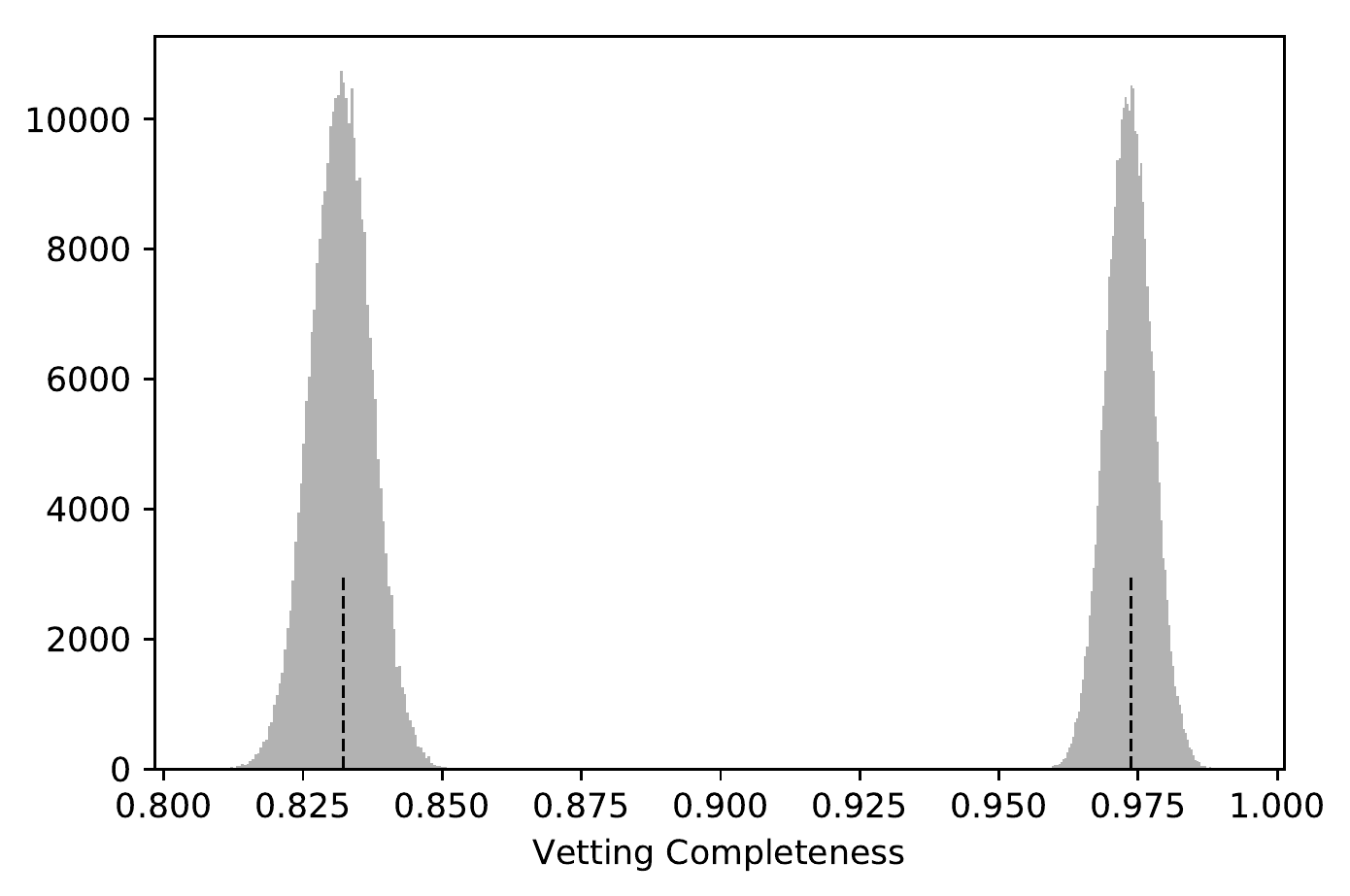}
   \caption{The vetting completeness rate function $\rho(p_i,m_j,\boldsymbol{\Bar{\theta}})$ evaluated with the posterior distribution. Right distribution: period = 50 days and expected MES = 25.  Left distribution: period = 365 days and expected MES = 10.  The dashed lines show the rates for the median $\boldsymbol{\Bar{\theta}}$.} \label{figure:compPostExamples}
\end{figure}

Figure~\ref{figure:compPostContours} shows the resulting rate function $\rho(p_i,m_j,\boldsymbol{\Bar{\theta}})$ evaluated at the median of the posteriors $\boldsymbol{\Bar{\theta}}$, along with the underlying rates for each grid cell.  Figure~\ref{figure:compPostExamples} shows two example positions on the expected MES-orbital period plane, illustrating both the dependence of vetting completeness on these parameters as well as the spread of vetting completeness due to the posterior $\boldsymbol{\theta}$ distribution.  We find that the approach described in this section is robust against changes in the grid.  Changing the grid resolution does not significantly change the results, so long as the resolution is sufficient to resolve features in the underlying data.

Figure~\ref{figure:compPostRealizations} shows the mean and standard deviation of 1000 realizations, drawn from the posterior $\boldsymbol{\theta}$ distribution, of the fraction of correctly vetted PCs for each cell. This figure should be compared with Figures~\ref{figure:nInjTCEs} and \ref{figure:injPCRate}.  As expected, where the number of TCEs per cell is low in Figure~\ref{figure:nInjTCEs}, the standard deviation is high.  Figure~\ref{figure:compPostResiduals} shows the residual of observed data in Figure~\ref{figure:injPCRate} from the mean rate in Figure~\ref{figure:compPostRealizations} in units of the standard deviation shown in Figure~\ref{figure:compPostRealizations}.  We see that while there are isolated large outliers, as well as larger residual values where the standard deviation is high, there is no indication of a bias in $\rho(p_i,m_j,\boldsymbol{\theta})$.  Additional details, including further characterization of the quality of our fit of $\rho(p_i,m_j,\boldsymbol{\theta})$, are found in the {\it htmlArchive} folders on the paper GitHub site\footref{footnote:github}.

\section{The Reliability Model} \label{section:reliabilityModel}
In this section we characterize the reliability of planet candidates in the Q1--Q17 DR25 KOI catalog.  We apply the probabilistic approach of \S\ref{section:vettingCompleteness} to the problem of characterizing the probability that a DR25 planet candidate is in fact a false alarm due to instrumental systematics, or some types of stellar variability.  We rely on the Q1--Q17 DR25 False Positive Probabilities table at the NASA Exoplanet Archive\footref{footnote:exoplanetArchive} \citep{Morton2016} to provide the probability that the planet candidate is a false positive due to astrophysical signals that imitate transits.  The {\it final reliability} for each planet candidate is the product of the false alarm reliability and the false positive probability.

\subsection{Vetting Reliability} \label{section:vetReliability}

\citet{Thompson2018} defined reliability as the ratio of the number of PCs which are true exoplanets, $T_{\mathrm{PC}}$, to the number of observed planet candidates $N_{\mathrm{PC}}$: 
\begin{equation} \label{eqn:reliability}
    R \equiv \frac{T_{\mathrm{PC}}}{N_{\mathrm{PC}}} 
        = 1 - \frac{N_{\mathrm{FP}}}{N_{\mathrm{PC}}} 
        \left( \frac{1-E}{E} \right), 
\end{equation}
where $N_{\mathrm{FP}}$ is the number of observed false positives and $ E \equiv \frac{N_{\mathrm{FP}}}{T_{\mathrm{FP}}}$ is the \emph{false positive effectiveness}, defined as the number of identified FPs, $N_{\mathrm{FP}}$, divided by the number of true FPs, $T_{\mathrm{FP}}$.  The second equality in equation~(\ref{eqn:reliability}) is exact when all quantities are from the same population, such as the observed data analyzed by the DR25 catalog.  Unfortunately, the true PCs and false positives, $T_{\mathrm{PC}}$ and $T_{\mathrm{FP}}$, are unknown for the observed data.  As explained in \citet{Thompson2018}, however, we can use the \emph{inverted} and \emph{scrambled} data sets\footref{footnote:simulatedData} described in \S\ref{section:introduction}, which are designed so that every detection is, by definition, a false alarm.

Astrophysical transit-like events such as KOIs or eclipsing binaries can trigger detections in the inverted and scrambled data, compromising the use of these data to measure false alarms. This happens in two ways: 1) the transits and eclipses add signals unlike the false alarms we are trying to measure, and 2) the Robovetter is not tuned to detect and remove these kinds of signals. \citet{Thompson2018} describes how the lists of inverted and scrambled detections were cleaned of signals from known transiting systems in \S2.3.3. Essentially, targets that are known binaries \citep{Kirk2016} and known KOIs \citep{Thompson2018} are removed from the list of detections, so they do not count as either a FP or a PC.  For the inverted set, because self-lensing and heartbeat star binaries type events can produce signals that look like inverted transits, 54 targets with significant periodic signals were also removed from the list.  The detections dropped from the inverted and scrambled data used in this study, as well as in \citet{Thompson2018}, are collected in the files {\it kplr\_droplist\_inv.txt} and {\it kplr\_droplist\_scr*.txt} (one for each scrambled data set) on the paper GitHub site\footref{footnote:github}. These stars are removed from the inverted/scrambled data before the analysis described in this section.

The inverted and scrambled data are designed to measure false alarms, not all false positives.  Thus, we must take care to restrict the formula for reliability in equation~(\ref{eqn:reliability}) to the population of false alarms.  This implies that equation~(\ref{eqn:reliability}) becomes 
\begin{equation} \label{eqn:FAreliability}
    R_{\mathrm{FA}} = 1 - \frac{N_{\mathrm{FA}}}{N_{\mathrm{notFA}}} 
        \left( \frac{1-E_{\mathrm{FA}}}{E_{\mathrm{FA}}} \right), 
\end{equation}
where FA indicates ``false alarm".  $N_{\mathrm{FA}}$ is the number of identified false alarms in the observed data (determined via the NTL flag in the KOI table) and $N_{\mathrm{notFA}}$ is the number of transit detections that are not vetted as false alarms.  $E_{\mathrm{FA}} = \frac{N_{\mathrm{FA}}}{T_{\mathrm{FA}}}$ is the {\it false alarm effectiveness}, the fraction of true false alarms $T_{\mathrm{FA}}$ (assumed to be all detections in the inverted/scrambled data) that are vetted as false positives in the scrambled and inverted data.  Equation~(\ref{eqn:FAreliability}) describes only false alarms measured by the inverted and scrambled data.  As described in \ref{section:totalReliability}, we will multiply this reliability against instrumental false alarms with the reliability against astrophysical false alarms (constructed using the Q1-Q17 False Positive Probabilities).

In order to apply the probabilistic approach developed in \S\ref{section:vettingCompleteness}, we rephrase the reliability formula in terms of rates rather than numbers by defining the fractions $F_\mathrm{FA} = N_{\mathrm{FA}}/N_{\mathrm{TCE}}$ and $F_\mathrm{notFA} = N_{\mathrm{notFA}}/N_{\mathrm{TCE}}$, where $N_{\mathrm{TCE}}$ is the number of TCEs in the observed data. We call $F_\mathrm{FA}$ the {\it observed false positive rate}.  We can substitute $N_{\mathrm{FA}}/N_{\mathrm{notFA}} = F_{\mathrm{FA}}/F_{\mathrm{notFA}}$ in equation~(\ref{eqn:FAreliability}).  We further assume that notFA and FA are a complete partition of all TCEs in the observed data, so $N_{\mathrm{FA}} + N_{\mathrm{notFA}} = N_{\mathrm{TCE}} \Rightarrow F_{\mathrm{FA}} + F_{\mathrm{notFA}} = 1$ and we can eliminate $N_{\mathrm{notFA}}$ from equation~(\ref{eqn:FAreliability}):
\begin{equation} \label{eqn:reliabilityRate}
    R_{\mathrm{FA}} = 1 - \frac{F_{\mathrm{FA}}}{1-F_{\mathrm{FA}}} 
        \left( \frac{1-E_{\mathrm{FA}}}{E_{\mathrm{FA}}} \right).
\end{equation}
We can now treat $F_\mathrm{FA}$ and $E_{\mathrm{FA}}$ as rates that determine the probability of drawing a TCE that will be vetted as a false alarm.  As in \S\ref{section:vettingCompleteness} we will treat $F_\mathrm{FA}$ and $E_{\mathrm{FA}}$ as rates in two separate binomial problems, with functional forms that depend on period and observed MES and whose coefficients are determined via an MCMC inference. 
Additional details, including further characterization of the quality of our rate fits, are found in the {\it htmlArchive} folders on the paper GitHub site\footref{footnote:github}.

\begin{figure*}[ht]
  \centering
  \includegraphics[width=\linewidth]{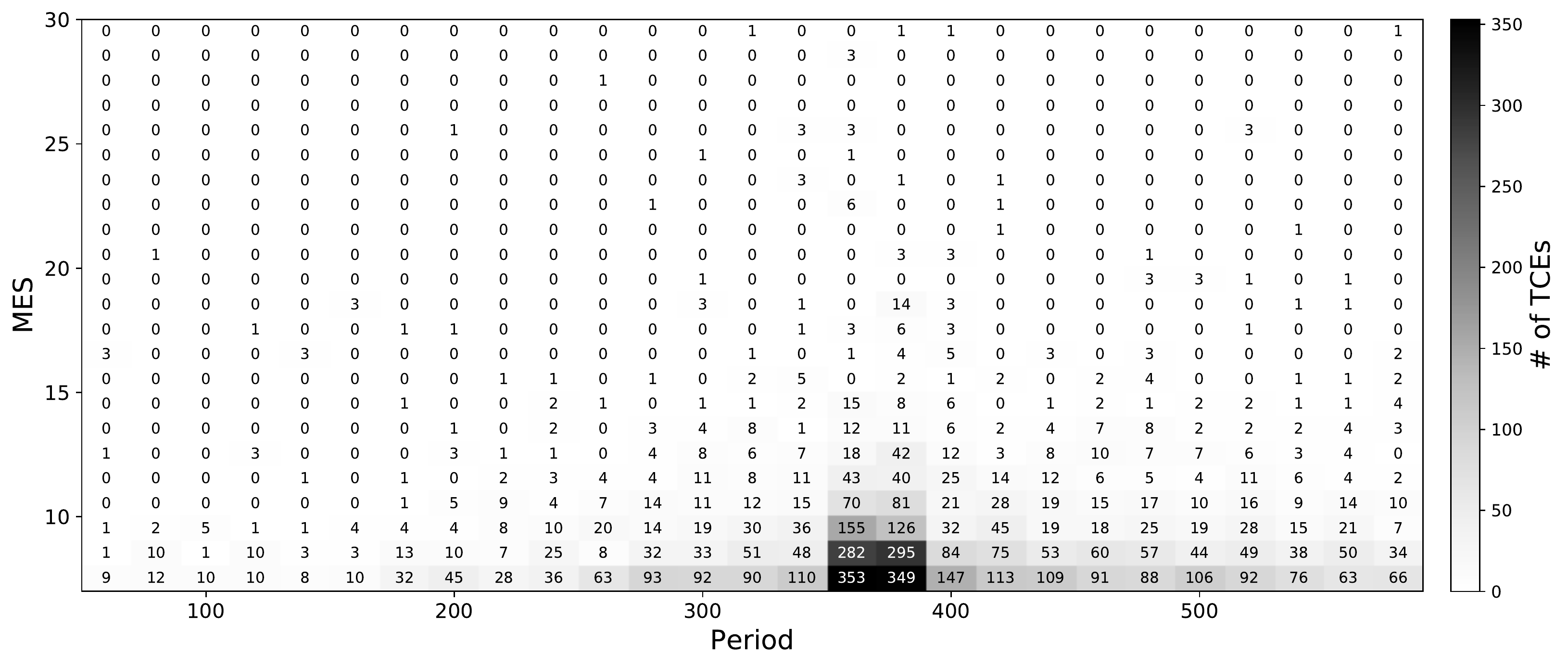} 
  \caption{The number of TCEs per cell found in the combined inverted/scrambled data.  The large number of TCEs at period $\approx 370$ days is the excess of detections due to instrumental false alarms shown in Figure~\ref{figure:periodSnrDist}, discussed in \S\ref{section:vettingIntro}.} \label{figure:nInvScrTCEs}
\end{figure*}

\begin{figure*}[ht]
  \centering
  \includegraphics[width=\linewidth]{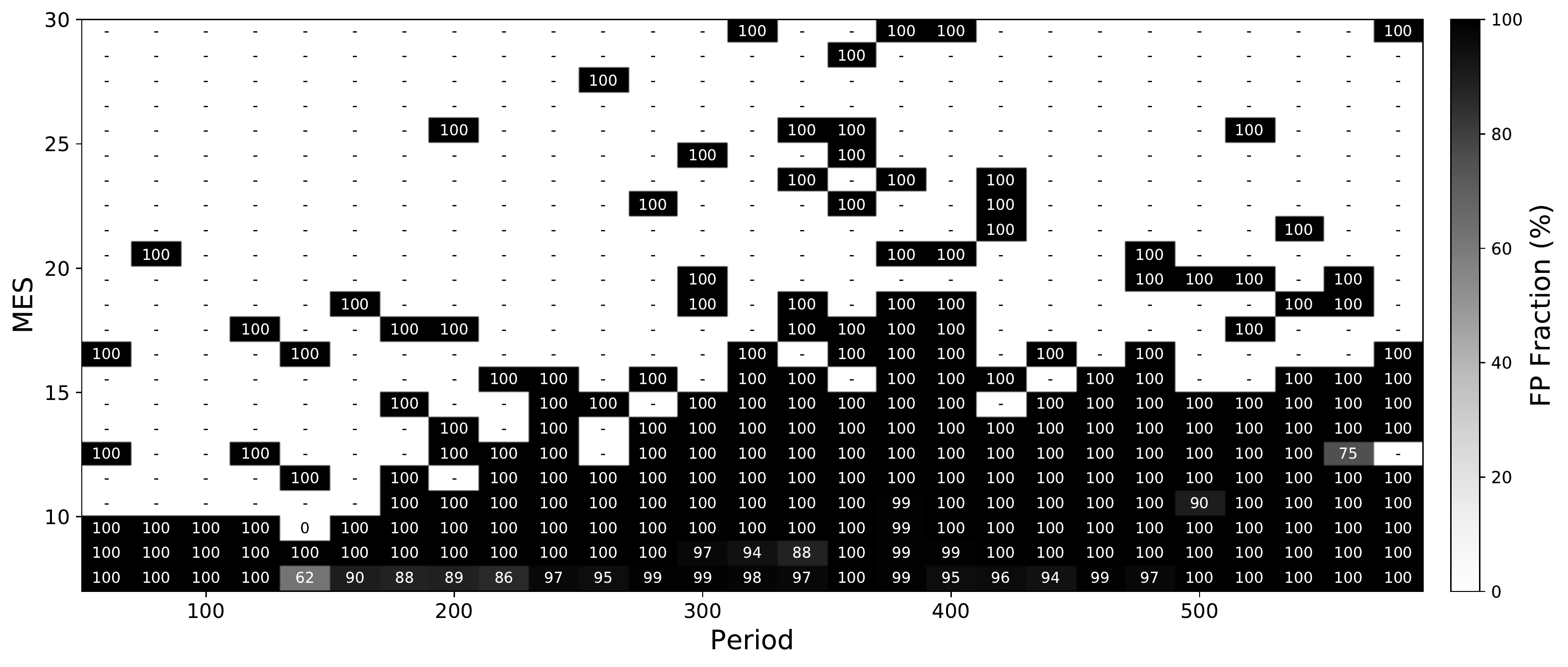} 
  \caption{The measured rate of correctly vetted inverted/scrambled FPs, which is a direct measurement of false alarm effectiveness $E_{\mathrm{FA}}$.  Cells with no detected TCEs are marked with `-'.} \label{figure:invScrFPRate}
\end{figure*}

\subsubsection{Characterization of False Alarm Effectiveness $E_{\mathrm{FA}}$} \label{section:FPEcharacterization}
To determine the false alarm effectiveness, $E_{\mathrm{FA}}$, we combine the inverted data with each of the three scrambled data sets \citep[see][for details on the properties of each set]{Coughlin2017} to create three data sets, called ``inverted/scrambled'', where every TCE should be considered a false alarm.  We proceed as in \S\ref{section:vettingCompleteness}, covering the period-(observed) MES plane with a regular grid, and measure the ratio of the number of false alarms to the number of TCEs in each cell. This problem is more challenging than the analysis of vetting completeness in \S\ref{section:vettingCompleteness} because the Robovetter has been tuned to do a very good job correctly identifying false alarms, resulting in relatively few cells with TCEs incorrectly vetted as PCs.  We therefore combine the three inverted/scrambled data sets through concatenation in order to produce a somewhat stronger signal.  This amounts to averaging the three data sets at the input level, avoiding small-number statistics issues that would arise if we fit the three data sets separately and averaging the resulting posteriors.  We refer to this concatenated data set as the combined inverted/scrambled data.

Figure~\ref{figure:nInvScrTCEs} shows the number of TCEs detected in the combined inverted/scrambled data.  We see that most detected TCEs in this data is for MES~$<15$ and period~$\geq$~250~days. Figure~\ref{figure:invScrFPRate} shows the fraction of correctly vetted false alarms, a measure of $E_{\mathrm{FA}}$, in each cell.  The signal we're measuring is small, and is dominated by smaller fractions at low MES and period $\leq 200$ days.

\begin{figure}[ht]
   \centering
   \includegraphics[width=\linewidth]{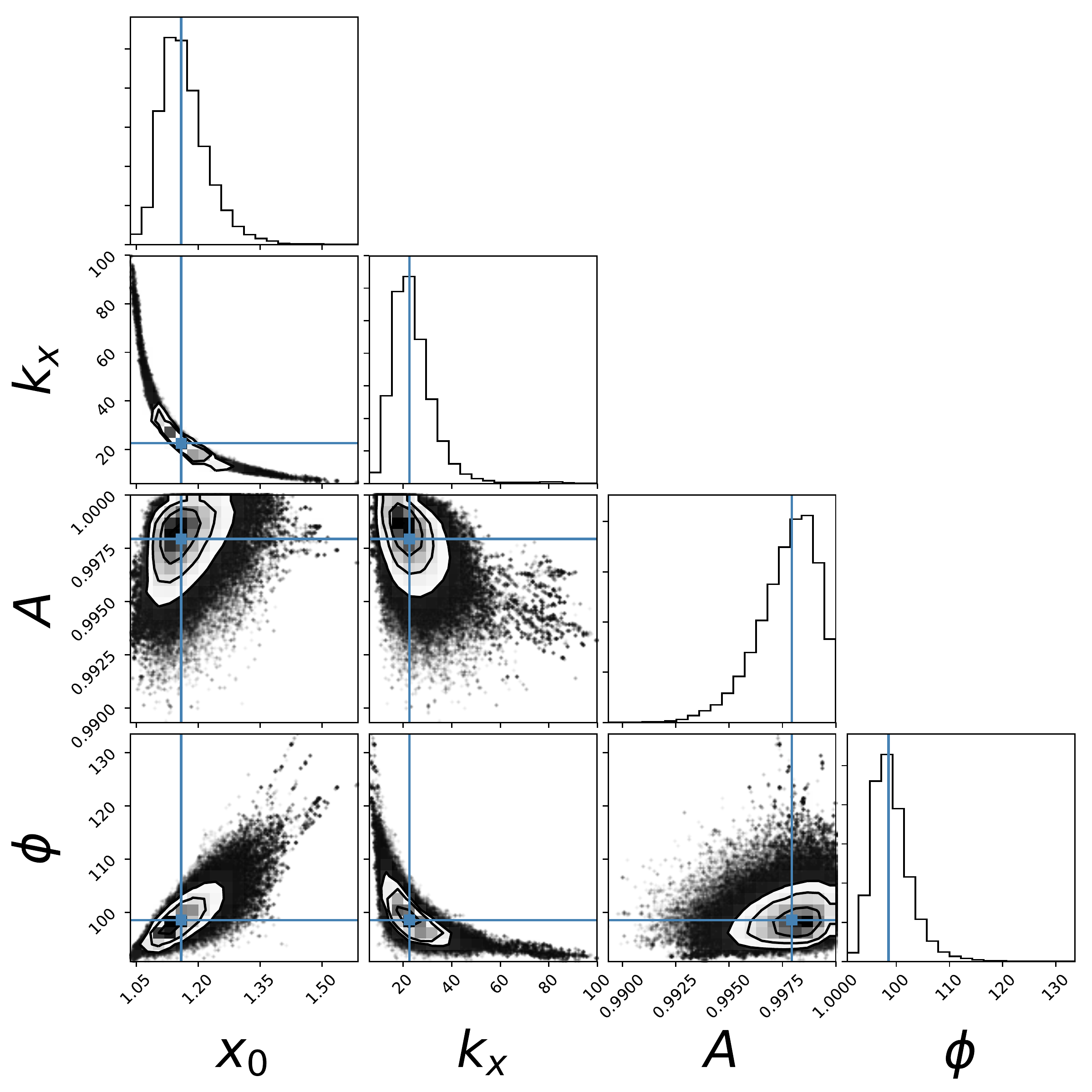}
   \caption{Posterior distributions for the false alarm effectiveness $E_{\mathrm{FA}}$ rate function parameters $\boldsymbol{\theta}$. The straight lines indicate the median values.} \label{figure:fpEPost}
\end{figure}

\begin{figure*}[hb]
  \centering
  \includegraphics[width=\linewidth]{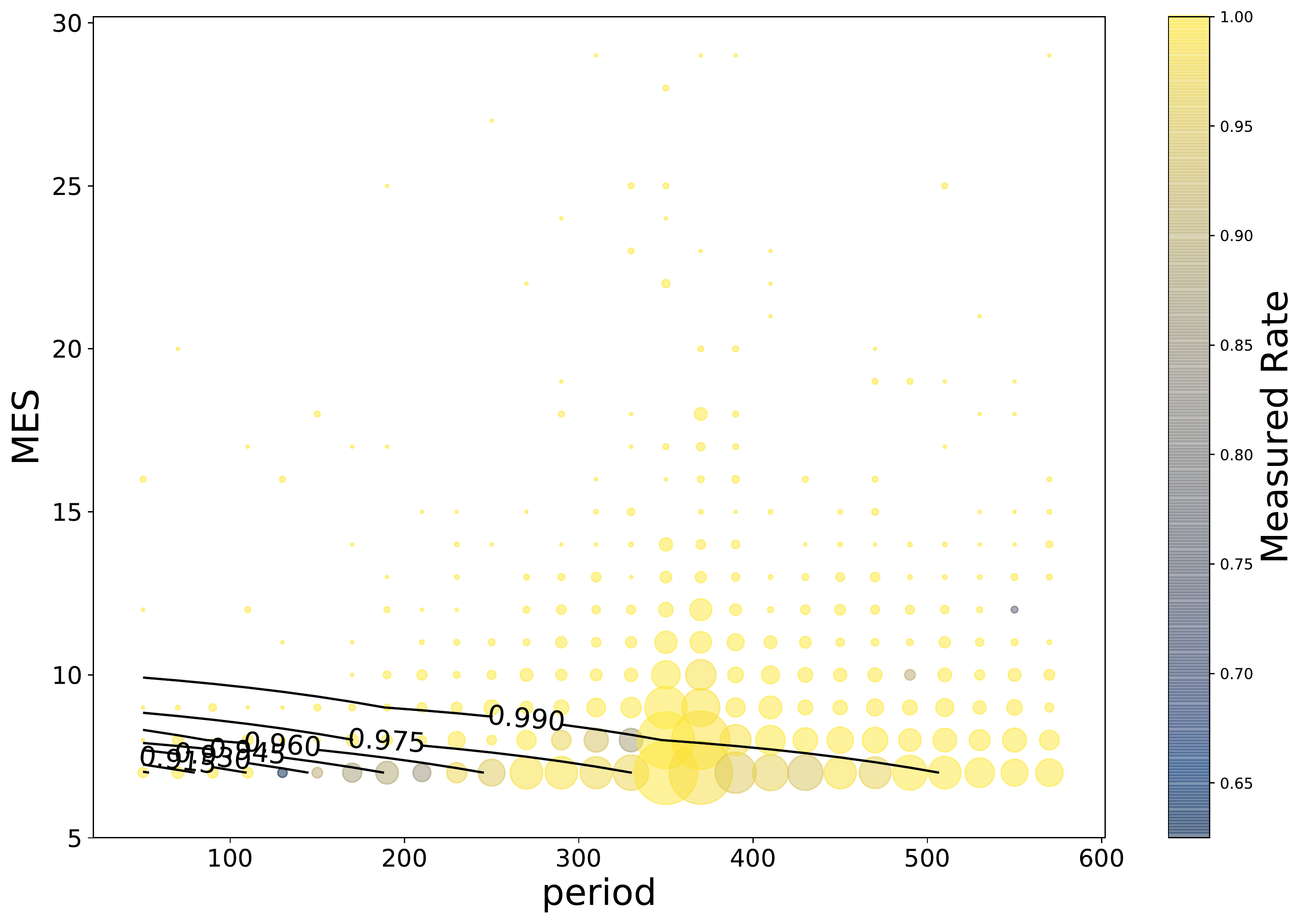}
  \caption{Contours of the false alarm efficiency rate function $E_{\mathrm{FA}}(p_i,m_j,\boldsymbol{\Bar{\theta}})$ for the median of the posteriors.  The colored shapes show the measured data in each grid cell, with the color indicating the measured rate, and the size indicating the number of TCEs in the cell.} \label{figure:fpEPostContours}
\end{figure*}

\begin{figure*}[ht]
   \centering
   \includegraphics[width=\linewidth]{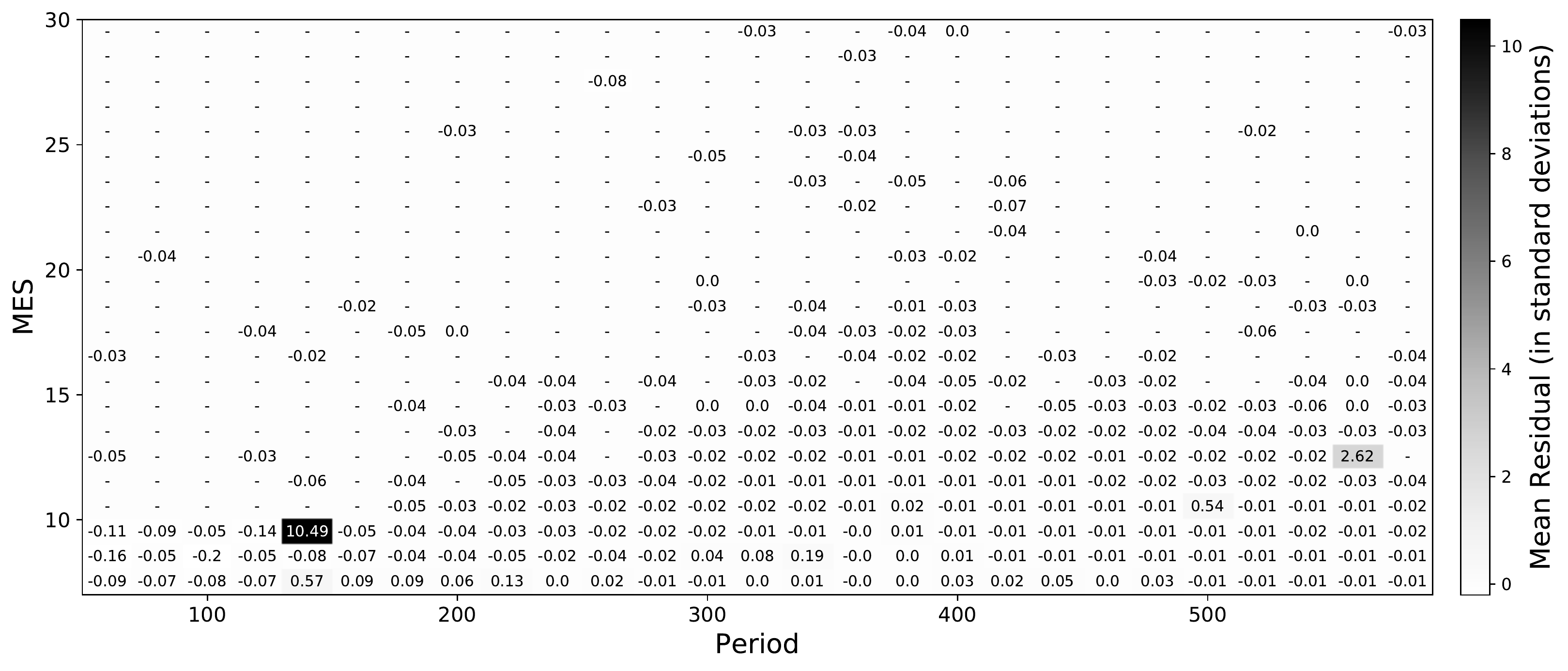} 
   \caption{The residuals of the measured $E_{\mathrm{FA}}$ rate in Figure~\ref{figure:invScrFPRate} from the mean normalized to the standard deviation showing no significant bias.} \label{figure:fpEPostResiduals}
\end{figure*}

We use the same likelihood as in \S\ref{section:vettingCompleteness}, equation~(\ref{eqn:reliability}), where in this case $E_{\mathrm{FA}}$ plays the role of $\rho$, $n_{i,j}$ is the number of TCEs detected in the combined inverted/scrambled data in cell $(i,j)$, and $c_{i,j}$ is the number of false positives identified  in cell $(i,j)$.  We perform the MCMC inference as described in \S\ref{section:vettingCompleteness}. We considered several functions, described in Appendix~\ref{app:fpEModelSelect}, and determined that a simple rotated logistic function best describes this data set.  For $\boldsymbol{\theta} = [x_0, k_x, \phi, A]$, $E_{\mathrm{FA}} \left(p, m, \boldsymbol{\theta}\right)$ is given by 
\begin{equation} \label{eqn:finalfpERho}
\begin{split}
    x =& \frac{\left( p - p_{\mathrm{min}}\right)}{\left( p_{\mathrm{max}} - p_{\mathrm{min}}\right)} \\
    y =& \frac{\left( m - m_{\mathrm{min}}\right)}{\left( m_{\mathrm{max}} - m_{\mathrm{min}}\right)} \\
    x_{\mathrm{rot}} =& (x - 0.5)*\cos(\phi) - (y - 0.5)*\sin(\phi) \\
    E_{\mathrm{FA}} =& A \, Y\left( x_{\mathrm{rot}} + 0.5, x_0, -k_x, 1 \right)
\end{split}
\end{equation}
where $p$ is the orbital period, $m$ is the observed MES, and $Y$ is the logistic function from equation~(\ref{eqn:logistic}).  We used the uniform priors $-1 \leq x_0 \leq 2$, $10^{-4} < k_x < 100$, $-180 < \phi < 180$, $0 < A < 1$.  The MCMC run used a hand-tuned initial condition because Python's optimize maximum likelihood solution was physically unreasonable ($A >> 1$, for example) and violated the prior. Our MCMC computation used 100 walkers, and ran for 5000 steps after 5000 steps of burn-in. Figure~\ref{figure:fpEPost} shows the resulting posteriors,  
giving $\boldsymbol{\theta} = [x_0, k_x, \phi, A]$ as 
\begin{equation} \label{eqn:fpeSolution}
\begin{split}
    x_0 &= 1.159^{+0.062}_{-0.044}, \qquad k_x = 22.587^{+8.811}_{-6.291}, \\
    \phi &= 98.551^{+3.834}_{-2.778}, \qquad A = 0.998^{+0.001}_{-0.002}.
\end{split} \nonumber
\end{equation}
The rate function $E_{\mathrm{FA}}(p_i,m_j,\boldsymbol{\Bar{\theta}})$ for the posterior median $\boldsymbol{\Bar{\theta}}$ is shown in Figure~\ref{figure:fpEPostContours}.  
As in \S\ref{section:vettingCompleteness}, 1000 realizations of the FP rate function were created, drawing from the posterior $\boldsymbol{\theta}$ distribution. The residuals of the observed false alarm fraction in Figure~\ref{figure:invScrFPRate} from the mean of these realizations in units of standard deviation is shown in Figure~\ref{figure:fpEPostResiduals}, demonstrating an overall good fit to the data.

\subsubsection{Characterization of Observed False Positive Rate $F_\mathrm{FA}$ } \label{section:obsFPcharacterization}

For $F_\mathrm{FA}$ we count the number of false positives found in the observed data, but we must be careful to consider only non-transit-like false alarms in order to be consistent with our characterization of effectiveness.  We identify such false alarms by selecting on the not-transit-like (NTL) flag = 0, indicating that the Robovetter identified this false positive as transit like, which identifies 21 candidate astrophysical false positives in our GK population inside $50 \leq \mathrm{period} \leq 600$ days and $0.5 \leq \mathrm{radius} \leq 15$ $R_{\oplus}$.  We manually examined these FPs and identified those that show a consistent astrophysical signal in all transits as astrophysical false positives.  Two TCEs with NTL=0 did not show such a consistent astrophysical signal and are deemed likely false alarms: 004371172-01 and 009394762-01.  The other 19 FPs with NTL=0 were identified as astrophysical and removed from the set of FPs used in the analysis of $F_\mathrm{FA}$.

\begin{figure*}[ht]
  \centering
  \includegraphics[width=\linewidth]{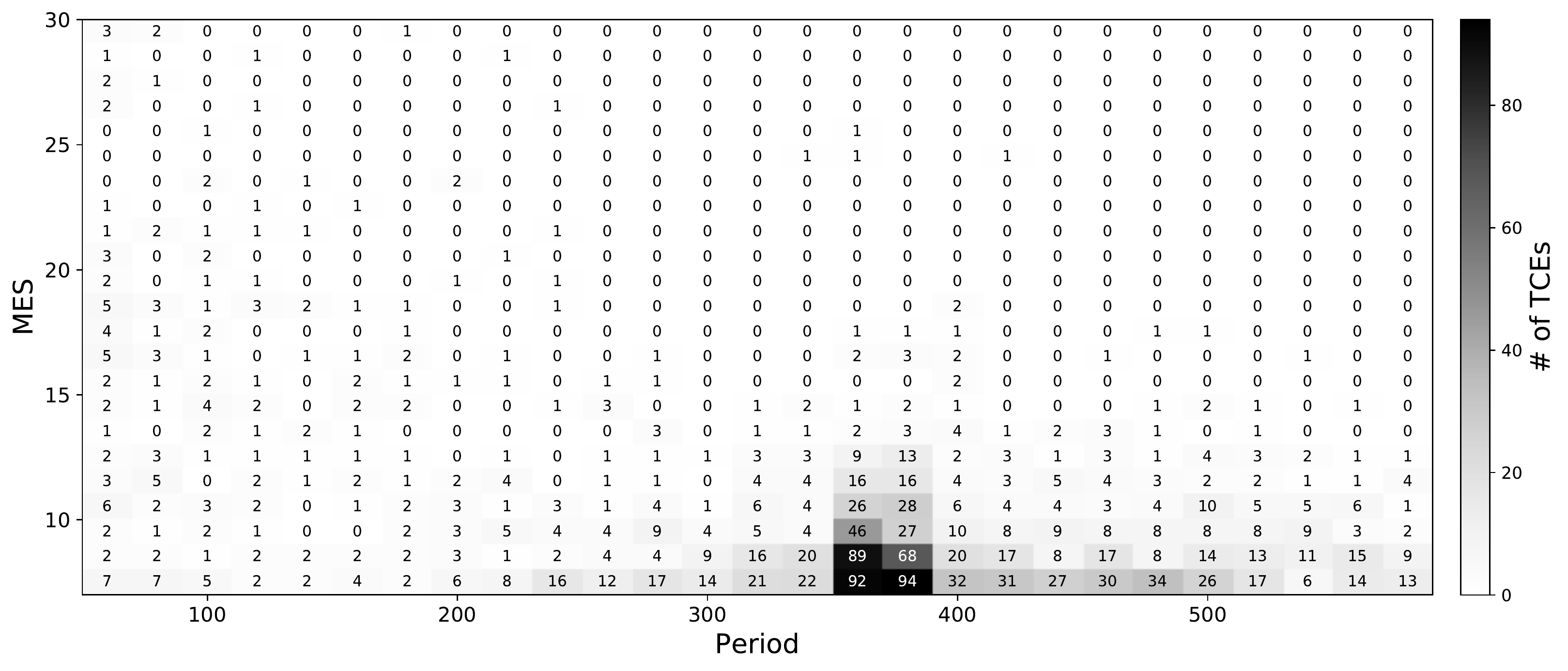} 
  \caption{The number of TCEs per cell found in the observed data. The large number of TCEs at period $\approx 370$ days is the excess of detections due to instrumental false alarms shown in Figure~\ref{figure:periodSnrDist}, discussed in \S\ref{section:vettingIntro}.} \label{figure:obsFpTCEs}
\end{figure*}

\begin{figure*}[ht]
  \centering
  \includegraphics[width=\linewidth]{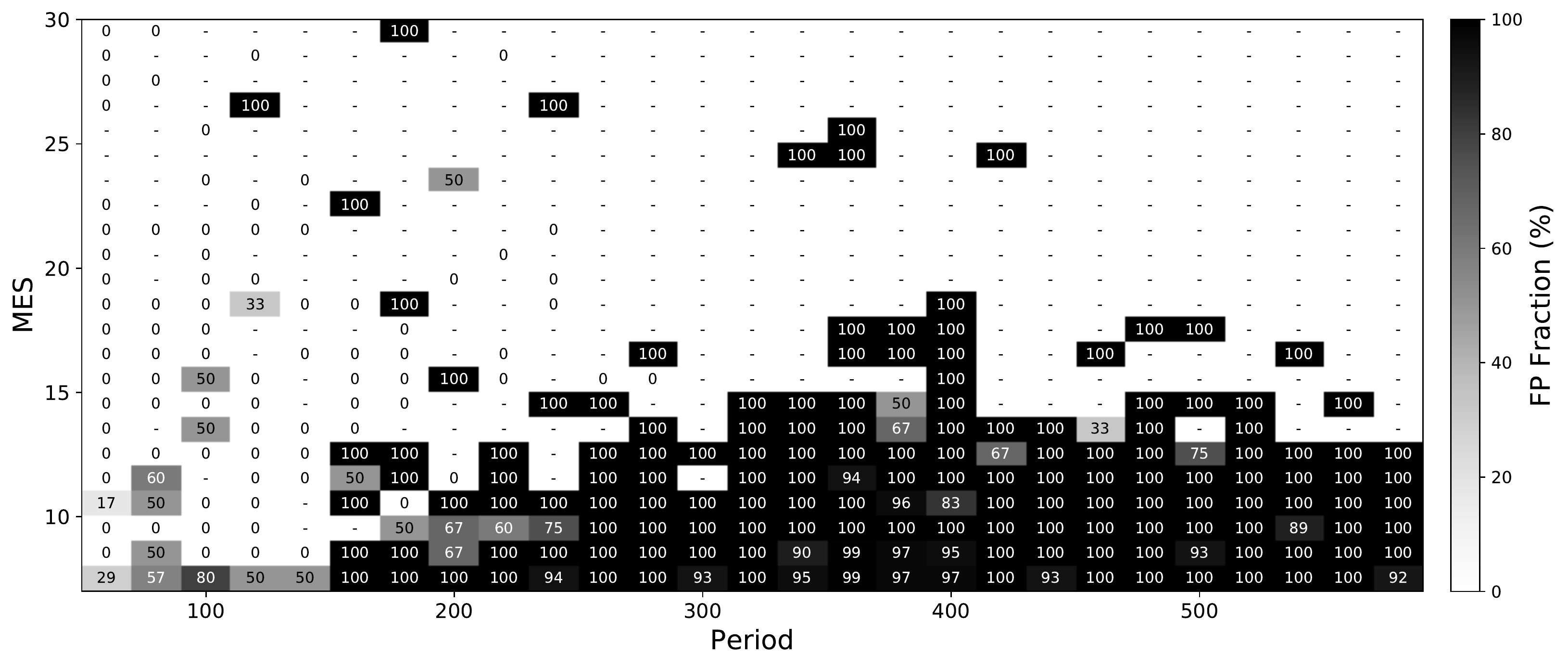} 
  \caption{The measured rate of identified false alarms in the observed data.  Cells with no detected TCEs are marked `-'.} \label{figure:obsFpRate}
\end{figure*}

Figure~\ref{figure:obsFpTCEs} shows the number of TCEs detected in the combined inverted/scrambled data.  We see that most detected TCEs in this data is for MES $<15$ and period $\geq 250$ days.  The close correspondence with Figure~\ref{figure:nInvScrTCEs} shows that most TCEs are false alarms. Figure~\ref{figure:obsFpRate} shows the fraction of identified false alarms (identified via the NTL flag as described above), a measure of $F_{\mathrm{FA}}$, in each cell.  

We proceed in a very similar manner to inferring $E_{\mathrm{FA}}$ in \S\ref{section:FPEcharacterization}.  We use equation~(\ref{eqn:reliability}) as the likelihood, with $F_{\mathrm{FA}}$ playing the role of $\rho$, $n_{i,j}$ is the number of TCEs detected in the observed data in cell $(i,j)$, and $c_{i,j}$ is the number of false alarms identified in cell $(i,j)$.  We perform the MCMC inference as described in \S\ref{section:vettingCompleteness}. We considered several functions, described in Appendix~\ref{app:obsFPModelSelect}, and determined that the same simple rotated logistic function as that used in \S\ref{section:FPEcharacterization} best describes this data set.  In equation~(\ref{eqn:finalfpERho}), $F_{\mathrm{FA}}$ replaces $E_{\mathrm{FA}}$, providing $F_{\mathrm{FA}} \left(p, m, \boldsymbol{\theta}\right)$ for $\boldsymbol{\theta} = [x_0, k_x, \phi, A]$.  

\begin{figure}[ht]
   \centering
   \includegraphics[width=\linewidth]{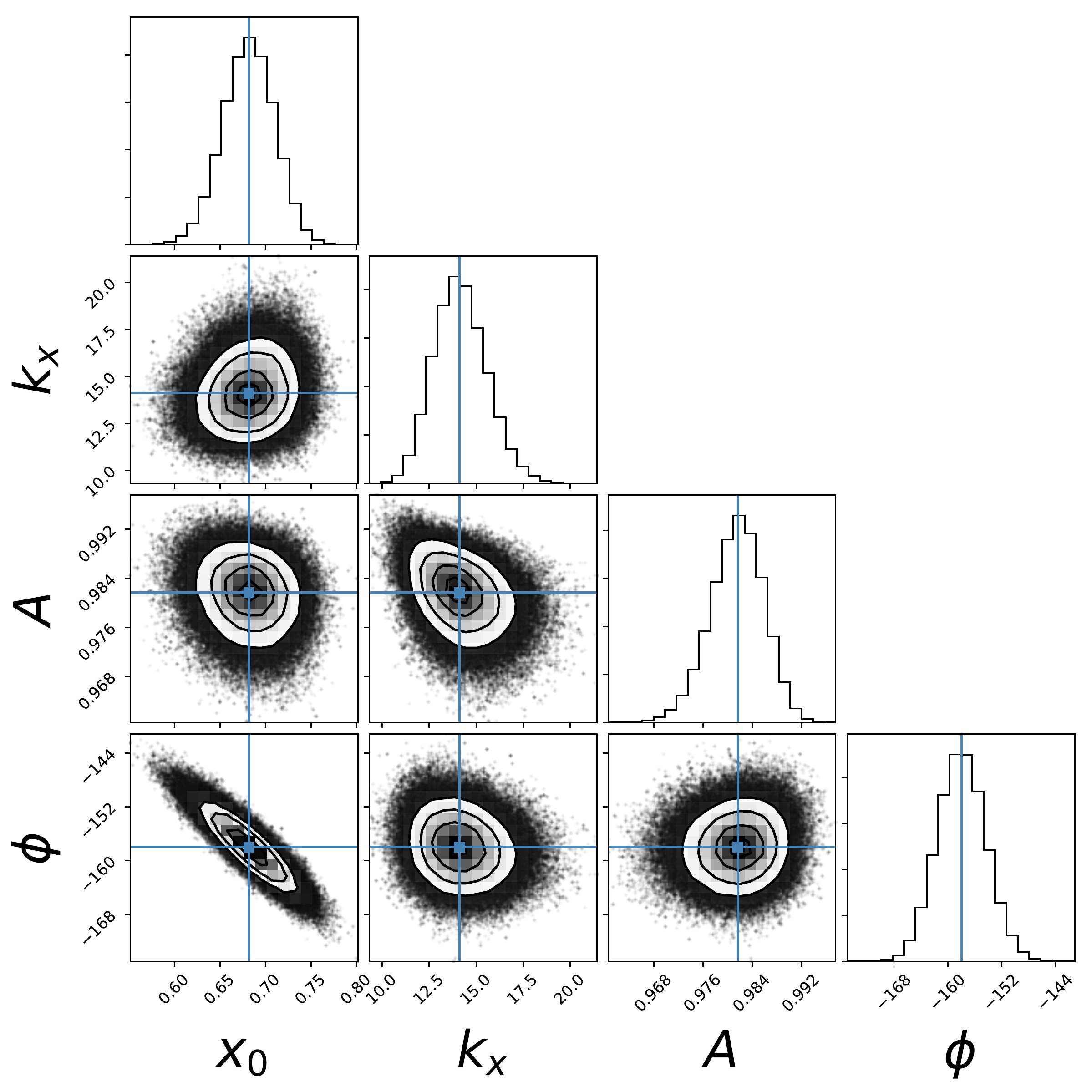}
   \caption{Posterior distributions for the observed false alarm rate $F_{\mathrm{FA}}$ parameters $\boldsymbol{\theta}$. The straight lines indicate the median values.} \label{figure:obsFpPost}
\end{figure}

\begin{figure*}[ht]
  \centering
  \includegraphics[width=\linewidth]{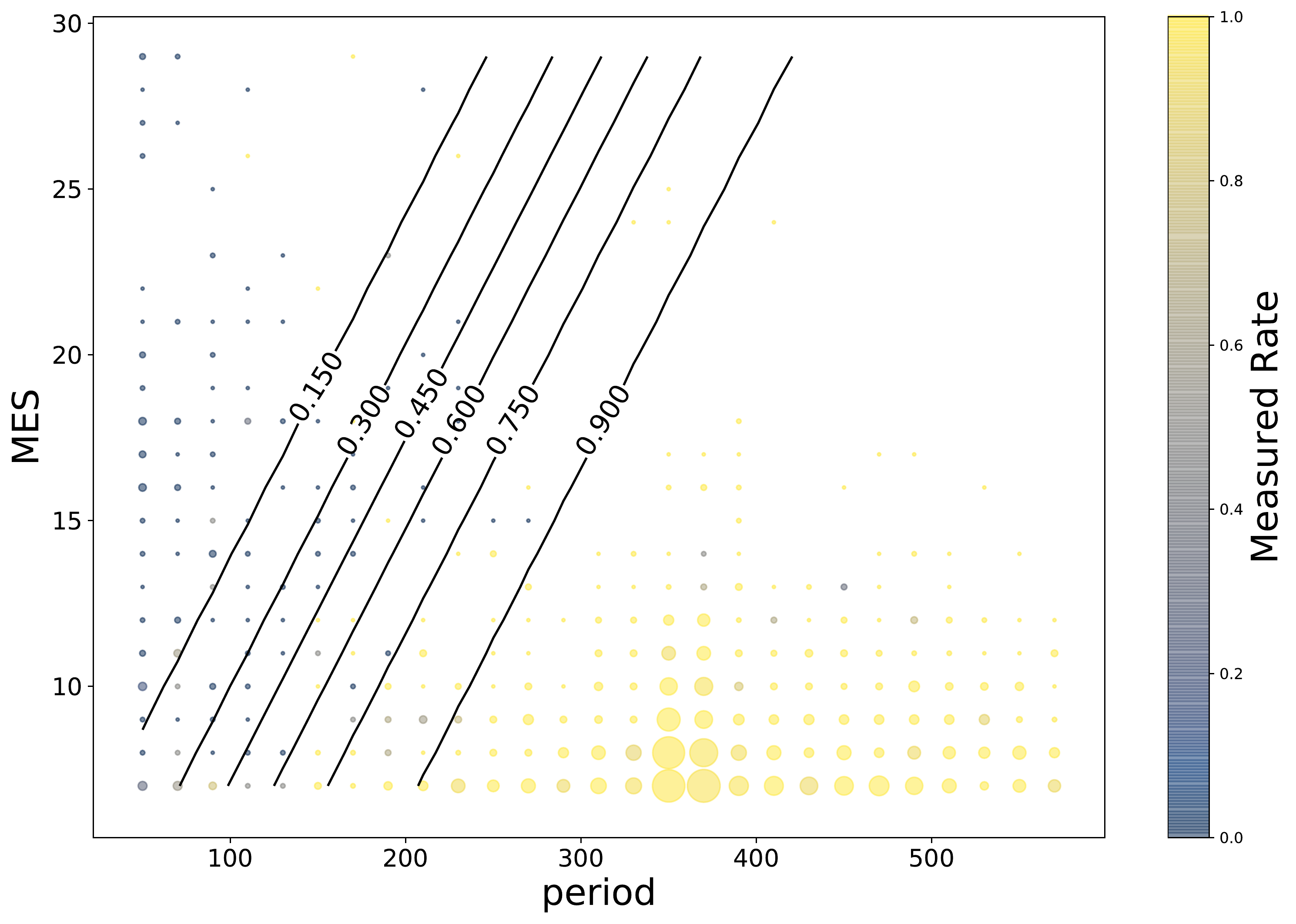}
  \caption{Contours of the observed false alarm rate $F_{\mathrm{FA}}(p_i,m_j,\boldsymbol{\Bar{\theta}})$ for the median of the posteriors.  The colored shapes show the measured data in each grid cell, with the color indicating the measured rate, and the size indicating the number of TCEs in the cell.} \label{figure:obsFpPostContours}
\end{figure*}

\begin{figure*}[ht]
   \centering
   \includegraphics[width=\linewidth]{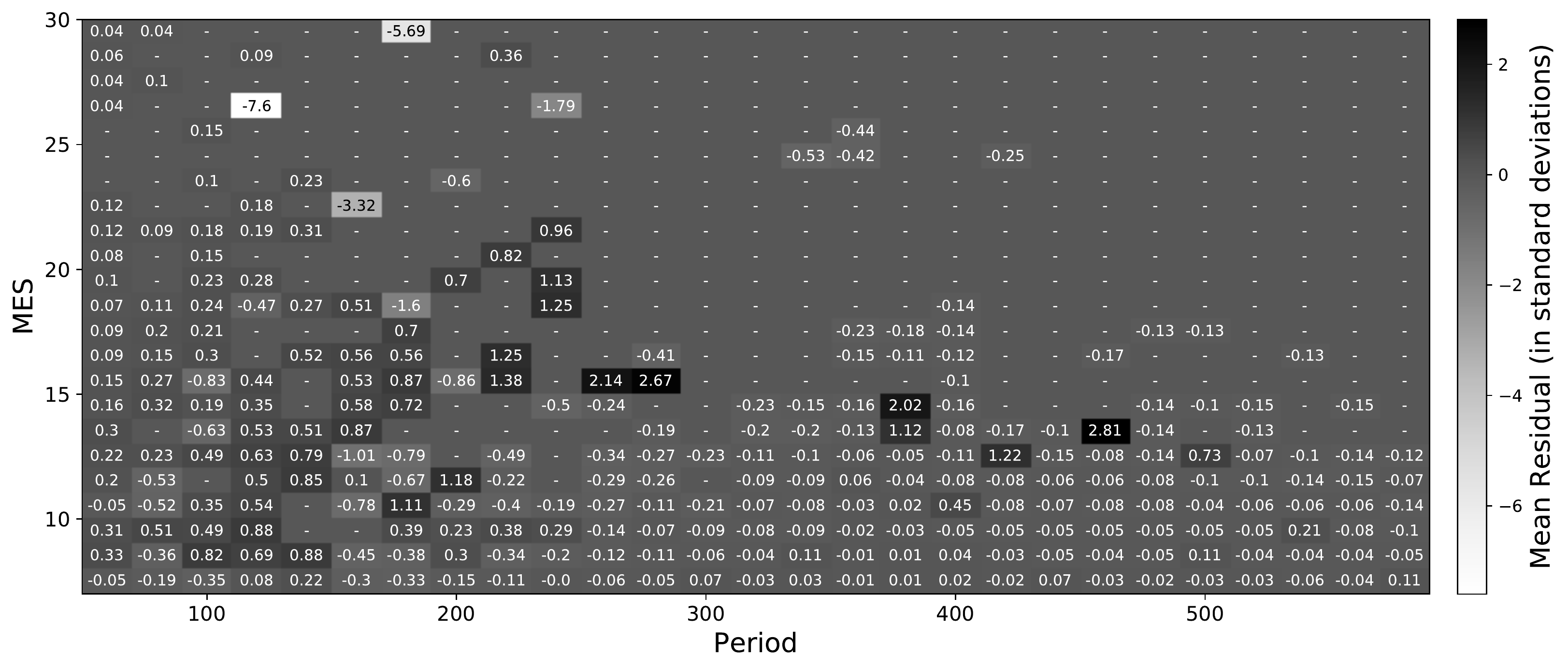} 
   \caption{The residuals of the measured $F_{\mathrm{FA}}$ rate in Figure~\ref{figure:obsFpRate} from the mean normalized to the standard deviation.  We see a small region with about a $1 \sigma$ bias, indicating an imperfect fit to the slope in the measured $F_{\mathrm{FA}}$ rate for period between 100 and 300 days.} \label{figure:obsFpPostResiduals}
\end{figure*}

We used the uniform priors $-1 \leq x_0 \leq 2$, $10^{-4} < k_x < 100$, $-180 < \phi < 180$, $0 < A < 1$, and initialized $\boldsymbol{\theta}$ by minimizing $-\log(L)$ using the Python optimize package. Our MCMC computation used 100 walkers, and ran for 5000 steps after 5000 steps of burn-in. Figure~\ref{figure:obsFpPost} shows the resulting posteriors, giving $\boldsymbol{\theta} = [x_0, k_x, \phi, A]$ as 
\begin{equation} \label{eqn:obsfpSolution}
\begin{split}
    x_0 &= 0.682^{+0.028}_{-0.029}, \qquad k_x = 14.120^{+1.469}_{-1.335}, \\
    \phi &= -157.967^{+3.608}_{-3.539}, \qquad A = 0.982^{+0.004}_{-0.004}.
\end{split} \nonumber
\end{equation}
The rate function $F_{\mathrm{FA}}(p_i,m_j,\boldsymbol{\Bar{\theta}})$ for the posterior median is shown in Figure~\ref{figure:obsFpPostContours}.  As in \S\ref{section:vettingCompleteness}, 1000 realizations of the FP rate function were created, drawing from the posterior $\boldsymbol{\theta}$ distribution. The residuals of the observed false alarm fraction in Figure~\ref{figure:obsFpRate} from the mean of these realizations in units of standard deviation is shown in Figure~\ref{figure:obsFpPostResiduals}, demonstrating an overall reasonable fit to the data. 

\subsubsection{Computing the False Alarm Reliability $R_\mathrm{FA}$ } \label{section:computeReliability}

Once we have the rate functions $F_{\mathrm{FA}}$ and $E_{\mathrm{FA}}$, we can compute the false alarm reliability $R_{\mathrm{FA}}\left(p, m \right)$ from equation~(\ref{eqn:reliabilityRate}).  In practice we evaluate $F_{\mathrm{FA}}$ and $E_{\mathrm{FA}}$ at a desired period and observed MES, either on a regular grid or for specific planet candidates.  

\begin{figure}[ht]
  \centering
  \includegraphics[width=\linewidth]{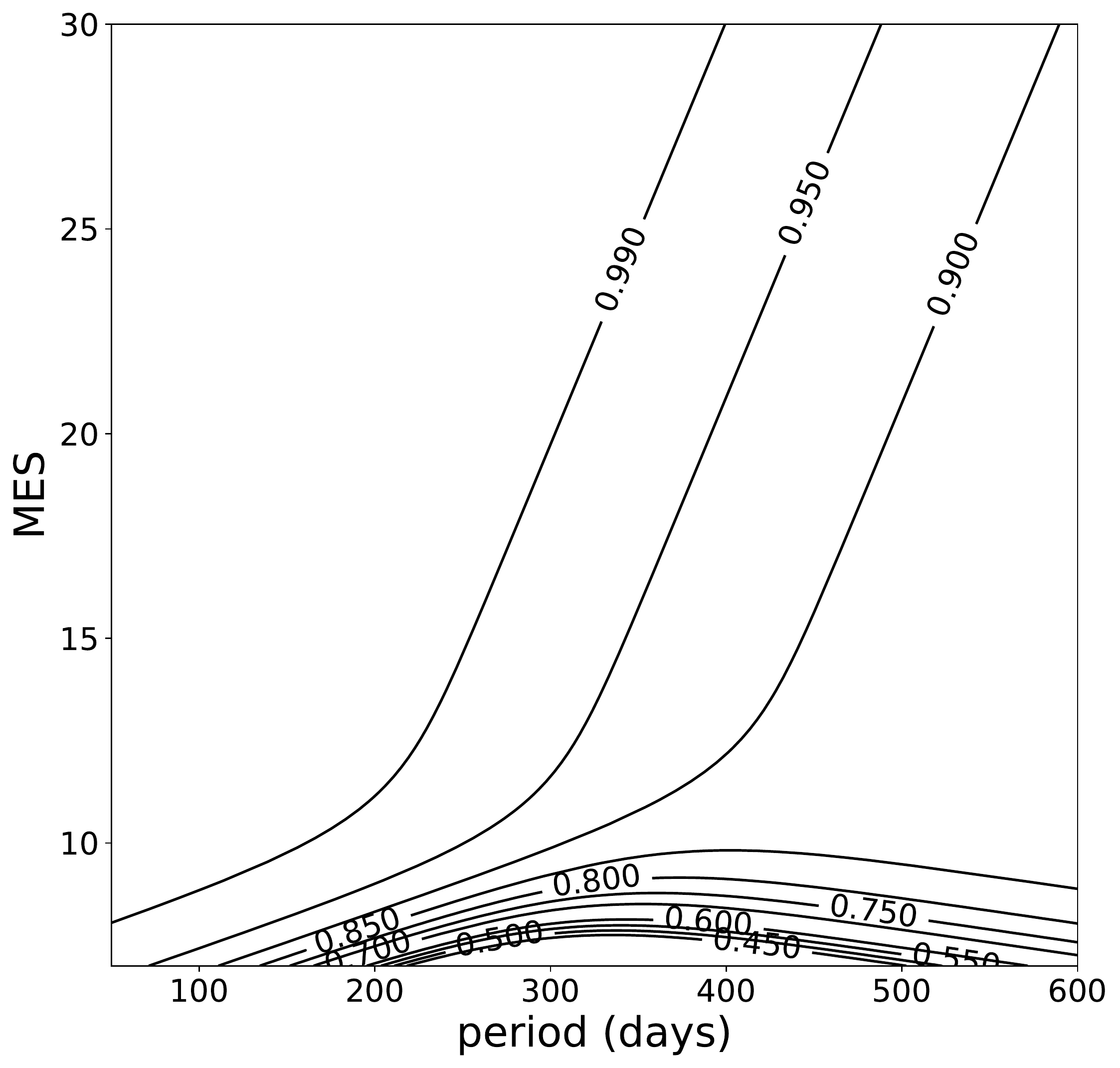}
  \caption{Contours of $R_{\mathrm{FA}}$ from the inferred $F_{\mathrm{FA}}$ from \S\ref{section:FPEcharacterization} and $E_{\mathrm{FA}}$ from \S\ref{section:obsFPcharacterization}.} \label{figure:reliabilityContours}
\end{figure}

\begin{figure}[ht]
   \centering
   \includegraphics[width=\linewidth]{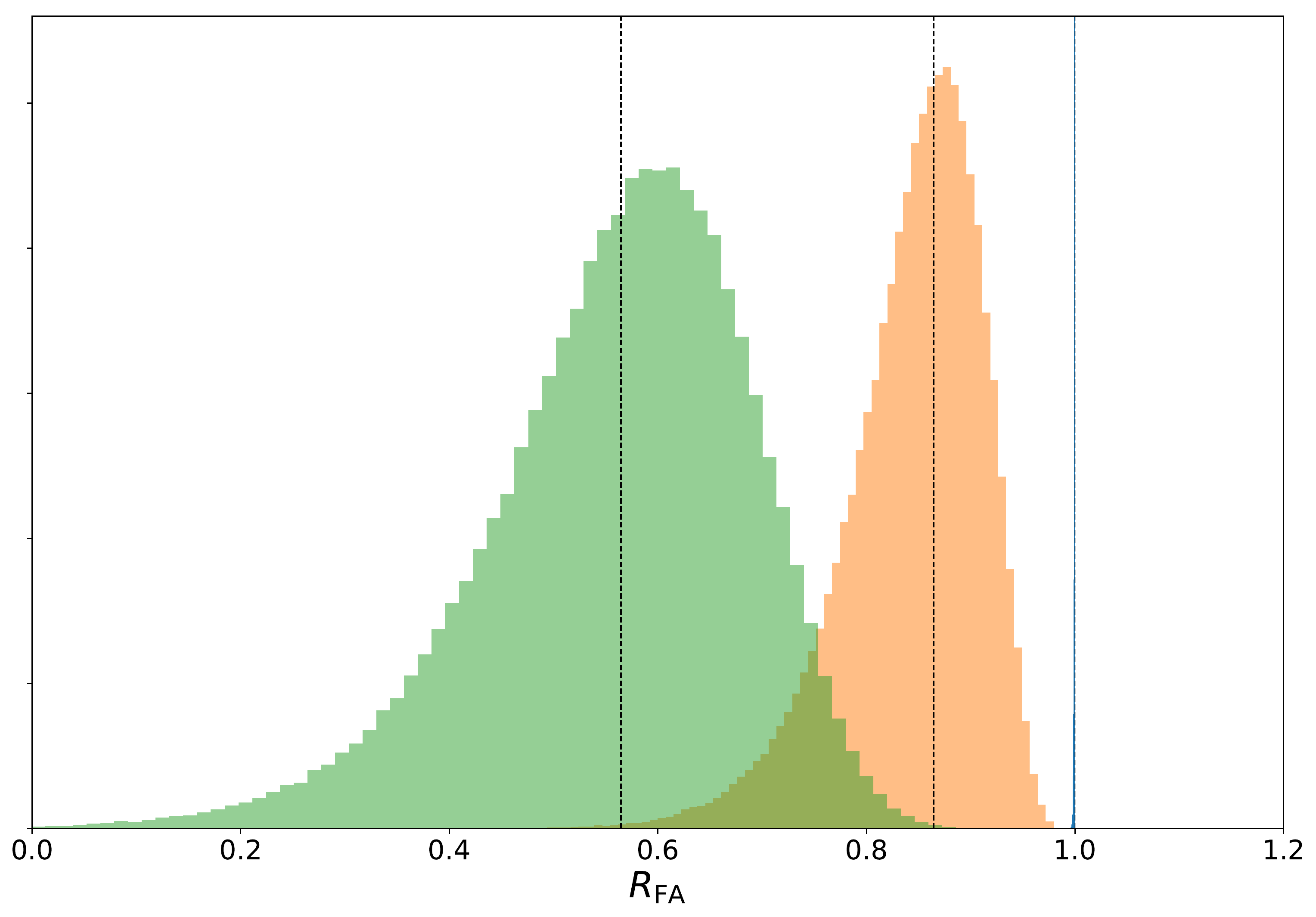}
   \caption{The false alarm reliability $R_{\mathrm{FA}}$ evaluated with the posterior distributions of $F_{\mathrm{FA}}$ and $E_{\mathrm{FA}}$ for three example periods and observed MES. Right distribution (very narrow and nearly coincident with the line $R_{\mathrm{FA}} = 1$): period = 200 days and MES = 25, with median reliability 1.0.  Middle distribution: period = 365 days and MES = 10, with median reliability 0.81.  Left distribution: period = 365 days and MES = 8, with median reliability 0.64.  The vertical lines show the rates for the median of the posteriors.} \label{figure:reliabilityPostExamples}
\end{figure}

Figure~\ref{figure:reliabilityContours} shows the resulting reliability function in the period-MES plane.  We see that for low MES there is decreased reliability around period 250 to 450 days, corresponding to the high number of TCEs in that range found in the inverted/scrambled data (see Figure~\ref{figure:nInvScrTCEs}), consistent with the excess of detections in Figure~\ref{figure:periodSnrDist}.  Figure~\ref{figure:reliabilityPostExamples} shows the reliability function evaluated over the full posteriors of $F_{\mathrm{FA}}$ and $E_{\mathrm{FA}}$ for three example periods and observed MES.  We see that for low MES near 1-year orbital periods the reliability drops to about 0.6 and has a large spread.  

\subsection{Astrophysical Reliability} \label{section:FPP}

The reliability function determined in \S\ref{section:vetReliability} only provides the probability that a planet candidate is not a false alarm. To determine the probability that a candidate is not an astrophysical false alarm such as a grazing or eclipsing eclipsing binary, we use the Q1--Q17 DR25 False Positive Probabilities\footref{footnote:exoplanetArchive} created using the technique developed in \citep{Morton2016}.  These probabilities were computed for all KOIs based largely on photometric data including  transit light curves and measured magnitudes.  We therefore assume that they are still valid even though we are using different stellar properties.  We define the astrophysical reliability of a planet candidate as 1 - the false positive probability of that candidate.

\subsection{Computing the Reliability for Each Planet Candidate}  \label{section:totalReliability}

We compute the reliability for each planet candidate by first evaluating $R_{\mathrm{FA}} \left(p, m \right)$ as described in \S\ref{section:computeReliability}, where $p$ is the observed orbital period and $m$ is the observed MES of the planet candidate from the KOI catalog.  Then we define the reliability $R = R_{\mathrm{FA}} \left(p, m \right) \cdot \left( 1 - \mathrm{FPP} \right)$ where FPP is the false positive probability for that planet candidate from the Q1--Q17 False Positive Probabilities table.

\section{Illustrative Occurrence Rates} \label{section:occurrence}

We present several illustrative occurrence rates, focusing on long-period, small planets where vetting completeness and reliability have the greatest impact.  We compute our occurrence rates with the method of \citet{Burke2015}, modeling occurrence rates as a Poisson point process with a rate given by a product of power laws in orbital period and planet radius.  We perform our occurrence rate analysis over the period and radius range of $50 \leq \mathrm{period} \leq 400$ days and $0.75 \leq \mathrm{radius} \leq 2.5$ $R_{\oplus}$, and integrate the resulting rate over two ranges considered by \citet{Burke2015}:
\begin{itemize}
    \item \boldmath $F_1$ \unboldmath: $50 \leq \mathrm{period} \leq 200$ days and $1 \leq \mathrm{radius} \leq 2$ $R_{\oplus}$, and
    \item \boldmath $\zeta_{\oplus}$ \unboldmath: within 20\% of Earth's orbital period and radius.
\end{itemize}

\begin{figure*}[ht]
  \centering
  \includegraphics[width=\linewidth]{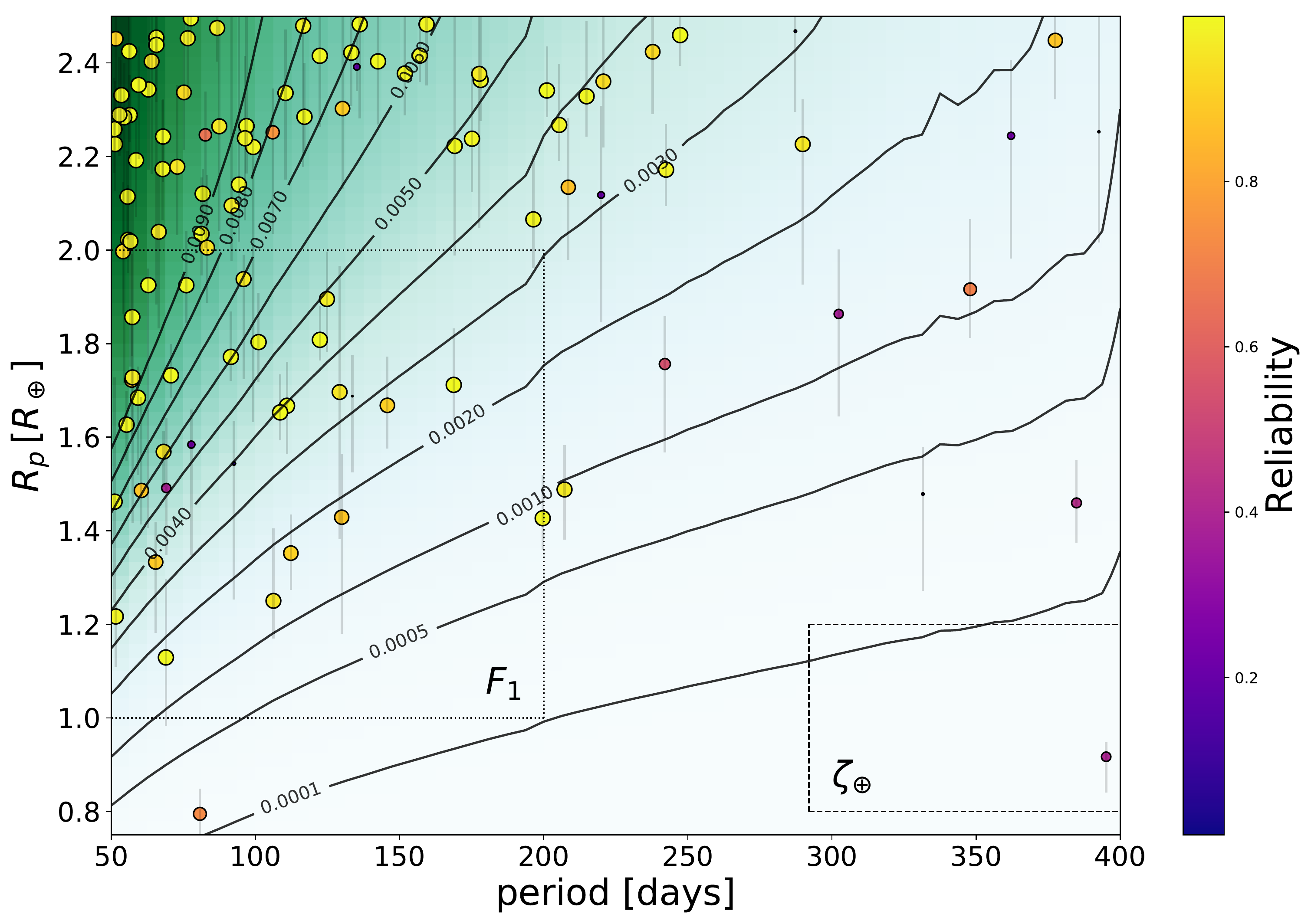}
  \caption{The baseline planet candidate population, colored and sized by reliability with planet radius error bars.  The background color map and contours indicate the summed completeness function $\eta(p, r)$.  The box on the left indicates the region integrated to obtain $F_1$, while the box on the right indicates the integration region for $\zeta_{\oplus}$.  The $\zeta_{\oplus}$ box extends out to 438 days.} \label{figure:planetPopulation}
\end{figure*}

Figure~\ref{figure:planetPopulation} shows our baseline planet candidate population, with the planet markers sized and colored by that planet's reliability and the background and contours showing the completeness function $\eta(p, r)$, including geometric transit probability. The $F_1$ and $\zeta_{\oplus}$ regions are indicated by boxes.  We see that while the $F_1$ region is reasonably well populated, it has a large completeness correction of $\sim 500$.  $\zeta_{\oplus}$, however, has only one low-reliability planet and a completeness correction $>10^4$, leading to large uncertainties in the estimate of $\zeta_{\oplus}$.  

\subsection{Methodology} \label{section:occurrenceMethodologies}

Following \citet{Youdin2011} and \citet{Burke2015}, we study the number of planets per star as a function of orbital period $p$ and planet radius $r$, $f(p, r)$, by inferring the population rate function $\lambda(p, r) \equiv \mathrm{d}^2 f/\mathrm{d}p \, \mathrm{d}r$ from a collection of planet detections at $(p_i, r_i)$ with a known completeness function $\eta(p, r)$ and reliability $R_{\mathrm{FA}}$.  If $\lambda(p, r, \boldsymbol{\theta})$ is a specific function parameterized by the parameter vector $\boldsymbol{\theta}$, then our problem is to determine $\boldsymbol{\theta}$.  We proceed by Bayesian inference: given a set of planet candidates with orbital period and radius $\{p_i, r_i\}$, $i = 1 \ldots N_p$ where $N_p$ is the number of planet candidates, by Bayes' theorem the probability of $\boldsymbol{\theta}$ is 
\begin{equation} \label{eqn:bayesTheta}
\begin{split}
    &P\left(\boldsymbol{\theta} | \{p_i, r_i\}, i = 1 \ldots N_p \right) \\
    &\propto P\left(\{p_i, r_i\}, i = 1 \ldots N_p |  \boldsymbol{\theta}\right) \pi \left(\boldsymbol{\theta}\right).
\end{split}
\end{equation}
where $\pi \left(\boldsymbol{\theta}\right)$ is a prior on $\boldsymbol{\theta}$.  Fixing $\pi \left(\boldsymbol{\theta}\right)$, finding the highest probability $\boldsymbol{\theta}$ amounts to maximizing the likelihood $P\left(\{p_i, r_i\}, i = 1 \ldots N_p |  \boldsymbol{\theta}\right)$.  

In Appendix~\ref{app:likelihoodDerivation} we show that maximizing the likelihood 
\begin{equation} \label{eqn:planetLilkelihood}
\begin{split}
    &P\left(\{p_i, r_i\}, i = 1 \ldots N_p |  \boldsymbol{\theta}\right) \\
    &= e^{- \Lambda(D)} \prod_{s=1}^{N_p}  \lambda(p_i, r_i, \boldsymbol{\theta})
\end{split}
\end{equation}
is equivalent to treating planet occurrence as a Poisson point process with rate $\lambda(p, r, \boldsymbol{\theta})$ that depends on period, planet radius, and parameters $\boldsymbol{\theta}$.  Here $\Lambda(D) = \int_{D} \eta(p, r) \lambda(p, r, \boldsymbol{\theta}) d p \, d r$ is the integral over the whole period-radius space $D$, where $\eta(p, r)$ is the summed completeness function from \S\ref{section:completeness}. However, we point out that equation~(\ref{eqn:planetLilkelihood}) is not itself a Poisson probability, as is sometimes implied in the literature.

The likelihood in equation~\ref{eqn:planetLilkelihood} accounts for completeness but not reliability.  Because this likelihood is derived from a Poisson distribution, which is defined only for discrete integer counts, we cannot account for reliability by weighting a planet's contribution by its reliability.  We address reliability by performing multiple Bayesian inferences of $\boldsymbol{\theta}$ using equation~(\ref{eqn:bayesTheta}), drawing from the planet candidates according to their reliability.  For example, a planet candidate with reliability 0.9 would be included in 90\% of these inferences, while another planet candidate with reliability 0.2 would be included in 20\% of these inferences.  Then the $\boldsymbol{\theta}$ posteriors of these inferences is concatenated to produce the posterior distribution of $\boldsymbol{\theta}$ accounting for reliability.  

Following \citet{Youdin2011} and \citet{Burke2015}, we model the planet candidate population rate $\lambda(p, r, \boldsymbol{\theta})$ as a product of power laws in period and radius.  Inspired by Foreman-Mackey's implementation of \citet{Burke2015}\footnote{\url{https://dfm.io/posts/exopop/}}, we adapt the form resulting from solving explicitly for the normalization $C_n$ from Burke's equation (8) and using it in his unbroken power law equation (7) \citep{Burke2015}: for $\boldsymbol{\theta} = \left(F_0, \alpha, \beta \right)$, 
\begin{equation} \label{eqn:powerLaw}
\begin{split}
    \lambda(p, r, \boldsymbol{\theta}) = F_0 \frac{(\alpha + 1) r^{\alpha}}{r_{\max}^{\alpha+1}-r_{\min}^{\alpha+1}} \frac{(\beta + 1) p^{\beta}}{p_{\max}^{\beta+1}-p_{\min}^{\beta+1}}
\end{split}
\end{equation}
This form ensures that $\int_D \lambda(p, r, \boldsymbol{\theta}) d p \, d r  = F_0$ so $F_0$ can be interpreted as the integrated planetary occurrence rate over the period-radius range used in the analysis.  

\subsection{Baseline Results} \label{section:baselineResults}

To perform our Bayesian MCMC inference, we use the {\it emcee} package. To measure the impact of correcting for reliability, we run inferences both without and with reliability correction.  For our inference without reliability correction, we use 16 walkers and run for 5000 steps after 1000 steps of burn-in. For our inference with reliability correction, we run 100 inferences as described in \S\ref{section:occurrenceMethodologies}, probabilistically sampling from the planet candidates according to their reliability, with each inference using 16 walkers and running for 2000 steps after 400 steps of burn-in. In both cases the walkers of each MCMC run are initialized in a small Gaussian distribution centered on the maximum-likelihood solution for that inference's planet population.  The posteriors from each of the 100 inferences with reliability correction were concatenated to produce the $\boldsymbol{\theta}$ posteriors. Table~\ref{table:baselineThetaResults} shows the median and 16th and 84th percentiles of these posterior distributions both with and without reliability correction.  We see that reliability has an overall impact of about 30\% in $F_0$, the integrated rate over our period and radius range of $50 \leq \mathrm{period} \leq 400$ days and $0.75 \leq \mathrm{radius} \leq 2.5$ $R_{\oplus}$.

\renewcommand{\arraystretch}{2}
\begin{table*}[ht]
\centering
\caption{Baseline occurrence rate results}\label{table:baselineThetaResults}
\begin{tabular}{ r c c c }
\hline
Parameter & No Reliability & With Reliability & FA-only Reliability \\
\hline
$F_0$ & $0.608^{+0.110}_{-0.090}$ & $0.432^{+0.089}_{-0.072}$ & $0.514^{+0.102}_{-0.083}$ \\
$\alpha$ & $0.304^{+0.519}_{-0.496}$ &  $0.796^{+0.635}_{-0.598}$ &  $0.500^{+0.558}_{-0.524}$ \\
$\beta$ & $-0.557^{+0.174}_{-0.169}$ &  $-0.823^{+0.202}_{-0.209}$ &  $-0.742^{+0.192}_{-0.196}$ \\
\hline
$\Gamma_{\oplus}$ & $0.212^{+0.111}_{-0.075}$ &  $0.094^{+0.066}_{-0.041}$ &  $0.139^{+0.086}_{-0.055}$ \\
$F_1$ & $0.190^{+0.035}_{-0.030}$ &  $0.144^{+0.032}_{-0.027}$ &  $0.171^{+0.034}_{-0.029}$ \\
$\zeta_{\oplus}$ & $0.034^{+0.018}_{-0.012}$ &  $0.015^{+0.011}_{-0.007}$ &  $0.023^{+0.014}_{-0.009}$ \\
\end{tabular}
\tablecomments{Baseline occurrence rate results comparing not accounting for reliability with accounting for reliability against both false positives and false alarms (middle column) and accounting for false alarm reliability only.  Central values and error bars are the median and 16th and 84th percentiles of the $\boldsymbol{\theta}$ posteriors of the Bayesian inference. $\Gamma_{\oplus} \equiv \mathrm{d}^2 f / \mathrm{d} \log p \, \mathrm{d} \log r = p_{\oplus} r_{\oplus} \lambda \left(p_{\oplus}, r_{\oplus}, \boldsymbol{\theta} \right)$, evaluated at Earth's period and radius, $F_1$ is the integrated planet rate over  $50 \leq \mathrm{period} \leq 200$ days and $1 \leq \mathrm{radius} \leq 2$ $R_{\oplus}$, and $\zeta_{\oplus}$ is the integrated rate within 20\% of Earth's orbital period and size.  
}
\end{table*}

\begin{figure*}[ht]
   \centering
   \includegraphics[width=\linewidth]{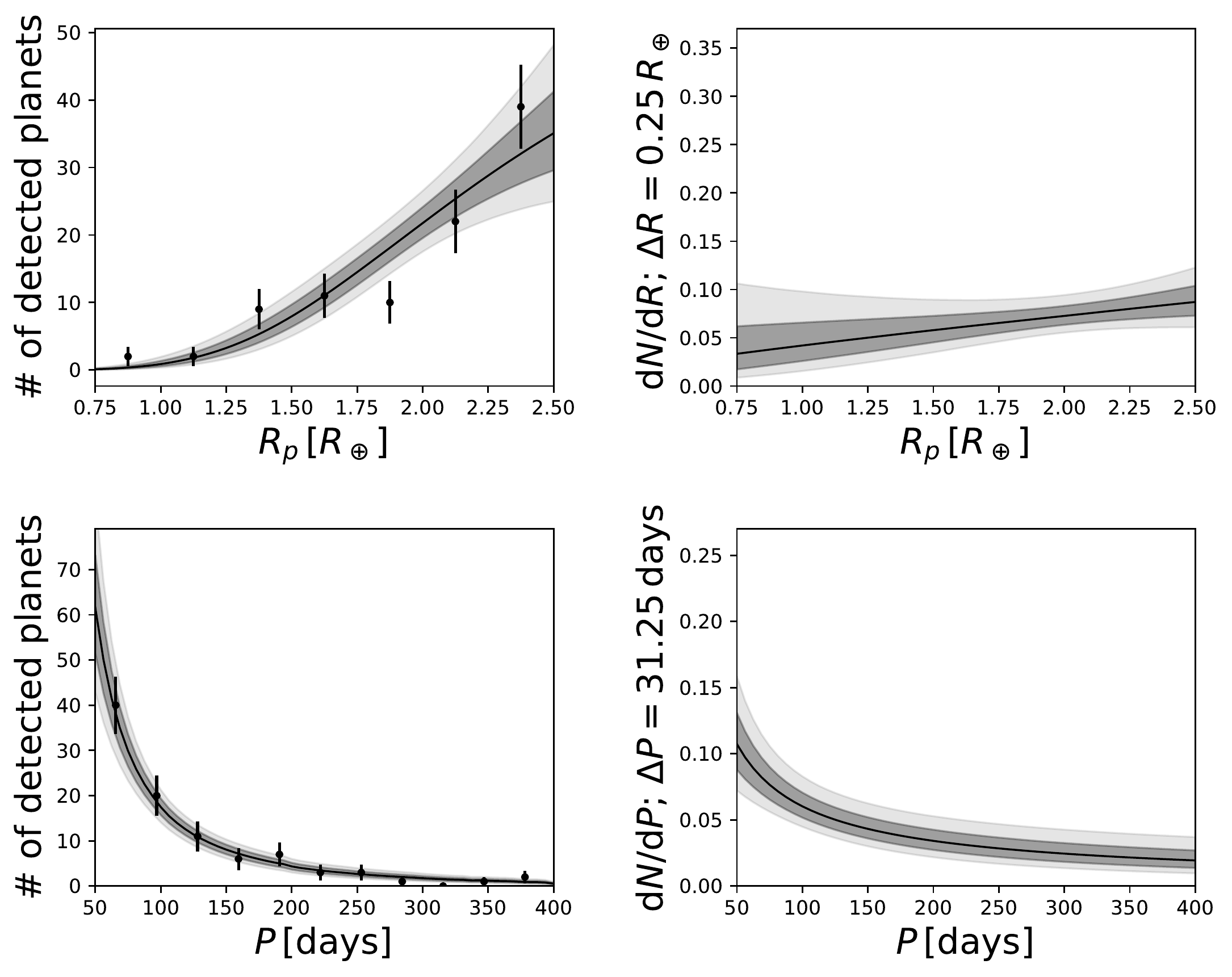}
   \caption{Marginal projections of the occurrence rate function $\lambda(p, r, \boldsymbol{\theta})$, accounting for reliability.  Left: the predicted number of planets compared with binned planet candidates.  Right: the marginalized rate function $\lambda(p, r, \boldsymbol{\theta})$.} \label{figure:occReliabilityMarginals}
\end{figure*}

Figure~\ref{figure:occReliabilityMarginals} shows the marginalized population rate function $\lambda(p, r, \boldsymbol{\theta})$ for the posterior $\boldsymbol{\theta}$ distribution, accounting for uncertainty.  This figure also compares the predicted number of planet detections with the binned planet candidates.  

\begin{figure}[ht]
   \centering
   \includegraphics[width=\linewidth]{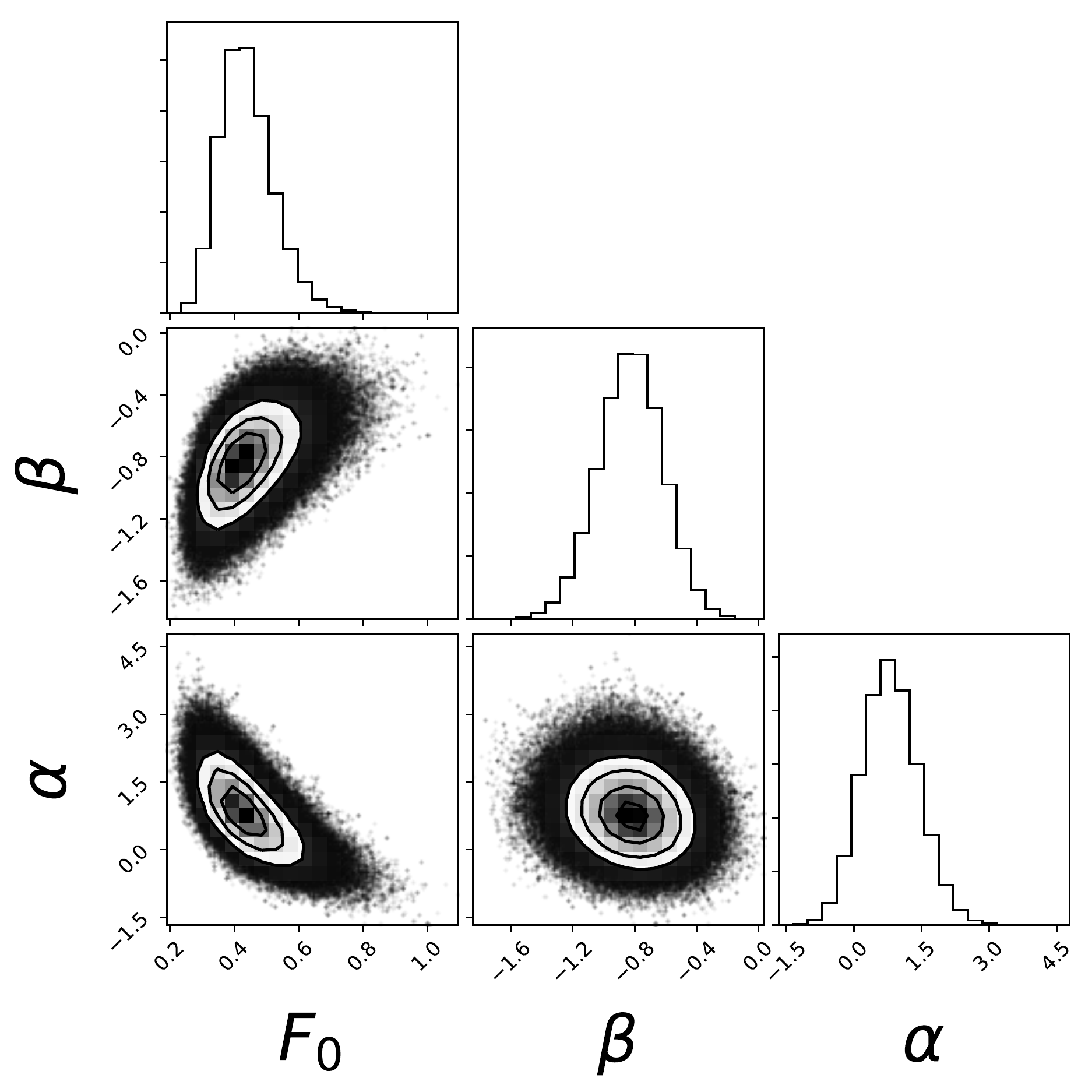}
   \caption{Posterior distributions for the occurrence rate parameters when correcting for reliability.} \label{figure:occReliabilityPost}
\end{figure}

\begin{figure*}[ht]
   \centering
   \includegraphics[width=0.45\linewidth]{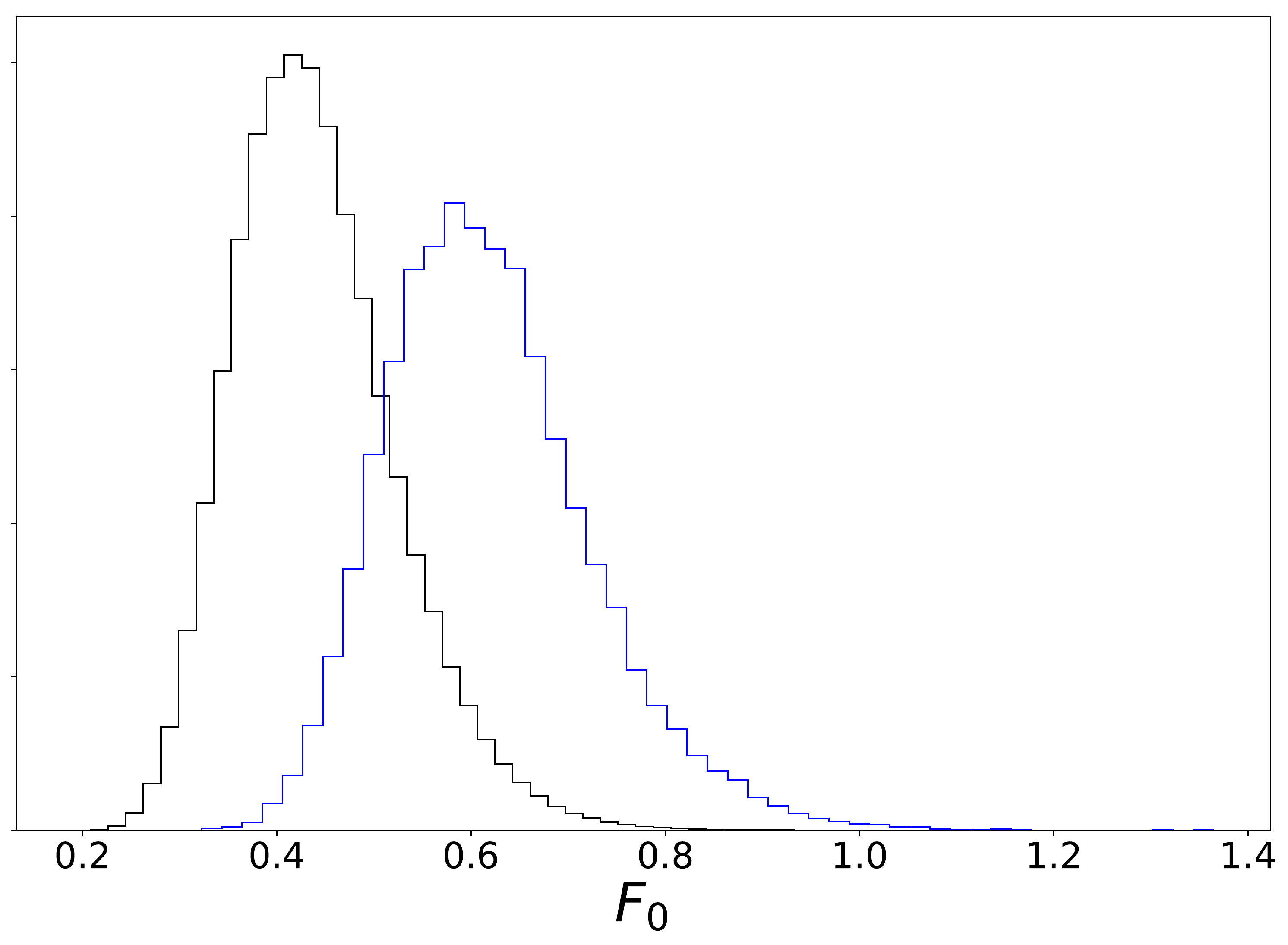} \qquad
   \includegraphics[width=0.45\linewidth]{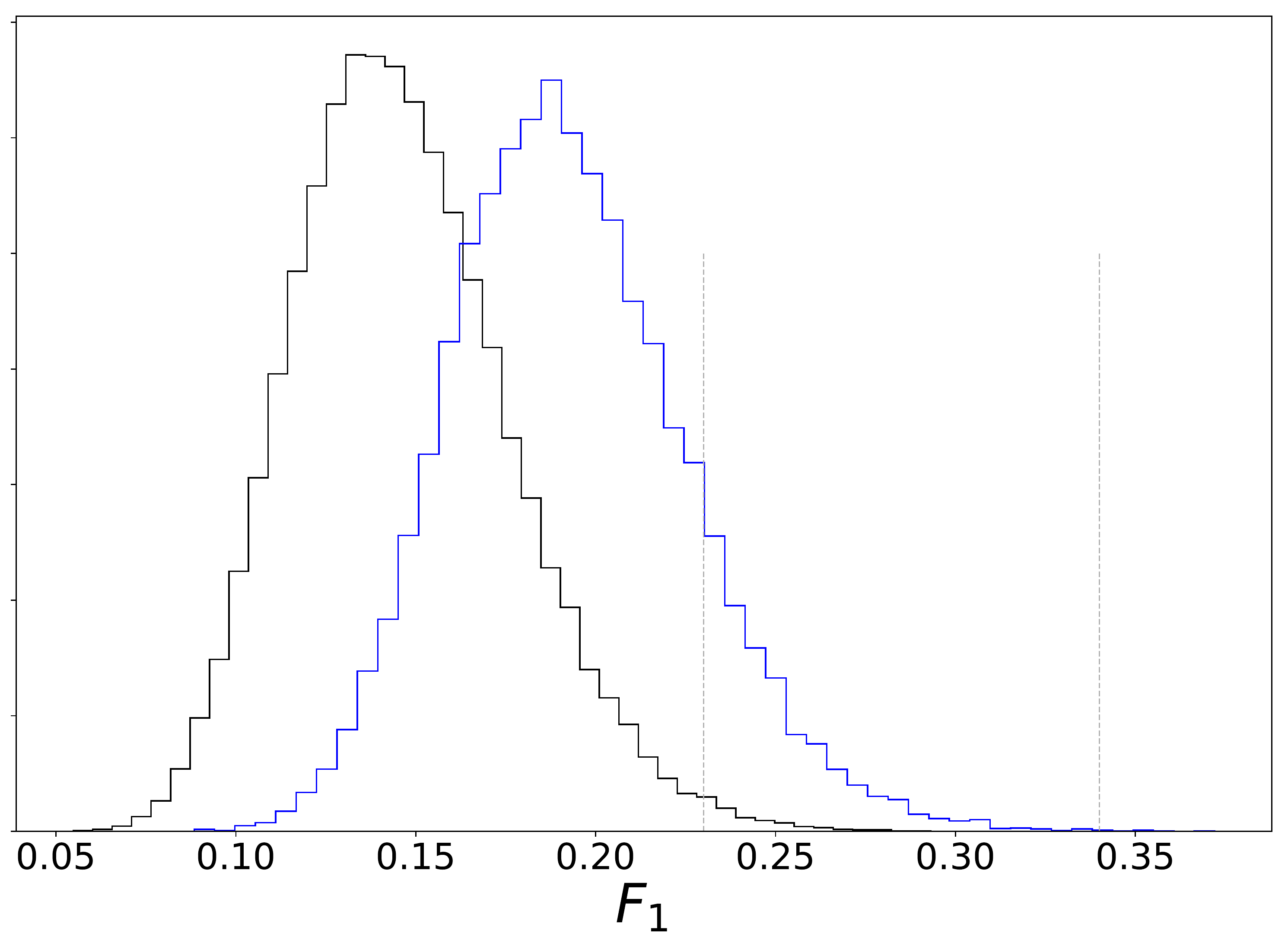} \\
   \includegraphics[width=0.45\linewidth]{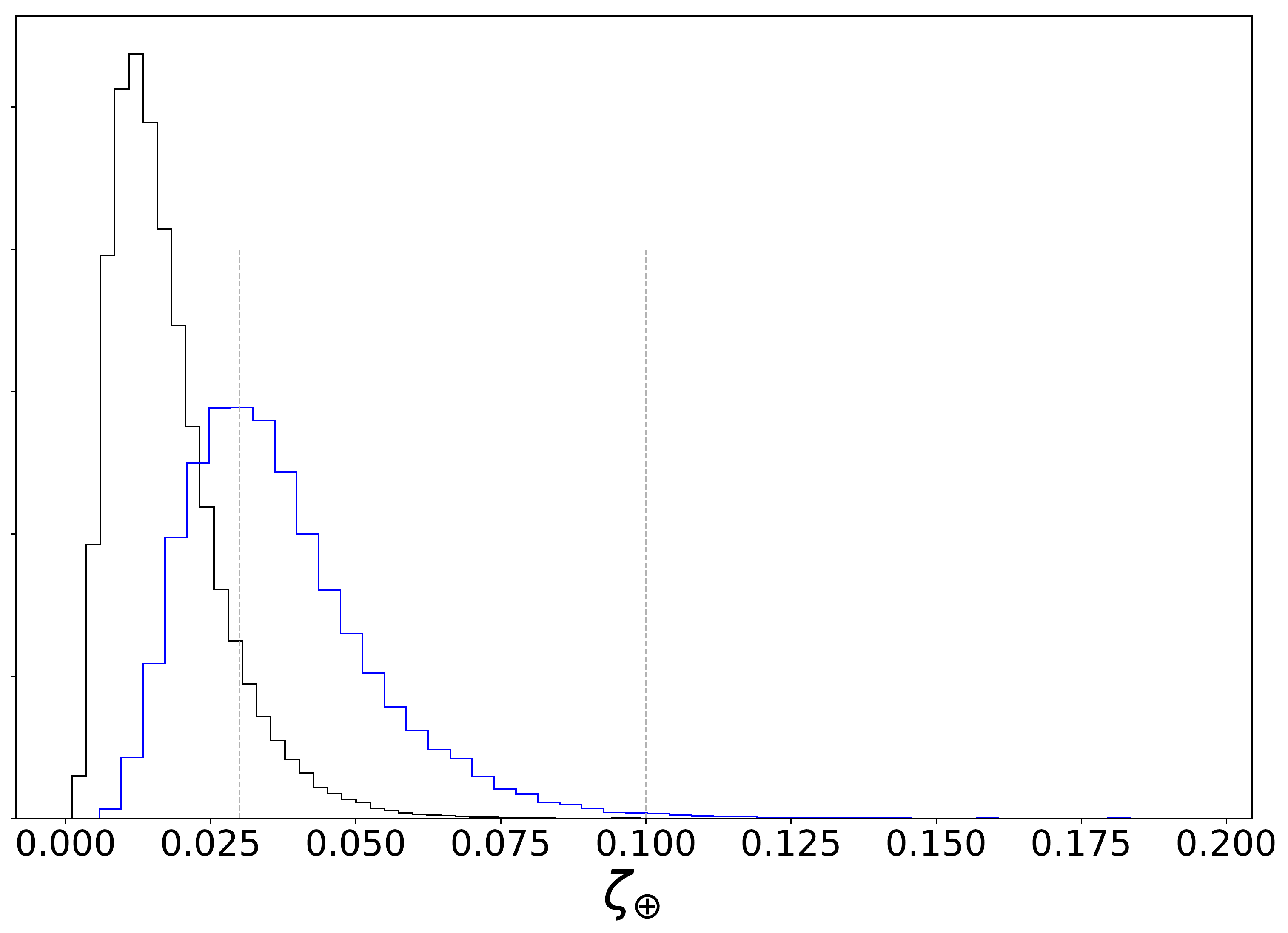} \qquad
   \includegraphics[width=0.45\linewidth]{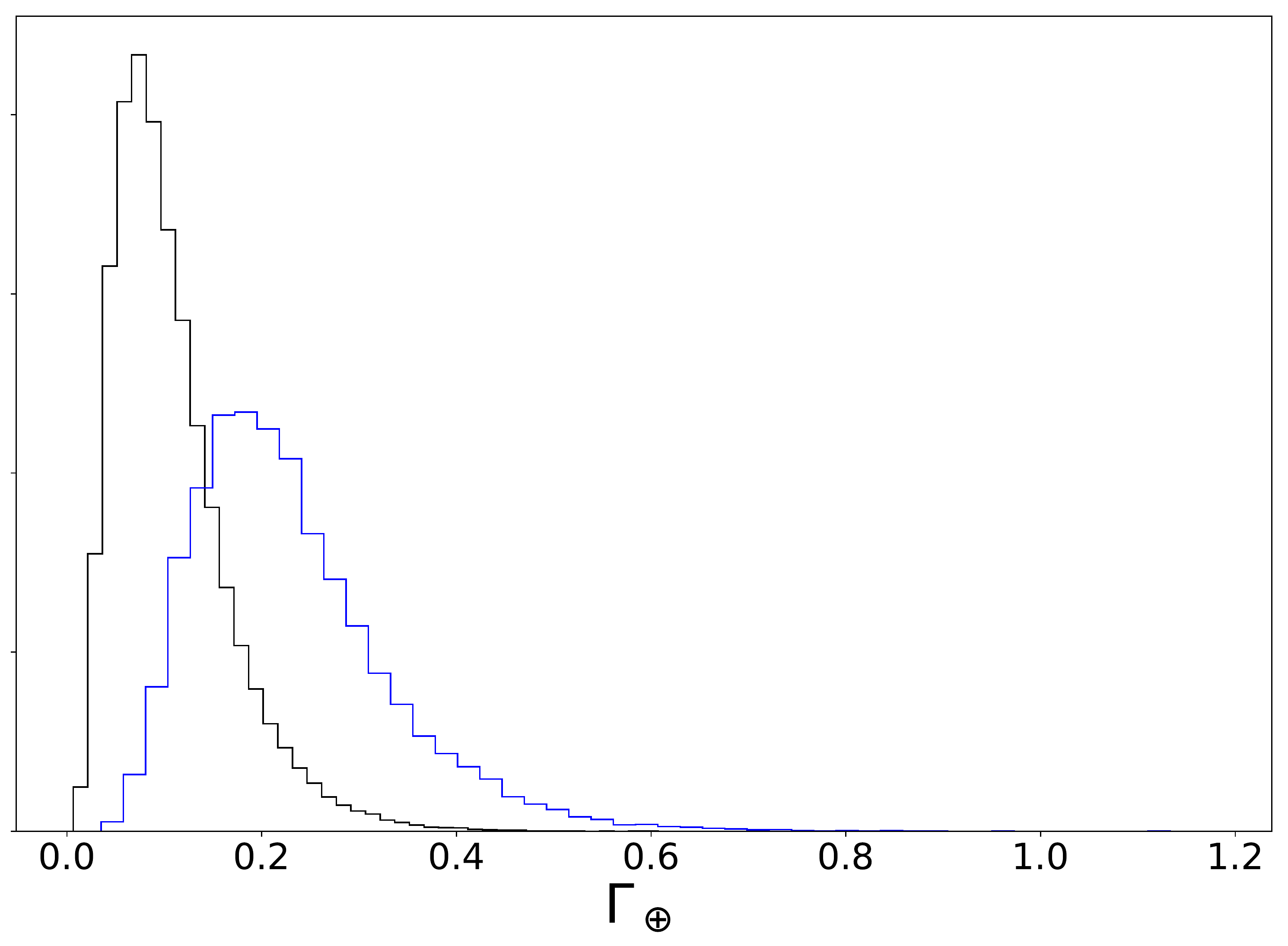}
   \caption{Comparison of various occurrence rates with and without reliability. In all panels, the right (blue) distribution is without accounting for reliability while the left (black) distribution is accounting for reliability.  {\bf Upper left:} $F_0$, the distribution of occurrence rates integrated over $50 \leq \mathrm{period} \leq 400$ days and $0.75 \leq \mathrm{radius} \leq 2.5 R_{\oplus}$.
   {\bf Upper Right:} $F_1$, the distribution of occurrence rates integrated over $50 \leq \mathrm{period} \leq 200$ days and $1 \leq \mathrm{radius} \leq 2$ $R_{\oplus}$ using all posterior values from the Bayesian inference.  Right dashed line: \citet{Burke2015} baseline $F_1$.  Left dashed line: \citet{Burke2015} "high reliability" $F_1$.
   {\bf Lower Left:} $\zeta_{\oplus}$, the distribution of occurrence rates integrated over 20\% of Earth's orbital period and size using all posterior values from the Bayesian inference.  Right dashed line: \citet{Burke2015} baseline $\zeta_{\oplus}$.  Left dashed line: \citet{Burke2015} ``high reliability'' $\zeta_{\oplus}$.
   {\bf Lower Right:} $\Gamma_{\oplus} \equiv \mathrm{d}^2 f / \mathrm{d} \log p \, \mathrm{d} \log r = p_{\oplus} r_{\oplus} \lambda \left(p_{\oplus}, r_{\oplus}, \boldsymbol{\theta} \right)$, evaluated at Earth's period and radius.  } \label{figure:occProdPost}
\end{figure*}

Figure~\ref{figure:occProdPost} and Table~\ref{table:baselineThetaResults} show $F_1$ and $\zeta_{\oplus}$, as well as $\Gamma_{\oplus} \equiv \mathrm{d}^2 f / \mathrm{d} \log p \, \mathrm{d} \log r = p_{\oplus} r_{\oplus} \lambda \left(p_{\oplus}, r_{\oplus}, \boldsymbol{\theta} \right)$, with and without accounting for reliability, evaluated over all posterior values of $\boldsymbol{\theta}$.  We see that even though there is significant overlap in the distributions with and without reliability, accounting for reliability has a strong impact: $\Gamma_{\oplus}$ and $\zeta_{\oplus}$ are are reduced by more than 50\%, which can be understood from the very small number of low-reliability planets in the $\zeta_{\oplus}$ region in Figure~\ref{figure:planetPopulation}.  $F_1$ is the integrated rate over a region of higher reliability, but reliability still has a strong effect.  $F_0$ is the integrated rate over our entire period-radius analysis range, but it is dominated by the fact that there are more high-reliability planet candidates, so reliability has an impact similar to $F_1$.  Table~\ref{table:baselineThetaResults} also shows the impact of accounting only for false alarm reliability, ignoring astrophysical false positive reliability, indicating that false alarm reliability accounts for about half the impact of the reliability correction.

We also computed occurrence for the SAG13 definition of $\eta_{\oplus}$\footnote{\label{footnote:sag13}\url{https://exoplanets.nasa.gov/exep/exopag/sag/\#sag13}}, $237 \leq \mathrm{period} \leq 860$ days and $0.5 \leq \mathrm{radius} \leq 1.5$ $R_{\oplus}$.  Without accounting for reliability, we find $\eta_{\oplus} = 0.302^{+0.181}_{-0.113}$, consistent with the results of \citet{zink2019}, while accounting for reliability yields $\eta_{\oplus} = 0.126^{+0.095}_{-0.055}$.  This result should be treated with caution because it involves extrapolation beyond the domain of both reliability and detection completeness characterization.

We find that the impact of accounting for reliability is significant for small planets in long-period orbits.  While one can note that the median values of occurrence rates in this regime are not much more than ``one $\sigma$'' apart, the observed shifts in the distributions on the order of 40\% are systematic, and clearly not due to statistical fluctuations.

\subsection{Simple Estimates of the Impact of Input Uncertainty} \label{section:uncertainties}
A full treatment of uncertainties in occurrence rates is beyond the scope of this paper.  Uncertainties in stellar properties would need to be accounted for in the selection of the parent stellar population, the modeling behind the detection completeness and impact on the Robovetter.  In this work we do, however, produce uncertainties in the false alarm reliability in \S\ref{section:vetReliability} through the MCMC posteriors of the fit functions, as well as planet radius uncertainties that follow from stellar radius transit fit uncertainties as described in \S\ref{section:planetCatalog}.  In this section we present simple experiments that examine the impact on our occurrence rates of the reliability and planet radius uncertainties.  We study the impact of planet radius uncertainties separately from the impact of reliability uncertainty.  

\subsubsection{Impact of Planet Radius Uncertainty} \label{section:rpUncertainties}
We study the impact of planet radius uncertainties without accounting for reliability.  We proceed in the same way that we study the impact of reliability, by performing several inference runs with a planet population in each run selected after applying the planet radius uncertainties.  Specifically, for each run, prior to the restriction of the planet candidate population to the radius range $0.75 \leq \mathrm{radius} \leq 2.5$ $R_{\oplus}$, we add to each planet's radius an error given by a draw from a Gaussian distribution with width equal to that planet's radius uncertainty.  Each planet is randomly assigned an upper or lower errorbar with 50\% probability.  The planet candidate population is then restricted to the range $0.75 \leq \mathrm{radius} \leq 2.5$ $R_{\oplus}$, and the inference is run.

\renewcommand{\arraystretch}{2}
\begin{table}[ht]
\caption{Impact of planet radius uncertainties (no reliability).}\label{table:rpUncertaintyResults}
\centering
\begin{tabular}{ r c c }
\hline
Parameter  & No Uncertainty & Planet Radius Uncertainty \\
\hline
$F_0$ & $0.608^{+0.110}_{-0.090}$ & $0.663^{+0.143}_{-0.112}$ \\
$\alpha$ & $0.304^{+0.519}_{-0.496}$ &  $-0.172^{+0.553}_{-0.530}$ \\
$\beta$ & $-0.557^{+0.174}_{-0.169}$ &  $-0.593^{+0.189}_{-0.193}$ \\
\hline
$\Gamma_{\oplus}$ & $0.212^{+0.111}_{-0.075}$ &  $0.276^{+0.157}_{-0.104}$ \\
$F_1$ & $0.190^{+0.035}_{-0.030}$ &  $0.216^{+0.041}_{-0.036}$ \\
$\zeta_{\oplus}$ & $0.034^{+0.018}_{-0.012}$ &  $0.045^{+0.026}_{-0.017}$ \\ 
\end{tabular}
\tablecomments{A study of the impact of planet radius uncertainties, not accounting for reliability.  The values without uncertainty are from Table~\ref{table:baselineThetaResults}.
See Table~\ref{table:baselineThetaResults} for an explanation of the rows.}
\end{table}

\begin{figure*}[ht]
   \centering
   \includegraphics[width=0.45\linewidth]{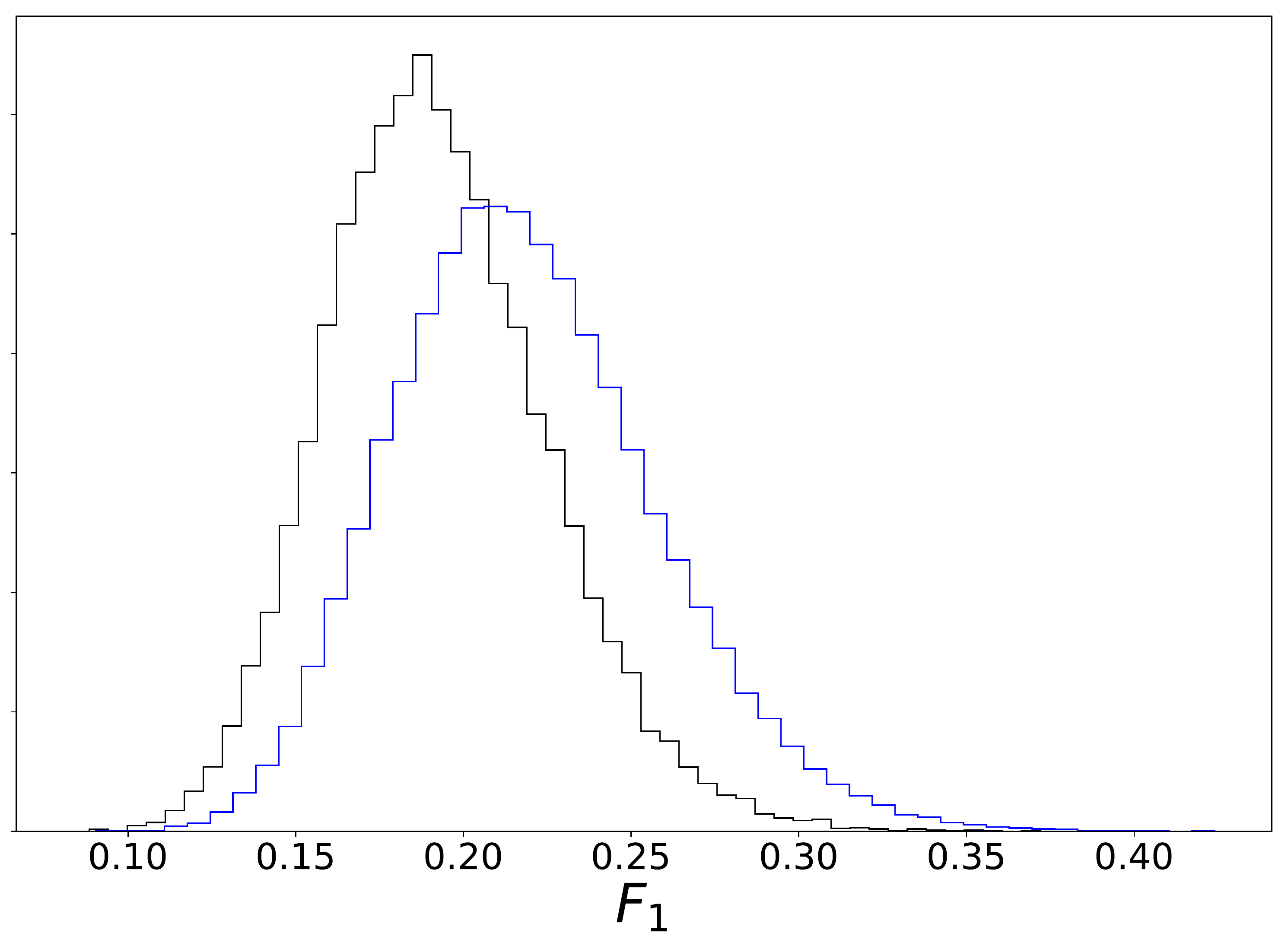} \qquad
   \includegraphics[width=0.45\linewidth]{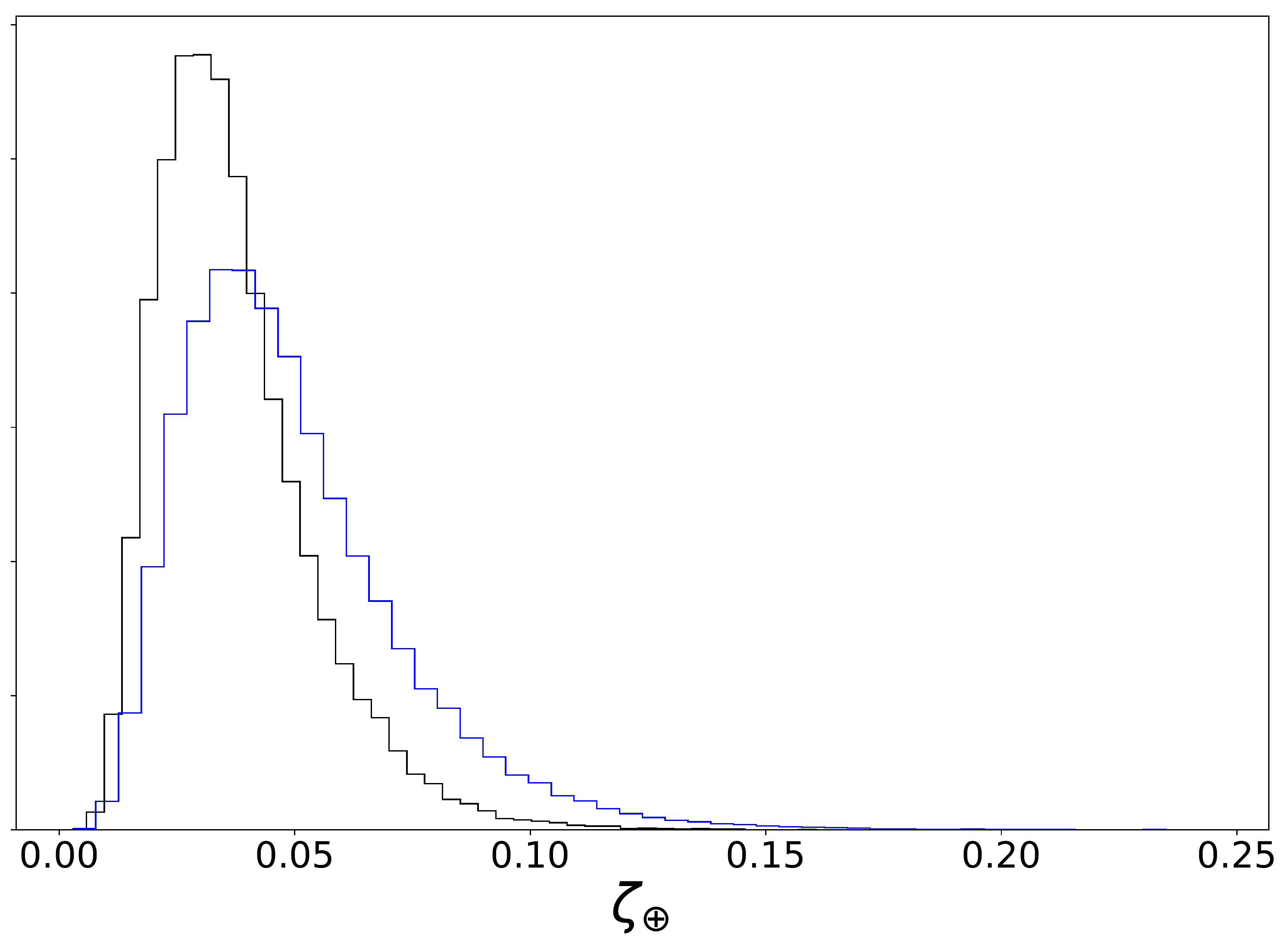} \\
   \includegraphics[width=0.45\linewidth]{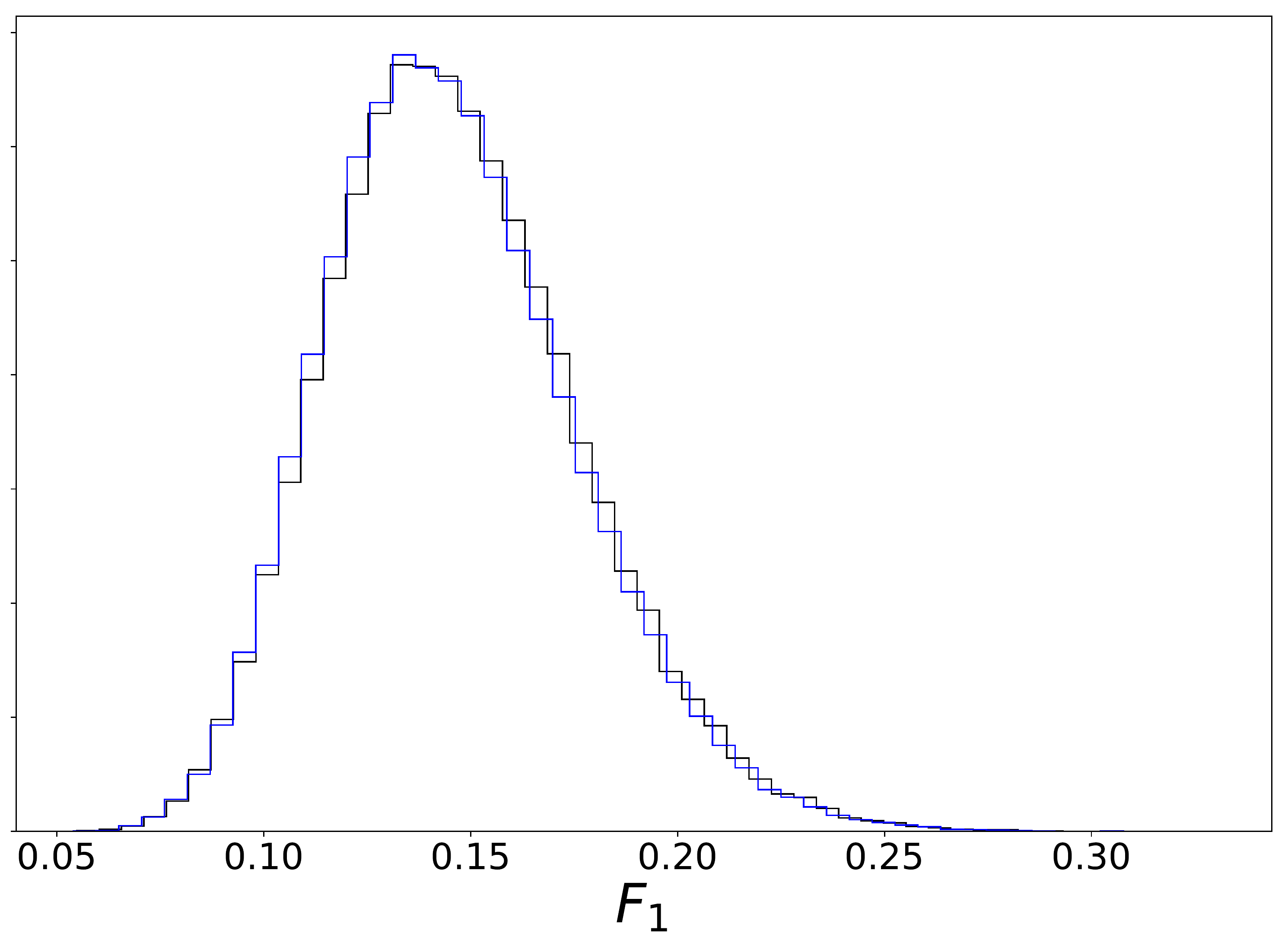} \qquad
   \includegraphics[width=0.45\linewidth]{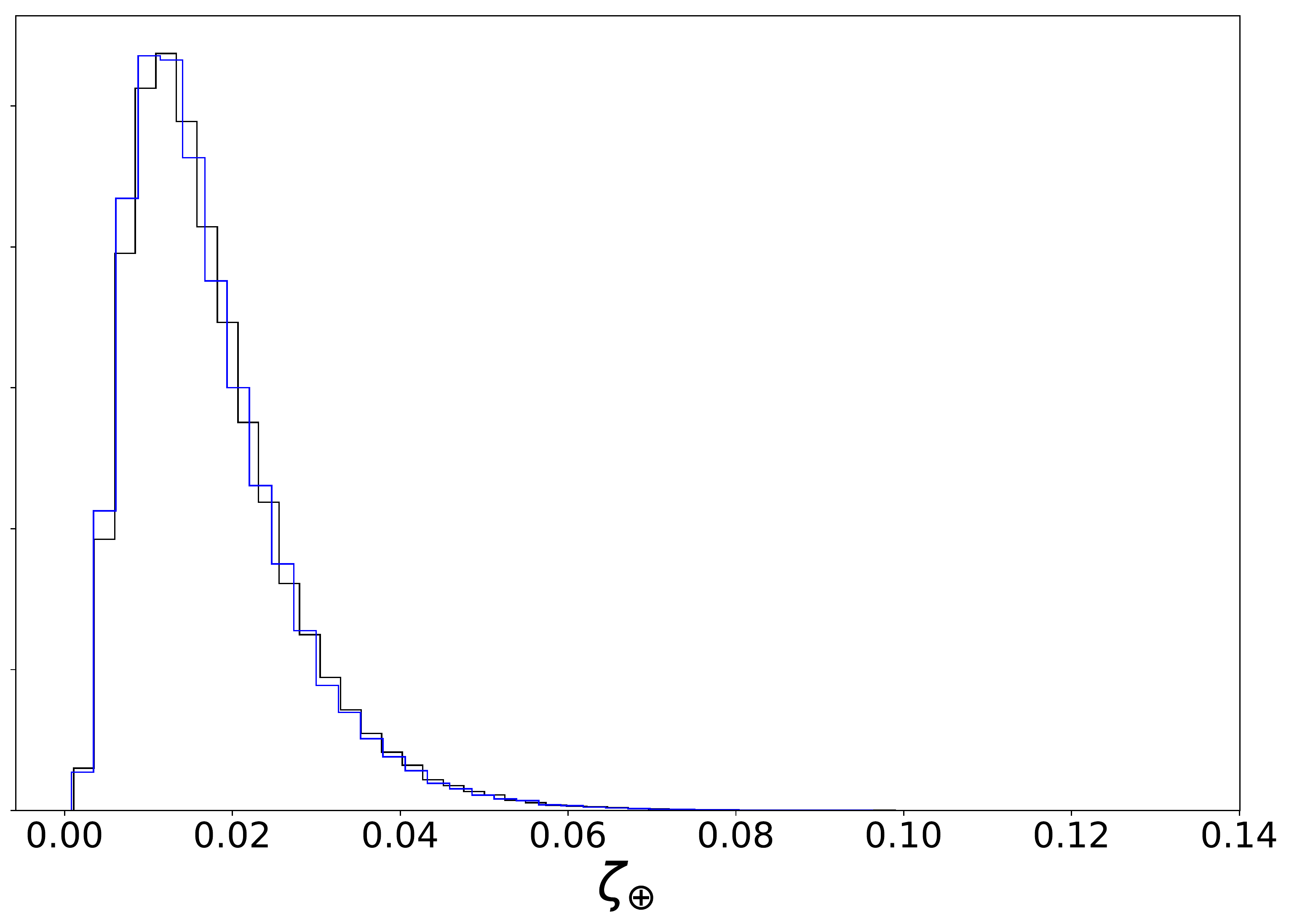}
   \caption{{\bf Top:} The impact of planet radius uncertainties (without including reliability) on $F_1$ (left) and $\zeta_{\oplus}$ (right).  Leftmost (black) distribution: no uncertainty. Rightmost (blue) distribution: including planet radius uncertainties.  This indicates that planet radius uncertainties have only a minor impact.
   {\bf Bottom:} The impact of planet reliability uncertainties on $F_1$ (left) and $\zeta_{\oplus}$ (right), indicating that planet reliability uncertainties have essentially no impact. Black distribution: no uncertainty. Blue distribution: including planet radius uncertainties.  } \label{figure:baselineUncertainty}
\end{figure*}

The impact of planet radius uncertainties resulting from 1000 inference runs is shown in Table~\ref{table:rpUncertaintyResults} and the top row of Figure~\ref{figure:baselineUncertainty}.  We see a small, consistent broadening of the width of the distributions and resulting error bars, and possibly a small systematic shift towards higher occurrence rates, but the overall impact is minor.  We believe this is due to the smaller uncertainties resulting from using Gaia stellar properties, and from the fact that near the boundary of our planet size range there are many planets, so planets are equally likely to exit and enter our range due to uncertainty.

\subsubsection{Impact of Reliability Uncertainty} \label{section:reliabilityUncertainties}

\renewcommand{\arraystretch}{2}
\begin{table}[ht]
\centering
\caption{Impact of reliability uncertainty (with reliability).}\label{table:reliabilityUncertaintyResults}
\begin{tabular}{ r c c }
\hline
Parameter & No Uncertainty & Reliability Uncertainty \\
\hline
$F_0$ & $0.432^{+0.089}_{-0.072}$ & $0.428^{+0.090}_{-0.072}$ \\
$\alpha$ & $0.796^{+0.635}_{-0.598}$ &  $0.807^{+0.631}_{-0.592}$ \\
$\beta$ & $-0.823^{+0.202}_{-0.209}$ &  $-0.832^{+0.207}_{-0.214}$ \\
\hline
$\Gamma_{\oplus}$ & $0.094^{+0.066}_{-0.041}$ &  $0.092^{+0.066}_{-0.040}$ \\
$F_1$ & $0.144^{+0.032}_{-0.027}$ &  $0.143^{+0.032}_{-0.027}$ \\
$\zeta_{\oplus}$ & $0.015^{+0.011}_{-0.007}$ &  $0.015^{+0.011}_{-0.006}$ \\
\end{tabular}
\tablecomments{A study of the impact of false alarm reliability uncertainties. The values without uncertainty are from Table~\ref{table:baselineThetaResults}. See Table~\ref{table:baselineThetaResults} for an explanation of the rows.}
\end{table}

We study the impact of uncertainties in reliability by modifying the method of computing occurrence rates with uncertainty described in \S\ref{section:baselineResults}.  Prior to each inference run, we draw from the posteriors of the parameter vectors for false alarm efficiency (\S\ref{section:FPEcharacterization}) and the observed false alarm rate (\S\ref{section:obsFPcharacterization}).  We use these draws evaluate the FA efficiency and observed rate functions at each planet candidate's period and observed MES, from which the false alarm reliability is computed.  The reliability is then computed as described in \S\ref{section:totalReliability}, and each planet is included in that run with probability given by this computed reliability.  In other words, the reliability rate function is realized for each run and applied to the planet candidate population.  

\begin{figure}[ht]
   \centering
   \includegraphics[scale=0.25]{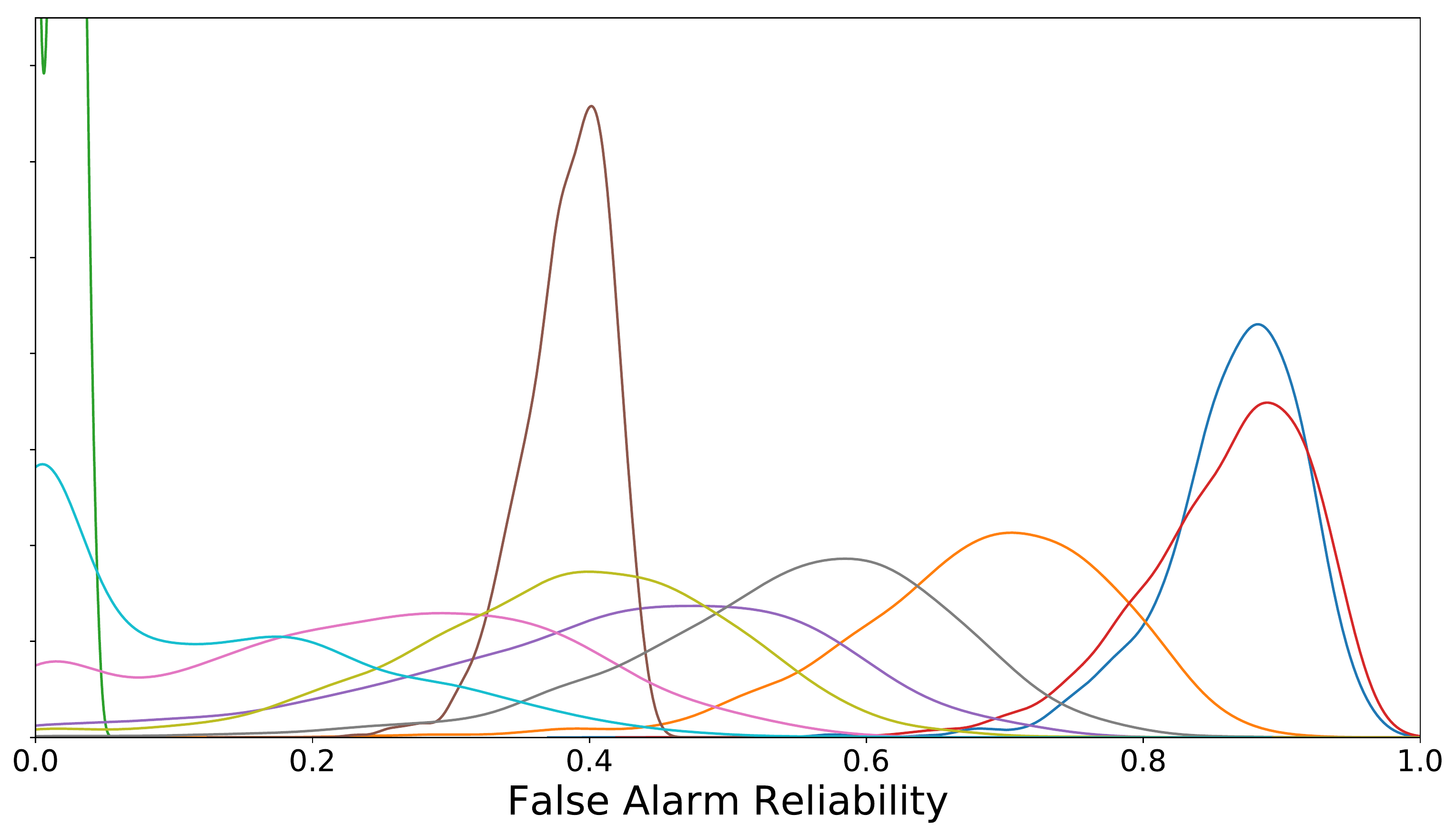}
   \caption{The distribution of reliabilities assigned to planet candidates with median reliability $< 0.9$ in the reliability uncertainty study.  As expected, the low-reliability candidates have broad distributions. } \label{figure:reliabilityDist}
\end{figure}

Table~\ref{table:reliabilityUncertaintyResults} and the bottom row of Figures~\ref{figure:baselineUncertainty} compare occurrence rates with reliability correction but no reliability uncertainty, computed with the median $\boldsymbol{\bar{\theta}}$ of \S\ref{section:FPEcharacterization} and \S\ref{section:obsFPcharacterization}, with the reliability distribution that results from using the respective full $\boldsymbol{\theta}$ posterior distributions.  We see that there is no significant impact due to reliability uncertainty apparent in 1000 inference runs.  This is in spite of the broad distributions of the low-reliability planet candidates shown in Figure~\ref{figure:reliabilityDist}, which shows the false alarm reliability values $R_{\mathrm RA}$ resulting from the $\boldsymbol{\theta}$ posterior distributions.  We believe this lack of impact on occurrence rates is due to the very small uncertainties for high-reliability targets (see Figure~\ref{figure:reliabilityPostExamples}) combined with the facts that the low-reliability targets occur with less frequency in each inference and that most distributions in Figure~\ref{figure:reliabilityDist} are close to symmetric, so are well-represented by their medians.

\subsection{Variations} \label{section:variations}
In this section we explore the impact of changing some of the inputs and assumptions in the baseline occurrence rate computed in \S\ref{section:baselineResults}.  Our motivation is to understand the dependencies of the occurrence rate on these inputs and assumptions.  In all cases except \S\ref{section:variationsScoreCut} the same models were found to be the best fit to vetting completeness, false alarm effectiveness and observed false alarm rate as in the baseline case, though the parameters of these models had different values for the different variations.

\begin{figure}[ht]
  \centering
  \includegraphics[width=\linewidth]{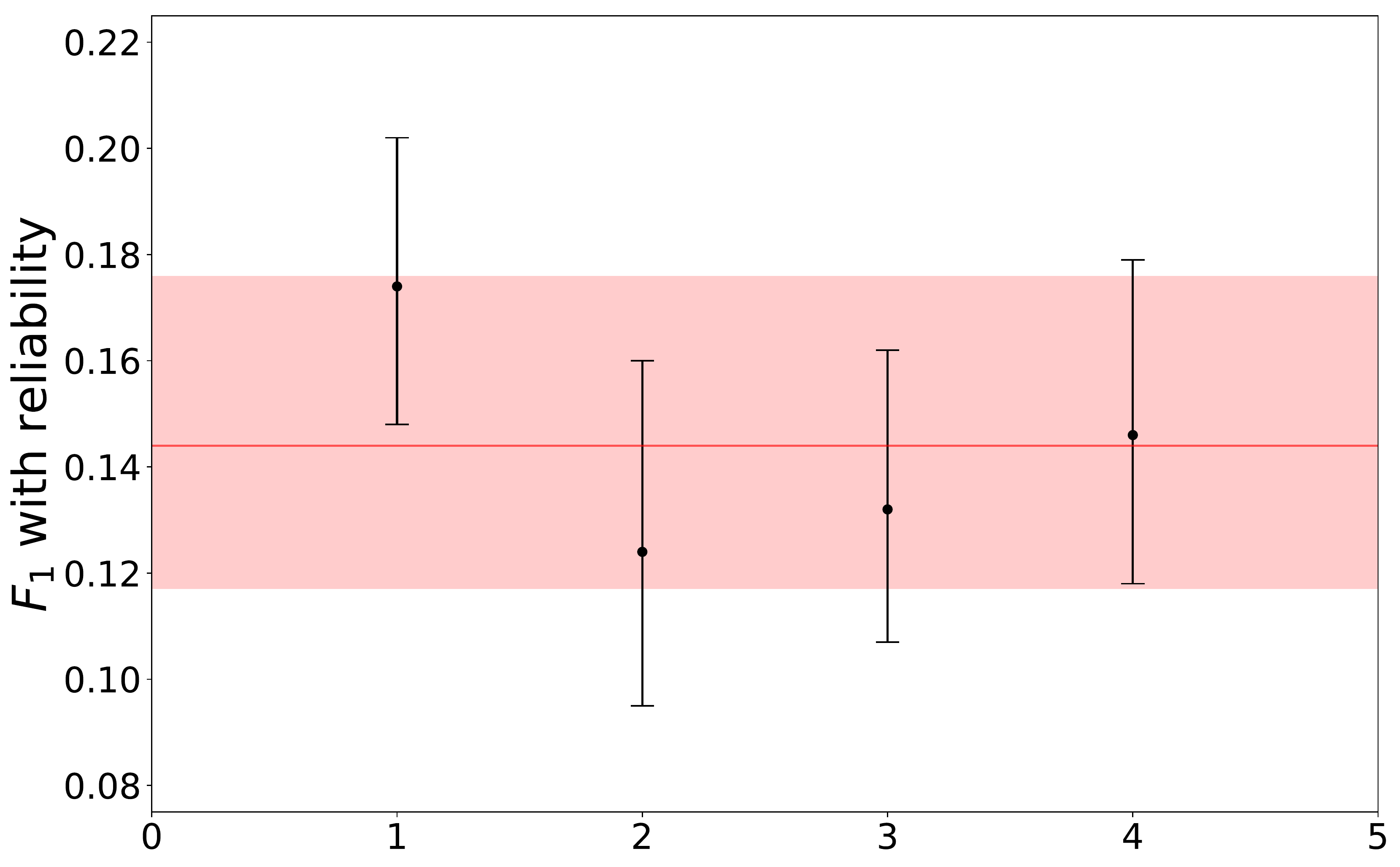} \\
  \includegraphics[width=\linewidth]{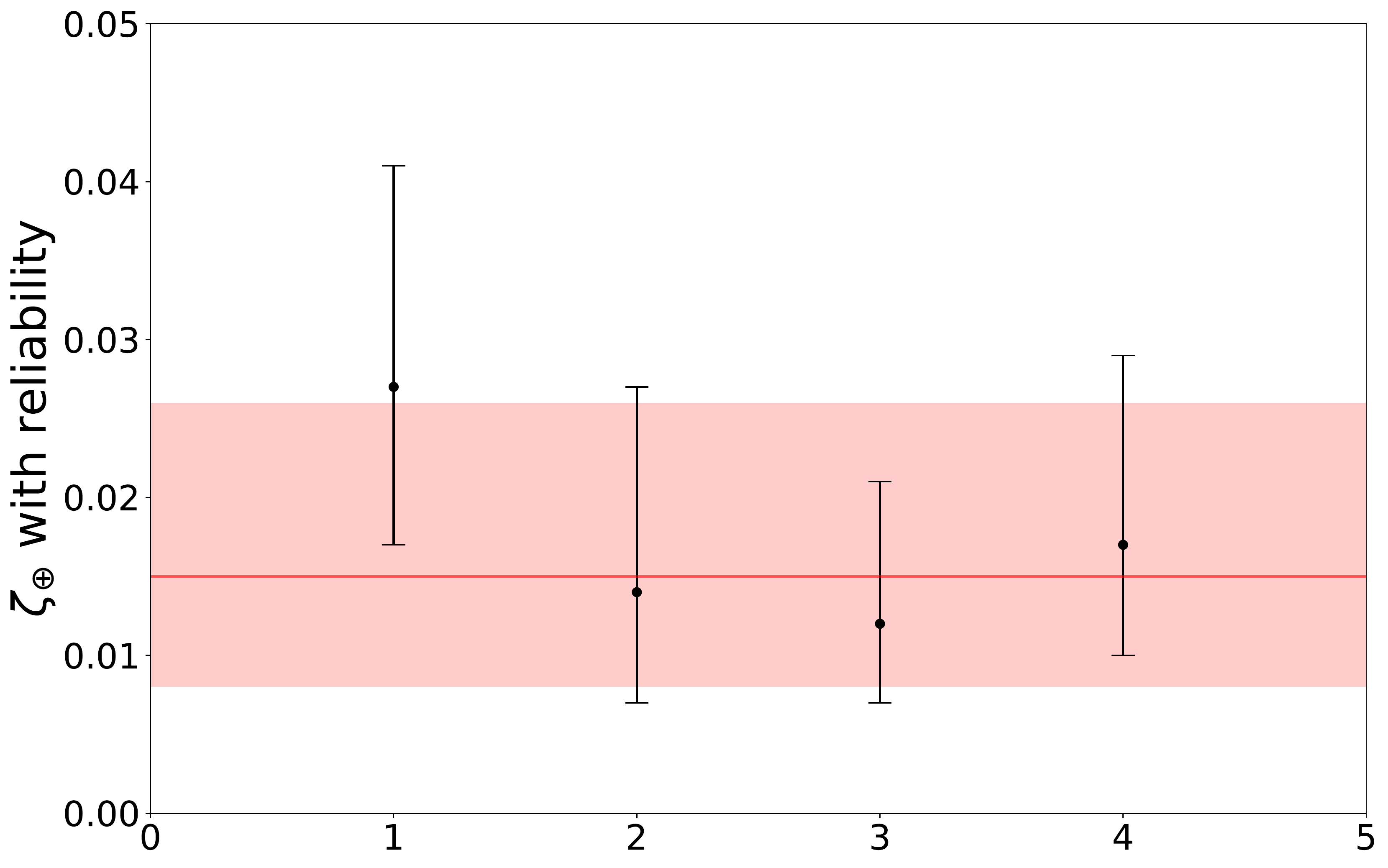}
  \caption{The impact of the variations considered in this section on $F_1$ (top) and $\zeta_{\oplus}$ (bottom) accounting for reliability.  The GK baseline from \S\ref{section:baselineResults} is shown with light red rectangle, with the horizontal line being the central value and the rectangle top and bottom showing the error bars.  The variations are shown at different $x$ locations: 1) using the Q1-Q17 DR25 stellar properties (\S\ref{section:variationsKic}); 2) including only planet candidates with Robovetter score $> 0.9$ (\S\ref{section:variationsScoreCut}); 3) without vetting completeness (\S\ref{section:variationsNoVetCompletness}); 4) without MES smearing (\S\ref{section:variationsNoMesSmear}).
 } \label{figure:variationPic}
\end{figure}

We present the resulting variation in occurrence rates in Table~\ref{table:occurrenceVariations}, which includes results from Table~\ref{table:baselineThetaResults} for comparison.  This comparison is shown graphically for $F_1$ and $\zeta_{\oplus}$ at the end of this section in Figure~\ref{figure:variationPic}.

\subsubsection{Using the Q1-Q17 DR25 Stellar Properties} \label{section:variationsKic}
Our baseline occurrence rates are substantially lower than several occurrence rates based on pre-Gaia stellar properties.  In this section we repeat our analysis, replacing the Gaia-based catalog of \citet{berger20} with the pre-Gaia Q1-Q17 DR25 stellar properties from the NASA Exoplanet Archive\footref{footnote:dr25Stellar}.  We perform the same cuts as described in \S\ref{section:stellarCatalog}, with the exception that there is no cut on binary or evolved flags (these do not exist in the Q1-Q17 DR25 stellar properties) and we remove all stars with radius $> 1.35 R_{\odot}$.  The final catalog contains 75,541 GK stars.

\begin{figure}[ht]
  \centering
  \includegraphics[width=\linewidth]{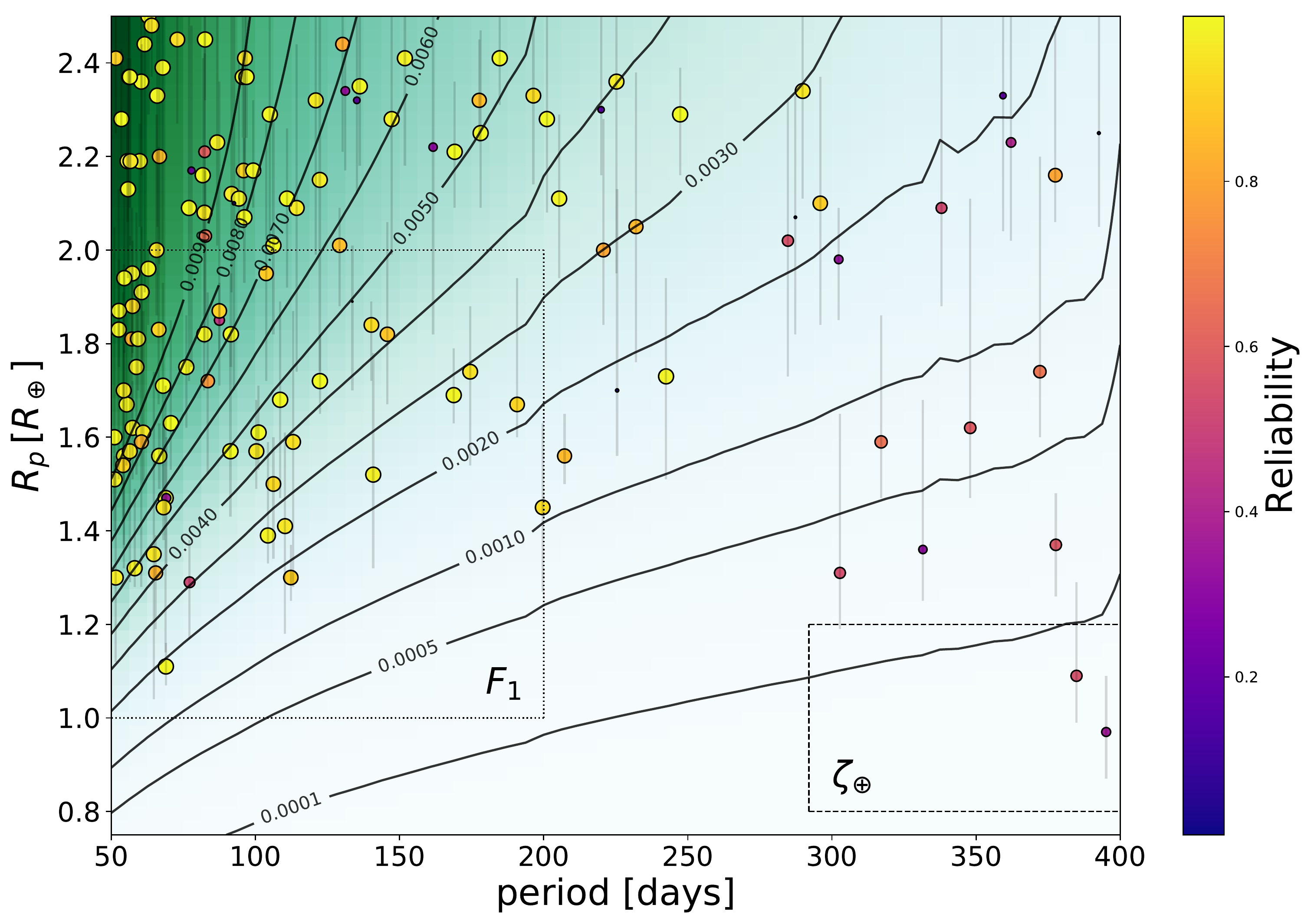}
  \caption{The planet candidate population when using the Q1-Q17 DR25 stellar properties, colored and sized by reliability.  Compared with the baseline population in Figure~\ref{figure:planetPopulation} there are substantially more planets in both the $F_1$ box (on the right) and at period $> 300$ days, leading to higher occurrence rates.  See Figure~\ref{figure:planetPopulation} for a description of the elements of this figure.} \label{figure:planetPopulationKic}
\end{figure}

We perform the same analysis as in the baseline case, starting from computing the vetting completeness for this stellar catalog, computing the summed completeness function $\eta$, the reliability and occurrence rates specified in \S\ref{section:baselineResults}.  Figure~\ref{figure:planetPopulationKic} shows the resulting planet population, summed completeness and reliability.  Comparison with Figure~\ref{figure:planetPopulation} shows that this catalog has more planet candidates in our period-radius range than when using \citet{berger20}.  This results in the higher occurrence rates shown in the ``DR25'' case in Table~\ref{table:occurrenceVariations}.  

The choice of catalog has a stronger impact on $\zeta_{\oplus}$ than on $F_1$: When not correcting for reliability, $F_1$ based on the DR25 stellar properties are  about 15\% higher than our baseline using \citet{berger20}, while the DR25-based $\zeta_{\oplus}$ is about 60\% higher.  When correcting for reliability, $F_1$ based on the DR25 stellar properties are  about 20\% higher, while the DR25-based $\zeta_{\oplus}$ is 80\% higher.  Computing the SAG13 definition of $\eta_{\oplus}$\footref{footnote:sag13}using the DR25 stellar properties without correcting for reliability yields $\eta_{\oplus} = 0.499^{+0.245}_{-0.164}$, while correcting for reliability gives $\eta_{\oplus} = 0.223^{+0.136}_{-0.087}$.  

The \citet{berger20} stellar catalog used in our baseline occurrence rates differs from the DR25 stellar catalog used in this section in both the values of the stellar properties themselves and in the cuts used to define the stellar parent population.  In Appendix~\ref{app:catalogComparison} we study the relative impact of the difference in stellar properties vs. the impact of the different population cuts on the difference in occurrence rates.  We find that the difference in occurrence rates is primarily due to the different stellar properties, primarily stellar radius and effective temperature (leading to different GK selections), and that the differing population cuts have a minor impact.

\subsubsection{Baseline with a Score Cut of 0.9} \label{section:variationsScoreCut}
The Robovetter outputs a score for each TCE, indicating the confidence with which the Robovetter vetted that TCE \citep{Thompson2018}. This score is not equivalent to reliability: for example the Robovetter confidently vetted several TCEs in the inverted/scrambled data incorrectly as PC with scores as high as  0.923.  But score is roughly correlated with reliability, and \citet{Thompson2018} suggests computing high-reliability occurrence rates by considering only planet candidates with Robovetter score above some threshold. This will result in a smaller planet candidate population with lower completeness, but the resulting larger completeness correction will, in principle, correct the occurrence rate.

In this variation we impose an aggressive score cut, rejecting any planet candidate with score $<0.9$. We use the \citet{berger20} catalog, and compute the completeness and reliability as in the baseline case, treating any TCE with score $<0.9$ as a false positive/alarm.  \citet{Mulders2018} uses this score cut in their analysis, but their analysis is on a very different period-radius range so is not directly comparable to our results.

\begin{figure}[ht]
  \centering
  \includegraphics[width=\linewidth]{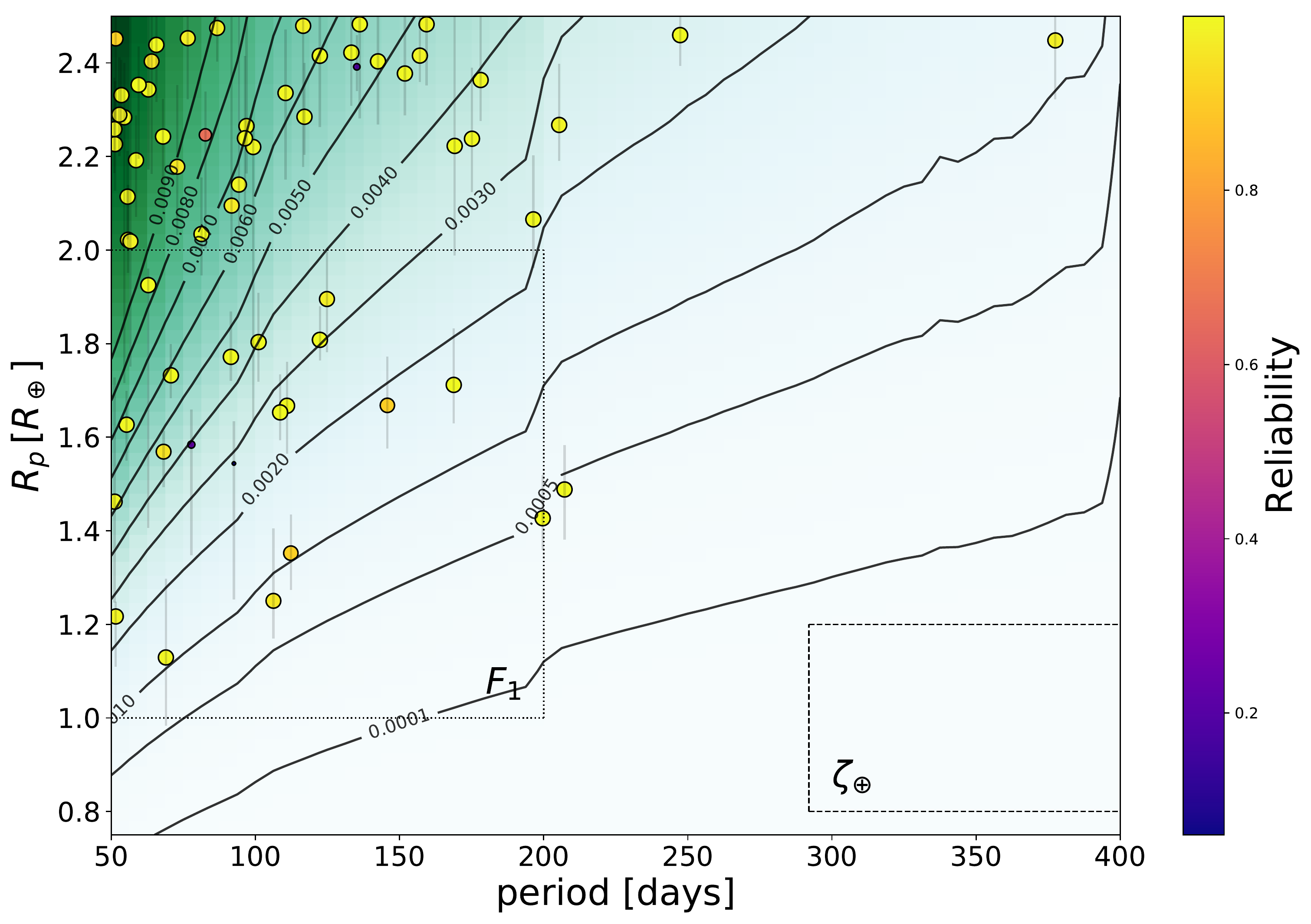}
  \caption{The planet candidate population when using the baseline \citet{berger20} stellar properties but only including planet candidates with a Robovetter score $\geq 0.9$, colored and sized by reliability.  Compared with the baseline population in Figure~\ref{figure:planetPopulation} there are substantially fewer planets in both the $F_1$ box (on the right) and at period $> 300$ days, but also lower completeness leading to larger completeness corrections.  Note the complete absence of small planets with orbital period $> 220$ days.  See Figure~\ref{figure:planetPopulation} for a description of the elements of this figure.} \label{figure:planetPopulationSc0p9}
\end{figure}

The result is a smaller, higher reliability planet candidate population, as shown in Figure~\ref{figure:planetPopulationSc0p9}, with noticeably lower completeness (compare the contours in Figure~\ref{figure:planetPopulation}).   In this case the false alarm vetting efficiency was best fit with the constant $=0.999$, resulting in a false alarm reliability very close to 1 for the entire period-radius range.  The few planet candidates with lower reliability in Figure~\ref{figure:planetPopulationSc0p9} are due to their astrophysical false positive probability, which results in $F_1$ and $\zeta_{\oplus}$ being slightly suppressed as shown in the ``Score $>0.9$'' case in Table~\ref{table:occurrenceVariations}. This is an illustration of the fact that score cuts cannot be relied on to provide a population that is high reliability with respect to astrophysical false positives.

The agreement in $\zeta_{\oplus}$ when using this score cut and the baseline given in Table~\ref{table:baselineThetaResults} is remarkable given the lack of planet candidates smaller than $2R_{\oplus}$ and orbital period $> 220$ days shown in Figure~\ref{figure:planetPopulationSc0p9} (compare Figure~\ref{figure:planetPopulation}).  We interpret this agreement as an indication that the baseline $\zeta_{\oplus}$ is dominated by extrapolation because, in the baseline population, long-period, small planets have low reliability, as discussed in \S\ref{section:baselineResults}.  Because the baseline results and those using requiring score $>0.9$ are essentially unconstrained extrapolations from radius $>2 R_{\oplus}$ and orbital period $< 220$ days to smaller planets and longer periods, we believe it is premature to conclude that using this score cut provides accurate occurrence rates for radius $<2 R_{\oplus}$ and orbital period $> 220$ days.  In \S\ref{section:discussion} we propose a strategy to explore this question.

\subsubsection{Baseline Without Vetting Completeness} \label{section:variationsNoVetCompletness}
This variation measures the impact of not including vetting completeness.  This will result in a smaller completeness correction where vetting completeness is low, so we expect somewhat lower long-period, small planet occurrence rates.  The ``No Vetting Efficiency'' case in Table~\ref{table:occurrenceVariations} shows a small suppression in $\Gamma_{\oplus}$, $F_1$ and $\zeta_{\oplus}$ when not accounting for vetting completeness.

\subsubsection{Baseline Without MES Smearing} \label{section:variationsNoMesSmear}
This variation measures the impact of not smearing the MES in the calculation of detection completeness, described in \S\ref{section:detectionCompleteness}.  The ``No MES Smear'' case in Table~\ref{table:occurrenceVariations}  indicates an increase in small-planet, long-period occurrence rates measured by increases in $\Gamma_{\oplus}$ and $\zeta_{\oplus}$, but not smearing the MES has essentially no impact on $F_1$.

\renewcommand{\arraystretch}{2}
\begin{table*}[ht]
\centering
\caption{Comparison of occurrence rate variations.} \label{table:occurrenceVariations}
\begin{tabular}{ l c c c c c c}
\hline
& \multicolumn{2}{c}{$F_0$} & \multicolumn{2}{c}{$\alpha$} & \multicolumn{2}{c}{$\beta$} \\
Case  & No Reliability & With Reliability & No Reliability & With Reliability & No Reliability & With Reliability \\
\hline
 \parbox[c]{2.1cm}{Baseline\\(from Table \ref{table:baselineThetaResults})} & $0.608^{+0.110}_{-0.090}$ & $0.432^{+0.089}_{-0.072}$ & $0.304^{+0.519}_{-0.496}$ & $0.796^{+0.635}_{-0.598}$ & 
 $-0.557^{+0.174}_{-0.169}$ & $-0.823^{+0.202}_{-0.209}$ \\
 DR25 & $0.675^{+0.115}_{-0.097}$ &  $0.474^{+0.090}_{-0.076}$ & $-0.517^{+0.402}_{-0.389}$ &  $-0.339^{+0.465}_{-0.444}$ & 
 $-0.552^{+0.153}_{-0.155}$ & $-0.888^{+0.188}_{-0.194}$ \\
 \parbox[c]{2.1cm}{Score $>0.9$} & $0.418^{+0.112}_{-0.084}$ &  $0.382^{+0.105}_{-0.079}$ & $0.616^{+0.708}_{-0.677}$ &  $0.780^{+0.775}_{-0.724}$ & 
 $-0.774^{+0.244}_{-0.253}$ & $-0.768^{+0.253}_{-0.263}$ \\
 \parbox[c]{2.1cm}{No Vetting\\Efficiency} & $0.554^{+0.101}_{-0.084}$ &  $0.389^{+0.079}_{-0.062}$ & $0.388^{+0.502}_{-0.502}$ &  $0.889^{+0.637}_{-0.603}$ & 
 $-0.609^{+0.177}_{-0.175}$ & $-0.889^{+0.203}_{-0.206}$ \\
 \parbox[c]{2.1cm}{No MES\\Smear} & $0.632^{+0.125}_{-0.101}$ &  $0.434^{+0.093}_{-0.073}$ & $0.104^{+0.508}_{-0.485}$ &  $0.666^{+0.632}_{-0.593}$ & 
 $-0.527^{+0.171}_{-0.178}$ & $-0.802^{+0.202}_{-0.209}$ \\
& \multicolumn{2}{c}{$\Gamma_{\oplus}$} & \multicolumn{2}{c}{$F_1$} & \multicolumn{2}{c}{$\zeta_{\oplus}$} \\
Case  & No Reliability & With Reliability & No Reliability & With Reliability & No Reliability & With Reliability \\
\hline
 \parbox[c]{2.1cm}{Baseline\\(from Table \ref{table:baselineThetaResults})} & $0.212^{+0.111}_{-0.075}$ & $0.094^{+0.066}_{-0.041}$ & $0.190^{+0.035}_{-0.030}$ & $0.144^{+0.032}_{-0.027}$ & 
 $0.034^{+0.018}_{-0.012}$ & $0.015^{+0.011}_{-0.007}$ \\
 DR25 & $0.334^{+0.134}_{-0.098}$ &  $0.164^{+0.084}_{-0.059}$ & $0.218^{+0.032}_{-0.029}$ &  $0.174^{+0.028}_{-0.026}$ & 
 $0.054^{+0.022}_{-0.016}$ & $0.027^{+0.014}_{-0.010}$ \\
 Score $>0.9$ & $0.103^{+0.089}_{-0.049}$ &  $0.087^{+0.081}_{-0.044}$ & $0.139^{+0.038}_{-0.030}$ &  $0.124^{+0.036}_{-0.029}$ & 
 $0.017^{+0.014}_{-0.008}$ & $0.014^{+0.013}_{-0.007}$ \\
 \parbox[c]{2cm}{No Vetting\\Efficiency} & $0.178^{+0.097}_{-0.064}$ &  $0.076^{+0.054}_{-0.033}$ & $0.176^{+0.031}_{-0.028}$ &  $0.132^{+0.030}_{-0.025}$ & 
 $0.029^{+0.016}_{-0.010}$ & $0.012^{+0.009}_{-0.005}$ \\
 \parbox[c]{2cm}{No MES\\Smear} & $0.246^{+0.132}_{-0.089}$ &  $0.103^{+0.072}_{-0.044}$ & $0.198^{+0.036}_{-0.033}$ &  $0.146^{+0.033}_{-0.028}$ & 
 $0.040^{+0.021}_{-0.014}$ & $0.017^{+0.012}_{-0.007}$ \\
\end{tabular}
\end{table*}

\section{Discussion} \label{section:discussion}

In this paper we show that a proper characterization of vetting completeness and reliability is important, particularly near the detection limit.  In particular, in \S\ref{section:baselineResults} we find that characterizing \Kepler\ reliability and completeness can impact occurrence rates by more than a factor of two near \Keplers\ detection limit (see Table~\ref{table:baselineThetaResults}). We introduce a new approach to characterizing vetting completeness and reliability for the \Kepler\ DR25 planet candidate population.  This approach casts the problem as one of binomial probabilities via parameterized rate functions fitted to the DR25 injection, inverted and scrambled data.  We develop parameterized models of completeness (described in \S\ref{section:completeness}), false alarm effectiveness (\S\ref{section:FPEcharacterization}) and the observed false alarm rate (\S\ref{section:obsFPcharacterization}).  The particular parametric models we choose are selected via the Akieke information criterion, which chooses the parametric model that maximizes the likelihood corrected for the number of model parameters (see Appendix~\ref{app:modelSelect}).  We do not claim that our parametric models are the best or in any sense ``true'', just that they are the best of the parametric models we considered, described in Table~\ref{table:modelDefinition}.  But our best models do a good job of accounting for the data, and are robust against choices such as grid resolution.  

We caution, however, that vetting completeness and false alarm reliability as defined in this paper are properties of the specific Robovetter metrics and vetting thresholds behind the DR25 planet candidate catalog, as well as our analysis method, rather than properties of the detections themselves.  For example, a different choice of Robovetter metrics may increase completeness while decreasing reliability or vice versa.  While a low reliability for a transit detection from the analysis in this paper is reason to be cautious about asserting that detection is due to a true planet, further analysis of \Kepler\ data can potentially result in higher confidence that a transit signal is due to a true planet.  For example, as described in \S\ref{section:vettingIntro}, at least one major source of false alarms, rolling bands, is highly dependent on focal plane position \citep{VanCleve2009,VanCleve2016}.  Though some of the DR25 vetting metrics, such as {\it skye} \citep{Thompson2018} are focal-plane dependent, the reliability analysis in this paper largely ignores focal plane dependence by averaging over the focal plane, and potentially underestimates the reliability of a detection in a focal plane position known to have a low occurrence of, for example, rolling bands.  Pixel-level analysis of transit events beyond that used in DR25 may be useful in distinguishing false alarms due to statistical fluctuations and cosmic ray events.  These observations can potentially be implemented as new Robovetter metrics, which could result in a higher-reliability, more complete planet candidate catalog.  

The four years of \Keplers\ observation of its primary field provides a data set unlikely to be excelled in the near future.  Full exploitation of this data for understanding exoplanet populations is only partially complete.  This paper is an attempt to fill in a significant step in that exploitation. We deliberately chose to limit the innovations in this paper to the characterization of and correction for completeness and reliability, and the use of the uniform Gaia-based stellar properties catalog of \citet{berger20}. We show the impact of these innovations by computing occurrence rates using standard methods from \citet{Burke2015} in order to facilitate comparison with previous occurrence rates based on similar methods.  The following discussion critically examines the assumptions underlying these occurrence rates, revealing weaknesses in both the DR25 catalog and the occurrence rate calculation method, and outlines some of the directions that we believe will prove fruitful in addressing these weaknesses.  

\subsection{Assumptions Underlying the Baseline Occurrence Rate} \label{section:baselineDiscussion}
We illustrate the impact of reliability by computing a variety of occurrence rates near the \Kepler\ detection limit (see \S\ref{section:occurrence}).  We chose our specific occurrence rate method, Bayesian inference using a dual power law population model in period and radius, because it is standard and well-understood.  {\bf We believe that our occurrence rates provide high-confidence insight into what the DR25 planet candidate catalog tells us about the exoplanet population for the period and radius range of $50 \leq \mathrm{period} \leq 400$ days and $0.75 \leq \mathrm{radius} \leq 2.5$ $R_{\oplus}$ under the following assumptions:} 
\begin{itemize}
    \item The parent stellar population is statistically well-described by \citet{berger20}.
    \item Detection and vetting completeness in the injected data, along with the analysis described in \S\ref{section:completeness}, capture the statistical behavior of detection and vetting completeness in the observed data.
    \item The false alarms in the inverted and scrambled data capture the statistical behavior of the false alarms in the observed data.
    \item The astrophysical false positives probability statistically captures the probability of astrophysical false positives in the planet candidates. 
    \item Exoplanets are distributed according to a Poisson point process.
    \item Dependence of planet occurrence on period and radius is modeled by a product of power laws in period and radius.
\end{itemize}
We discuss each of these assumptions in turn.

{\bf The parent stellar population:} As stated in \S\ref{section:stellarCatalog}, we choose the \citet{berger20} stellar properties because they are informed by Gaia radii and uniformly treat the stellar properties across the parent population.  While detailed observations of individual stars may provide more accurate stellar properties for individual stars, our method requires statistically uniform analysis. This is provided by the isochrone fitting approach of \citet{berger20}.  Therefore we believe that this stellar catalog is very well suited to our analysis.  We showed in \S\ref{section:variationsKic}, however, that the occurrence rate depends critically on the stellar properties in the parent population.  Inaccuracies, in particular biases in stellar radius estimates, can have a strong impact on occurrence rates.

{\bf Detection and vetting completeness:} The injected data and analysis from \S\ref{section:completeness} makes many assumptions.  The detection completeness analysis makes several empirical approximations (described in \citet{BurkeJCat2017}) that may not apply well to individual planet candidate host stars.  As described in \S\ref{section:catalogs}, some of our stellar and planet candidate population approaches the restrictions stated in \citet{BurkeJCat2017} with respect to transit duration.  Because our occurrence rates include regions with very few planet candidates, it is possible that the detection completeness for long-period planets is not as well modeled as that for short-period planets.  Regarding vetting completeness, we are assuming that the Robovetter vets the injected detections with the same statistical accuracy as real transiting planets in the observed data.  While we have confidence in this assumption, it is possible that some true planet transit signals have properties not captured by injection which confound the Robovetter, such as asymmetric transit shapes due to non-zero eccentricity, transit timing variations, and out-of-transit flux variations.

{\bf False alarm characterization:} As we discussed in \S\ref{section:reliabilityModel}, inverted and scrambled data is believed to statistically model three identified classes of false alarms: rolling bands, statistical fluctuations combined with cosmic ray-induced pixel events that conspire to imitate long-period small transiting planets and stellar variability.  The evidence for this belief is that the distribution of detected TCEs in the inverted and scrambled data closely matches the clearly anomalous distribution of detections in the observed data centered on the \Kepler\ orbital period (see \citet{Thompson2018}), and tuning the Robovetter to eliminate this distribution from the PC population in the inverted and scrambled data also eliminates the anomalous distribution in the observed data.  While using the inverted and scrambled data to model the false alarm population is clearly effective, there are likely other types of false alarms not represented by the inverted/scrambled data, though these seem to be a minor component compared to those represented by inverted/scrambled data.  Such unmeasured false alarms would cause an overestimate of false alarm reliability.

{\bf Astrophysical false positive characterization:} The false positive probabilities \citep{Morton2016} are computed making strong assumptions about the lack of evidence for stellar multiplicity associated with the transit host star.  While these false positive probabilities model stellar multiplicity as candidate hypotheses, the prior used in this model strongly assumes a lack of evidence for stellar multiplicity.  As described in \citet{Hirsch2017} and \citet{Ciardi2015}, there is evidence that a non-trivial fraction, possibly 20\%, of \Kepler\ target stars have unknown stellar companions.  Such companions could cause an overestimate of the reliability of a subset of the PC population.  

{\bf Poisson Likelihood:} The use of the Poisson likelihood (Equation~(\ref{eqn:planetLilkelihood})) for the distribution of exoplanets is a standard choice, but may not be correct.  For example, the assumption that the probability of different planets on the same star are independent of one another (an assumption behind Equation~(\ref{eqn:planetLilkelihood})) is almost certainly not correct, as indicated by existence of many packed exoplanet systems.  There is also evidence that the detection of one planet on a star can prevent the detection of other planets on the same star \citep{zinkChristiansen2019}. Likelihood-free methods, such as approximate Bayesian computation as applied to occurrence rates in \citet{Hsu2018} or the population sampling method used in \citet{zink2019} may yield more accurate occurrence rates.

{\bf The power law population model:} Evidence is mounting against the use of a simple product of power laws in period and radius when modeling exoplanet population statistics.  This is already apparent in the top-left panel of Figure~\ref{figure:occReliabilityMarginals}, where the power law is a poor fit to the observed planet population as a function of radius.  This is likely due to the Fulton gap \citep{Fulton2017}, though the orbital periods in our analysis are somewhat longer than in Fulton's analysis.  Further, \citet{petigura2018} presents evidence that host star metallicity is an important parameter in exoplanet population statistics.  As pointed out by \citet{Hsu2018}, model mis-specification is unlikely to lead to accurate results. Several authors have avoided the use of parameterized models in occurrence rate computations (for examples, see \citet{ForemanMackey2014,Hsu2018,Howard2012}), which is likely to lead to more accurate occurrence rates.  

We believe that whatever method of statistical analysis is applied to the planet candidate catalog, characterizing and correcting for vetting completeness and reliability is critical.  In the long-period, small-planet regime we have shown that reliability can reduce occurrence rates by a factor of two.  We expect that this will be the case regardless of the statistical method and model, because reliability is a property of the planet candidate catalog.  The effect of vetting completeness is less dramatic in the DR25 planet candidate catalog (as opposed to detection completeness, which is very important), but should not be neglected.  Other planet catalogs may increase vetting reliability at the expense of vetting completeness, in which case vetting completeness can be more significant.

For habitability studies, the common practice of grouping together a wide class of stars and computing occurrence rates as functions of period and radius is potentially misleading.  For example, the large range of stellar luminosities in our GK population shown in Figure~\ref{figure:stellarLum} means that not all stars share the same habitable zones expressed as orbital periods.  But grouping such a wide class of stars is necessary to provide the required statistics due to the small number of long-period, small planet detections and the sparseness of false alarms described in \S\ref{section:FPEcharacterization}.  Occurrence rates computed as functions of insolation flux and planet radius for the same class of stars would provide the needed statistics, and are likely more informative for habitable exoplanet population studies.  Recent improvements in stellar characterization of the parent stellar sample, represented by \citet{berger20}, make insolation-flux based occurrence rates a viable alternative.  

\subsection{Improving the Planet Candidate Catalog} \label{section:catalogDiscussion}
The discussion in \S\ref{section:baselineDiscussion} outlines the assumptions behind extracting our occurrence rate from the DR25 planet candidate catalog, and how those assumptions may fail.  The DR25 catalog itself can likely be improved upon, particularly in the long-period small planet regime.  There is evidence that several long-period, small radius detections were incorrectly classified as false positives: the \Kepler\ False Positive Working Group (FPWG) \citep{Bryson2015} has identified several TCEs vetted as false positive in DR25 that are viable planet candidates, identified with {\it fpwg\_disp\_status} = POSSIBLE PLANET in the \Kepler\ certified false positive table at the NASA Exoplanet archive\footref{footnote:exoplanetArchive}.  This is expected because the DR25 vetting process deliberately balanced statistical uniformity and accuracy for individual objects, which was required for the study in this paper but compromised accurate vetting for some objects.  In principle, the lowered completeness resulting from mis-classifying true planets as false positives is corrected by characterizing vetting completeness.  But in the long-period, small-planet regime there are very few, low-reliability detections and very large completeness corrections (see Figure~\ref{figure:planetPopulation}), which is vulnerable to large errors due to small statistics.  

Accurate characterization of completeness and reliability as developed in this paper opens an intriguing approach to addressing the problem of few detections at long period and small radius: planet candidate catalogs that have lower reliability and higher completeness.  This would mitigate the small-statistics problem by providing more detections with a smaller completeness correction resulting in better statistical constraints on the extrapolations discussed in \S\ref{section:variationsScoreCut}. We recommend an exploration of Robovetter thresholds that increase the number of detections, lowering reliability and increasing completeness.  The methods to measure reliability described in this paper are a crucial step towards being able to extract more accurate occurrence rates from such a catalog.  

We believe that the reliability of the planet candidate catalog can be improved by the development of metrics beyond those described in \citet{Thompson2018}.  We provide two examples that may prove fruitful.  
\begin{itemize}
    \item As described above, different regions of the \Kepler\ focal plane have different different false alarm characteristics, which can be leveraged to more accurately evaluate the likelihood that a transit signal is due to a false alarm.  
    \item Pixel-level analysis can be developed beyond the DR25 vetting metrics, based on the expectation that false alarms are likely to be significantly different from star-like transit signals at the pixel level, particularly in difference images  \citep{Bryson2013}.
\end{itemize}
We expect that such improved vetting metrics will address the small statistics problem by increasing the reliability of planet candidates near the detection limit, so they have stronger statistical weight in occurrence rate calculations.

Followup observation can potentially play a role in validating the reliability characterization developed in this paper.  Ground- or space-based observations of a significant number of DR25 PCs, confirming them as planets or determining them to be false positives, could provide a ground truth of the number of PCs that are true planets.  This ground truth can be used to independently compute the reliability of the DR25 PC population.  We caution against using such followup observations to modify the planet candidate catalog, however, as that is likely to violate the uniformity assumptions behind the completeness correction.  

\section{Conclusion}
This paper presents a new, probabilistic approach to statistically characterizing the vetting completeness and reliability \Kepler\ DR25 exoplanet catalog.  Using a standard occurrence rate calculation, we demonstrated that correcting for reliability can have a significant impact on occurrence rates, particularly near the \Kepler\ detection limit at orbital periods longer than 200 days and planet radius $<$1.5~R$_{\oplus}$.  We also showed that the choice of stellar properties for the searched stellar sample has a significant effect on occurrence rates.  The results in this paper were made possible by the uniform detection and vetting methods behind DR25 that lend themselves to statistical characterization.  We believe that the results presented in this paper are directly applicable to other exoplanet surveys such as K2, TESS, and PLATO so long as they create their catalogs in a similarly uniform way and expend the effort to create test data sets that measure completeness and false positives.  

\acknowledgments
We thank NASA, \Kepler\ management, and the Exoplanet Exploration Office for continued support of and encouragement for the analysis of \Kepler\ data. We thank Bill Borucki and the \Kepler\ team for the excellent data that makes these studies possible.  We thank Daniel Foreman-Mackey for his Python Jupyter notebooks, which kicked off and inspired the work presented in this paper. T.A.B and D.H. acknowledge support by the National Science Foundation (AST-1717000).

%

\vspace{5mm}
\facilities{\Kepler}


\software{Jupyter, emcee \citep{ForemanMackeyEmcee}, KeplerPorts \citep{BurkeJCat2017}}



\clearpage

\appendix

\section{Vetting Completeness and Reliability Model Selection} \label{app:modelSelect}

We investigated a variety of models for vetting completeness (\S\ref{section:vettingCompleteness}), false alarm efficiency (\S\ref{section:FPEcharacterization}) and observed false alarm rate (\S\ref{section:obsFPcharacterization}).  We selected the models based largely on the lowest Akiake Information Criterion $\mathrm{AIC} = 2d - 2\ln(L)$, where $d$ is the number of degrees of freedom in the model and $L$ is the likelihood of the model.  When comparing two models with likelihoods $L_1$ and $L_2$, the relatively likelihood of model 1 relative to model 2 is given by $\exp((\mathrm{AIC}_2 - \mathrm{AIC}_1)/2$.  The AIC criterion is not always successful, however, particularly when the model contains parameters that do not converge.  In addition, the lowest AIC sometimes results in models that are obviously not physical.  In such cases we made judgement calls when making the selections, as described below.

Details of each model's analysis is found on the GitHub website\footref{footnote:github} in the directory {\it GKbaseline/htmlArchive}.

The models we considered are defined in Table~\ref{table:modelDefinition}.  All models have as input orbital period $p$, MES (either expected for vetting completeness, or observed for reliability) $m$, orbital period range $[ p_{\min}, p_{\max} ]$, MES range $[ m_{\min}, m_{\max} ]$ and a parameter vector $\boldsymbol{\theta}$.  $\boldsymbol{\theta}$ has different elements for different models.  The function evaluations start with scaling the period and MES input to $(x,y) \in [0, 1] \times [0, 1]$ in order to facilitate rotations.  This scaling is done for consistency even if there are no rotations in the model.  In some models, rotations are applied using angles $\phi_x$ and $\phi_y$ from $\boldsymbol{\theta}$.  If there is only one rotation angle $\phi$ in $\boldsymbol{\theta}$, then $\phi_x = \phi_y = \phi$.  
\begin{equation} \label{eqn:funcXform}
\begin{split}
    x =& \frac{\left( p - p_{\min}\right)}{\left( p_{\mathrm{max}} - p_{\min}\right)} \\
    y =& \frac{\left( m - m_{\min}\right)}{\left( m_{\mathrm{max}} - m_{\min}\right)} \\
    x_{\mathrm{rot}} =& (x - 0.5)*\cos(\phi_x) - (y - 0.5)*\sin(\phi_x) \\
    y_{\mathrm{rot}} =& (y - 0.5)*\cos(\phi_y) - (x - 0.5)*\sin(\phi_y) \\
\end{split}
\end{equation}
Many of the functions are constructed from the logistic function $Y\left(x, x_0, k, \nu \right)$ defined in equation~(\ref{eqn:logistic}).  We also use the broken power law and non-normalized two-dimensional Gaussian
\begin{equation} \label{eqn:funcComp}
    B(x, b, \alpha, \beta) = 
    \begin{cases}
        \left( \frac{x+1}{b+1}  \right)^\alpha & x < b \\
        \left( \frac{x+1}{b+1}  \right)^\beta & x \geq b
    \end{cases}
    \qquad  G(x, x_0, y, y_0, \sigma_x, \sigma_y) = 
    \exp \left(-\frac{(x-x_0)^2}{2 \sigma_x^2}-\frac{(y-y_0)^2}{2 \sigma_y^2} \right).
\end{equation}

\renewcommand{\arraystretch}{1.5}
\begin{table}[ht]
\centering
\caption{Model definitions} \label{table:modelDefinition}
\footnotesize
\begin{tabularx}{\textwidth}{ r c X X  }
\hline
Model Name & $\boldsymbol{\theta}$ & Model Definition & Description\\
\hline
constant & $[c]$ & $c$ & Constant function \\
gaussian & $[x_0, y_0, \sigma_x, \sigma_y, b]$ 
    & $A \, G(x, x_0, y, y_0, \sigma_x, \sigma_y) + b$ & Gaussian + background\\
dualBrokenPowerLaw & $[b_x, b_y, \alpha_x, \beta_x, \alpha_y, \beta_y, A]$ 
    & $A \, B(x, b_x, \alpha_x, \beta_x) \newline
        \times B(y, b_y, \alpha_y, \beta_y)$  & Broken power law in $x$ and $y$\\
logisticY0 & $[y_0, k, A]$ & $A \, Y \left(y, y_0, k, 1 \right)$ & $y$ Logistic \\
logisticY & $[y_0, k, A, b]$ & $A \, Y \left(y, y_0, k, 1 \right) + b$ & $y$ Logistic + constant \\
rotatedLogisticY & $[y_0, k, \phi, A, b]$ & $A \, Y \left(y_{\mathrm{rot}} + 0.5, y_0, k, 1 \right) + b$ & Rotated $y$ logistic \newline + constant \\
rotatedLogisticX0 & $[x_0, k, \phi, A]$ & $A \, Y\left(x_{\mathrm{rot}} + 0.5, x_0, -k, 1 \right)$ & Rotated $x$ logistic \\
rotatedLogisticX02 & $[x_0, k, \phi, \nu, A]$ & $A \, Y\left(x_{\mathrm{rot}} + 0.5, x_0, -k, \nu \right)$ & Rotated $x$ logistic \newline
    w/ shape parameter \\
rotatedLogisticX0+gaussian & $[x_0, k, \phi, a_0, b_0, \sigma_x, \sigma_y, \gamma, A]$ 
    & $A \, Y\left(x_{\mathrm{rot}} + 0.5, x_0, -k, 1 \right) \newline
        + \gamma \, G(x, a_0, y, b_0, \sigma_x, \sigma_y)$ & Rotated $x$ logistic \newline + Gaussian \\
logisticX0xlogisticY0 & $[x_0, y_0, k_x, k_y, A]$ 
    & $A \, Y\left(x, x_0, -k, 1 \right) \newline
        \times Y \left(y, y_0, k, 1 \right)$ & $x$ logistic times $y$ logistic \\
logisticX0xlogisticY02 & $[x_0, y_0, k_x, k_y, \nu_x, \nu_y, A]$ 
    & $A \, Y\left(x, x_0, -k, \nu_x \right) \newline
        \times Y \left(y, y_0, k, \nu_y \right)$ & $x$ logistic times $y$ logistic  \newline
        w/ shape parameters \\
logisticX0xRotatedLogisticY0 & $[x_0, y_0, k_x, k_y, \phi, A]$ 
    & $A \, Y\left(x, x_0, -k, 1 \right) \newline
        \times Y \left(y_{\mathrm{rot}} + 0.5, y_0, k, 1 \right)$ & $x$ logistic times rotated $y$ logistic \\
logisticX0xRotatedLogisticY02 & $[x_0, y_0, k_x, k_y, \nu, \phi, A]$ 
    & $A \, Y\left(x, x_0, -k, 1 \right) \newline
        \times Y \left(y_{\mathrm{rot}} + 0.5, y_0, k, \nu \right)$ & $x$ logistic times \newline rotated $y$ logistic   \newline
        w/ shape parameter \\
rotatedLogisticYXLogisticY & $[y_0, y_1, k_0, k_1, \phi, A, b]$ 
    & $A \, Y\left(y_{\mathrm{rot}} + 0.5, y_0, k_0, 1 \right) \newline
        \times Y \left(y, y_1, k_1, 1 \right) + b$ & $y$ logistic times rotated $y$ logistic  + constant\\
rotatedLogisticX0xlogisticY0 & $[x_0, y_0, k_x, k_y, \phi_x, \phi_y, A]$ 
    & $A \, Y\left(x_{\mathrm{rot}} + 0.5, x_0, -k, 1 \right) \newline
        \times Y \left(y_{\mathrm{rot}} + 0.5, y_0, k, 1 \right)$ & Rotated $x$ logistic \newline
            times rotated $y$ logistic \\
rotatedLogisticX0xlogisticY02 & $[x_0, y_0, k_x, k_y, \nu_x, \nu_y, \phi_x, \phi_y, A]$ 
    & $A \, Y\left(x_{\mathrm{rot}} + 0.5, x_0, -k, \nu_x \right) \newline
        \times Y \left(y_{\mathrm{rot}} + 0.5, y_0, k, \nu_y \right)$ & Rotated $x$ logistic \newline times rotated $y$ logistic \newline 
        w/ shape parameters \\
rotatedLogisticYXFixedLogisticY & $[y_0, k, \phi, A, b]$ 
    & $A \, Y\left(y_{\mathrm{rot}} + 0.5, y_0, k_0, 1 \right) \newline
        \times Y \left(y, 0.25, 33.331, 1 \right) + b$ & fixed $y$ logistic times  \newline rotated 
        $y$ logistic \newline  + constant
\end{tabularx}
\tablecomments{The function Y is defined in equation~(\ref{eqn:logistic}), and the functions G and B are defined in equation~(\ref{eqn:funcComp}). }
\end{table}

\subsection{Vetting Completeness Model Selection} \label{app:vetCompModelSelect}
The functions considered for fitting the observed vetting completeness rate in \S\ref{section:vettingCompleteness} are given in Table~\ref{table:vetCompModelSelection}, along with AIC values and relative likelihoods.  We chose logisticX0xRotatedLogisticY0 because it had the highest relative likelihood, best convergence behavior, and appears to be a good fit to the data.

\renewcommand{\arraystretch}{1.5}
\begin{table}[ht]
\centering
\caption{Candidate vetting completeness rate functions} \label{table:vetCompModelSelection}
\begin{tabular}{ r c c c  }
\hline
Model Name & Median AIC & Minimum AIC & Relative Likelihood \\
\hline
logisticY0 & 2819.03 & 2819.03 & 2.68e-137 \\
dualBrokenPowerLaw & 2246.87 & 2245.45 & 4.70e-13 \\
logisticX0xlogisticY0 & 2223.80 & 2223.79 & 4.82e-08 \\
logisticX0xlogisticY02 & 2227.20 & 2226.76 & 8.78e-09 \\
logisticX0xRotatedLogisticY0 & 2190.10 & 2190.11 & 1.00 \\
logisticX0xRotatedLogisticY02 & 2190.06 & 2189.83 & 1.02 \\
rotatedLogisticX0xlogisticY0 & 2192.08 & 2192.11 & 0.372 \\
rotatedLogisticX0xlogisticY02 & 2193.69 & 2193.19 & 0.166 \\
\end{tabular}
\tablecomments{Candidate vetting completeness rate functions considered for the analysis in \S\ref{section:vettingCompleteness}, with their AIC values and relative likelihoods based on the median AIC values.  The relative likelihoods are with respect to the selected model logisticX0xRotatedLogisticY0.}
\end{table}

\subsection{False Alarm Effectiveness Model Selection} \label{app:fpEModelSelect}
The functions considered for fitting the observed false alarm effectiveness rate in \S\ref{section:FPEcharacterization} are given in Table~\ref{table:fpEffModelSelection}, along with AIC values and relative likelihoods.  We chose rotatedLogisticX0 because it has a high relative likelihood, had the fewest parameters, and gave the most reasonable convergence results compared with other high-likelihood models.  

\begin{table}[ht]
\centering
\caption{Candidate false alarm effectiveness rate functions} \label{table:fpEffModelSelection}
\begin{tabular}{ r c c c  }
\hline
Model Name & Median AIC & Minimum AIC & Relative Likelihood \\
\hline
rotatedLogisticX0 & 214.39 & 214.06 & 1 \\
rotatedLogisticX02 & 212.49 & 210.93 & 1.70 \\
constant & 270.55 & 270.55 & 6.35e-13 \\
dualBrokenPowerLaw & 250.08 & 244.08 & 1.77e-08 \\
gaussian & 247.10 & 243.48 & 7.87e-08 \\
rotatedLogisticX0xlogisticY0 & 220.44 & 220.08 & 4.84e-02 \\
rotatedLogisticX0+gaussian & 228.03 & 216.74 & 1.09e-03 \\
rotatedLogisticY & 215.80 & 212.36 & 2.35e-02 \\
rotatedLogisticYXLogisticY & 220.60 & 216.53 & 4.47e-02 \\
logisticY & 238.92 & 238.45 & 4.69e-06 \\
rotatedLogisticYXFixedLogisticY & 215.95 & 212.58 & 0.458 \\
\end{tabular}
\tablecomments{Candidate false alarm effectiveness rate functions considered for the analysis in \S\ref{section:FPEcharacterization}, with their AIC values and their relative probabilities based on the median AIC values. The relative likelihoods are with respect to the selected model rotatedLogisticX0.}
\end{table}

\subsection{Observed False Alarm Rate Model Selection} \label{app:obsFPModelSelect}
The functions considered for fitting the observed false alarm rate in \S\ref{section:obsFPcharacterization} are given in Table~\ref{table:fpObsModelSelection}, along with AIC values and relative likelihoods.  We chose rotatedLogisticX0 because it gave the most reasonable results compared with other high-relative-likelihood models.  We rejected rotatedLogisticX02 because one of its parameters did not converge.

\begin{table}[ht]
\centering
\caption{Candidate observed false alarm rate functions} \label{table:fpObsModelSelection}
\begin{tabular}{ r c c c  }
\hline
Model Name & Median AIC & Minimum AIC & Relative Likelihood \\
\hline
rotatedLogisticX0 & 307.20 & 307.19 & 1.00 \\
rotatedLogisticX0xlogisticY0 & 313.20 & 310.49 & 4.98e-02 \\
dualBrokenPowerLaw & 396.48 & 394.01 & 4.10e-20 \\
rotatedLogisticX02 & 302.87 & 301.10 & 8.71 \\
rotatedLogisticX0xlogisticY02 & 310.98 & 303.80 & 0.151 \\
rotatedLogisticX0+gaussian & 307.02 & 305.55 & 1.10 \\
\end{tabular}
\tablecomments{Candidate observed false alarm rate functions considered for the analysis in \S\ref{section:obsFPcharacterization}, with their AIC values and their relative probabilities based on the median AIC values. The relative likelihoods are with respect to the selected model rotatedLogisticX0.}
\end{table}

\clearpage
\section{Derivation of the Likelihood from the Poisson Probability} \label{app:likelihoodDerivation}
We briefly summarize Bayesian inference using a Poisson likelihood.  We will work in the period-radius parameter space.  

If our planet population is described by a point process with a period and radius-dependent rate $\lambda(p,r)$, then the probability that $n_i$ planets occur around an individual star in some region $B_i$ (say a grid cell) of period-radius space is
\begin{equation*}
    P \{ N \left( B_i \right) = n_i \} = \frac{\left( \Lambda(B_i) \right)^{n_i}}{n_i !} e^{-\Lambda(B_i)}
\end{equation*}
where
\begin{equation*}
    \Lambda(B_i) = \int_{B_i} \lambda(p, r) d p \, d r.
\end{equation*}
We now cover our entire period-radius range $D$ with a sufficiently fine regular grid with spacing $\Delta p$ and $\Delta r$ so that each grid cell $i$ centered at  period and radius $(p_i, r_i)$ contains at most one planet.  Then in cell $i$
\begin{equation*}
P \{ N \left( B_i \right) = n_i \} \approx 
\begin{cases}
    \lambda(p_i,r_i) \Delta p \Delta r e^{-\Lambda(B_i)} & n_i = 1 \\
    e^{-\Lambda(B_i)} & n_i = 0.
\end{cases}
\end{equation*}
We now ask: what is the probability of a specific number $n_i$ of planets in each cell $i$?  We assume that the probability of a planet in different cells are independent, so 
\begin{equation}
\begin{split}
    &P \{ N \left( B_i \right) = n_i, i=1,\ldots,K\} \\
        &= \prod_{i=1}^K \frac{\left( \Lambda(B_i) \right)^{n_i}}{n_i !} e^{-\Lambda(B_i)} \\
        &\approx \left( \Delta p \Delta r \right)^{K_1} \, e^{-\sum_{i=1}^K \Lambda(B_i)} \prod_{i=1}^{K_1} \lambda(p_i, r_i) \\
        &= \left( \Delta p \Delta r \right)^{K_1} \, e^{-\int_{D} \lambda(p, r) d p \, d r} \prod_{i=1}^{K_1} \lambda(p_i, r_i) \label{equation:poisson1}
\end{split}
\end{equation}
because the $B_i$ cover $D$ and are disjoint.  Here $K$ is the number of grid cells and ${K_1}$ is the number of grid cells that contain a planet = the number of planet candidates.  So the grid has disappeared, and we only need to evaluate $\lambda(p, r)$ at the planet locations $(p_i, r_i)$ and integrate the rate function $\lambda$ over the entire domain.

We do not observe all the planets, however.  We account for incompleteness, including geometric transit probability, by replacing $\lambda(p, r)$ with $\eta_s (p, r) \lambda(p, r)$ in equation~(\ref{equation:poisson1}), where $\eta_s (p, r)$ is the completeness function for this star $s$ measured in \S\ref{section:completeness}.  The result is the probability

\begin{equation}
    P \{ N \left( B_i \right) = n_i, i=1,\ldots,K \} 
        = \left( \Delta p \Delta r \right)^{K_1} \, e^{-\int_{D} \eta_s(p, r) \lambda d p(p, r) \, d r} \prod_{i=1}^{K_1} \eta_s(p_i, r_i) \lambda(p_i, r_i). \label{equation:poisson2}
\end{equation}

We now consider the probability of detecting planets around a set of $N_*$ stars.  Assuming that the planet detections on different stars are independent of each other, then the joint probability of a specific set of detections specified by the set $\{n_i, i=1,\ldots,N_*\}$ in cell $i$ on on all stars indexed by $s$ is given by 

\begin{equation}
\begin{split}
    P \{ N_s \left( B_i \right) = n_{s,i}, s=1,\ldots,N_*, i=1,\ldots,K \} 
        &= \prod_{s=1}^{N_*} \left( \Delta p \Delta r \right)^{K_1} \, e^{-\int_{D} \eta_s(p, r) \lambda(p, r) d p \, d r} \prod_{i=1}^{K_1} \eta_s(p_i, r_i) \lambda(p_i, r_i) \\
        &= V \, e^{- \int_{D} \eta(p, r) \lambda(p, r) d p \, d r} \prod_{s=1}^{N_*}  \prod_{i=1}^{K_1} \eta_s(p_i, r_i) \lambda(p_i, r_i)
        \label{equation:poisson3}
\end{split}
\end{equation}
where $V = \left( \Delta p \Delta r \right)^{(K_1 N_*)}$ and $\eta(p, r) = \sum_{s=1}^{N_*} \eta_s (p, r)$ is the sum of the completeness functions over all stars.

We now let the rate function $\lambda(p, r, \boldsymbol{\theta})$ depend on a parameter vector $\boldsymbol{\theta}$, and consider the problem of finding the $\boldsymbol{\theta}$ that maximizes the likelihood
\begin{equation}
\begin{split}
    P \{ N_s \left( B_i \right) = n_{s,i}, s=1,\ldots,N_*, i=1,\ldots,K | \boldsymbol{\theta} \} 
        &= V \, e^{- \int_{D} \eta(p, r) \lambda(p, r, \boldsymbol{\theta}) d p \, d r} \prod_{s=1}^{N_*}  \prod_{i=1}^{K_1} \eta_s(p_i, r_i) \lambda(p_i, r_i, \boldsymbol{\theta}) \\ 
        &= V \, \left( \prod_{s=1}^{N_*} \eta_s(p_i, r_i) \right) e^{- \int_{D} \eta(p, r) \lambda(p, r, \boldsymbol{\theta}) d p \, d r} \prod_{i=1}^{K_1}  \lambda(p_i, r_i, \boldsymbol{\theta}). \\ 
        \label{equation:poisson4}
\end{split}
\end{equation}
Because we are maximizing with respect to $\boldsymbol{\theta}$, we can ignore all terms that do not depend on $\boldsymbol{\theta}$.  Therefore maximizing equation~(\ref{equation:poisson4}) is equivalent to maximizing 
\begin{equation}
    P \{ N_s \left( B_i \right) = n_{s,i}, s=1,\ldots,N_*, i=1,\ldots,K | \boldsymbol{\theta} \} 
    = e^{- \int_{D} \eta(p, r) \lambda(p, r, \boldsymbol{\theta}) d p \, d r} \prod_{i=1}^{K_1}  \lambda(p_i, r_i, \boldsymbol{\theta}). \\ 
        \label{equation:poisson5}
\end{equation}

\clearpage
\section{Comparison of Catalog Cuts} \label{app:catalogComparison}

In \S\ref{section:variationsKic} we found that using the DR25 stellar properties catalog results in larger occurrence ratse than the baseline of \S\ref{section:baselineResults}, which uses the stellar properties from \citet{berger20}.  This difference can result from both the difference in the stellar properties themselves, and the fact that different stellar population cuts were used.  In particular, as described in \S\ref{section:stellarCatalog}, the \citet{berger20} catalog contains only stars with good Gaia noise characteristics, and we impose further Gaia fit quality requirements by removing stars with  qualityFlag = highRUWE.  

In this appendix we explore the relative impact of the difference in stellar properties between the two catalogs compared with the impact of the different cuts.  We consider the \citet{berger20} catalog with and without the cuts specific to this catalog.  We then consider the same stars as \citet{berger20}, but using the DR25 stellar properties and cuts.  Finally we consider the DR25 stellar catalog and its restriction to those stars contained in the supplemental catalog of \citet{Mathur2016}.  In all cases all steps of the occurrence rate computation are recomputed, including detection/vetting completeness and reliability.  

We compute the occurrence rates $F_1$ and $\zeta_{\oplus}$, defined in \S\ref{section:occurrence}.  For two cases using the stellar properties of \citet{berger20} which differ in the population cuts:
\begin{itemize}
    \item Case 1: the baseline of \S\ref{section:baselineResults}, starting with the \citet{berger20} catalog, with all the cuts described in \S\ref{section:stellarCatalog}, and planet radii corrected for Gaia stellar radii as described in \S\ref{section:planetCatalog}.  Starts with 186,548 stars and ends up with 58,974 GK stars after cuts.
    \item Case 2: Same as case 1, starting with the \citet{berger20} catalog, except without the highRUWE, Bin or Evol cuts described in \S\ref{section:stellarCatalog}, replacing these cuts with the cut on stellar radius removing stars with $R_* > 1.35 R_{\odot}$ described in \S\ref{section:variationsKic}.  Starts with 186,548 stars, and ends up with 66,956 GK stars after cuts.
\end{itemize}
We examine three cases using the DR25 stellar properties, which differ in the population cuts:
\begin{itemize}
    \item Case 3: The same cuts as case 2, starting with the \citet{berger20} catalog, except using DR25 stellar properties and original DR25 planet radii.  Starts with 186,548 stars and ends up with 71,168 GK stars after cuts.
    \item Case 4: The DR25 stellar catalog as described in section 4.1 using original DR25 planet radii.  Starts with 200,038 stars and ends up with 75,541 GK stars after cuts.
    \item Case 5: The DR25 stellar catalog as in case 4, restricted to those stars in \citet{Mathur2016} and using original DR25 planet radii.  Starts with 197,096 stars and ends up with 74,989 GK stars after cuts.
\end{itemize}
Cases 1 through 3 start with the same stars, and differ in the cuts and the use of Gaia-based vs. DR25 stellar properties.

\begin{figure}[ht]
  \centering
  \includegraphics[width=0.48\linewidth]{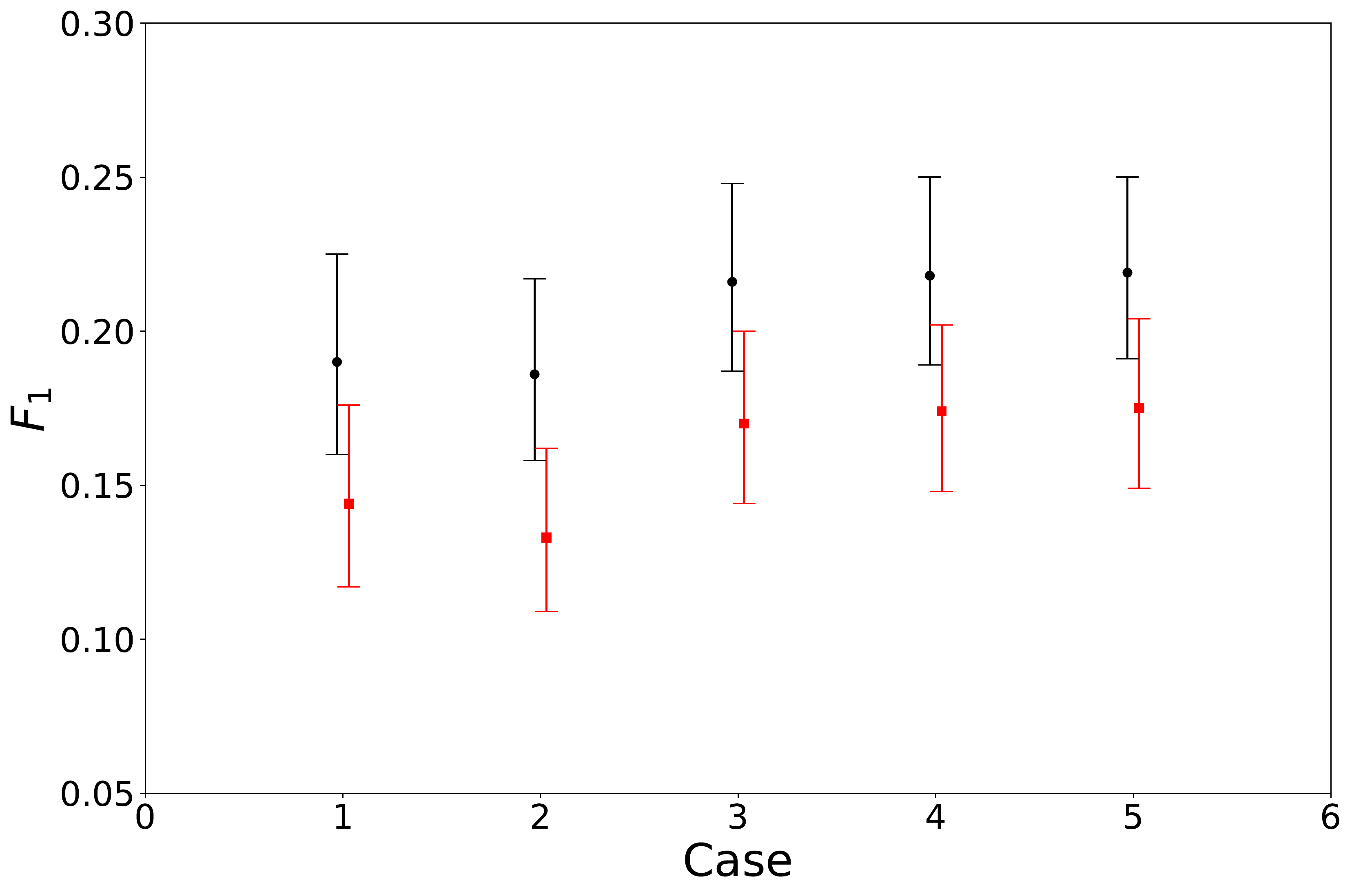} 
  \includegraphics[width=0.48\linewidth]{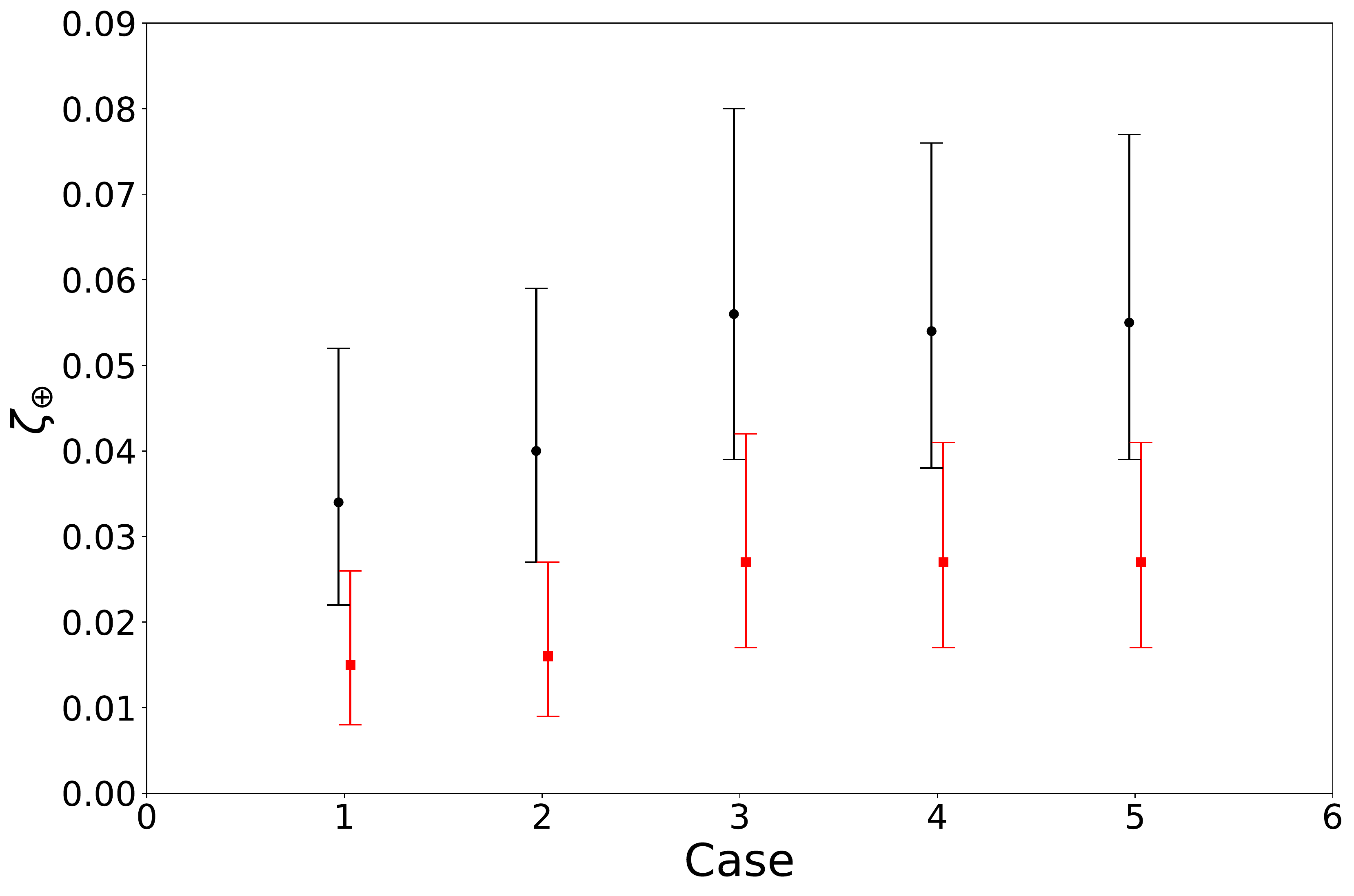}
  \caption{The impact of the different catalog cases considered in this appendix on $F_1$ (left) and $\zeta_{\oplus}$ (right).  Results without correcting for reliability are shown with black dots, and corrected for reliability with red squares.  Cases 1 and 2 use the stellar properties of \citet{berger20} with differing catalog cuts, while cases 3, 4 and 5 use DR25 stellar properties with differing catalog cuts.  
 } \label{figure:catalogComparison}
\end{figure}

The occurrence rates $F_1$ and $\zeta_{\oplus}$ for the various cases are given in Table~\ref{table:catalogComparisonResults} and shown in Figure~\ref{figure:catalogComparison}.  We see that using the same stellar properties gives similar occurrence rates, with a noticeable difference in occurrence rates computed using different stellar properties.  Differing cuts using the same stellar properties apparently has a much smaller impact.  We therefore conclude that stellar properties (including differences in GK classification due to differences in effective temperature) is the dominant cause of the different occurrence rates, and the population cuts play a minor role.  In all cases correcting for reliability has a significant impact on $\zeta_{\oplus}$.

\renewcommand{\arraystretch}{2}
\begin{table}[ht]
\centering
\caption{Comparison of occurrence rates using different catalogs and cuts.} \label{table:catalogComparisonResults}
\begin{tabular}{ r c c c c c c}
\hline
& \multicolumn{2}{c}{$F_0$} & \multicolumn{2}{c}{$F_1$} & \multicolumn{2}{c}{$\zeta_{\oplus}$} \\
Case  & No Reliability & With Reliability & No Reliability & With Reliability & No Reliability & With Reliability \\
\hline
 1 & $0.608^{+0.110}_{-0.090}$ & $0.432^{+0.089}_{-0.072}$ & $0.190^{+0.035}_{-0.030}$ & $0.144^{+0.032}_{-0.027}$ & 
 $0.034^{+0.018}_{-0.012}$ & $0.015^{+0.011}_{-0.007}$ \\
 2 & $0.609^{+0.112}_{-0.091}$ &  $0.393^{+0.083}_{-0.065}$ & $0.186^{+0.031}_{-0.028}$ &  $0.133^{+0.029}_{-0.024}$ & 
 $0.040^{+0.019}_{-0.013}$ & $0.016^{+0.011}_{-0.007}$ \\
 3 & $0.680^{+0.121}_{-0.099}$ &  $0.470^{+0.098}_{-0.079}$ & $0.216^{+0.032}_{-0.029}$ &  $0.170^{+0.030}_{-0.026}$ & 
 $0.056^{+0.024}_{-0.017}$ & $0.027^{+0.015}_{-0.010}$ \\
 4 & $0.675^{+0.115}_{-0.097}$ &  $0.474^{+0.090}_{-0.076}$ & $0.218^{+0.032}_{-0.029}$ &  $0.174^{+0.028}_{-0.026}$ & 
 $0.054^{+0.022}_{-0.016}$ & $0.027^{+0.014}_{-0.010}$ \\
 5 & $0.678^{+0.117}_{-0.096}$ &  $0.476^{+0.094}_{-0.077}$ & $0.219^{+0.031}_{-0.028}$ &  $0.175^{+0.029}_{-0.026}$ & 
 $0.055^{+0.022}_{-0.016}$ & $0.027^{+0.014}_{-0.010}$ \\
\end{tabular}
\tablecomments{Case 1 values are from Table~\ref{table:baselineThetaResults}, and case 4 values are from Table~\ref{table:occurrenceVariations}. }
\end{table}

We can get some insight into the change in occurrence rates by examining the impact of stellar properties on the PC population in the parameter space $50 \leq \mathrm{period} \leq 400$ days and $0.75 \leq \mathrm{radius} \leq 2.5$ $R_{\oplus}$. We examine the difference between case 2 and case 3 because these cases start with the same parent stellar population and apply the same cuts.  In case 2 these cuts are applied using the \citet{berger20} stellar properties, while in case 3 these cuts are applied using the DR25 stellar properties.  In case 2 there are 107 PCs in the period and radius range out of 67,306 stars in the parent GK population, which yields 0.666 per star after dividing by the average completeness of 0.00239.  In case 3 there are 116 PCs out of 71,168 stars in the parent GK population, which yields 0.64 planets per star after dividing by the average completeness of 0.00255.  These simple estimates are consistent with the values of $F_0$ found in Table~\ref{table:catalogComparisonResults}.  To understand the changes in $F_1$ and $\zeta_{\oplus}$, we need to examine in more detail how the stellar properties effect the PC population. 

\begin{figure}[ht]
  \centering
  \includegraphics[width=0.95\linewidth]{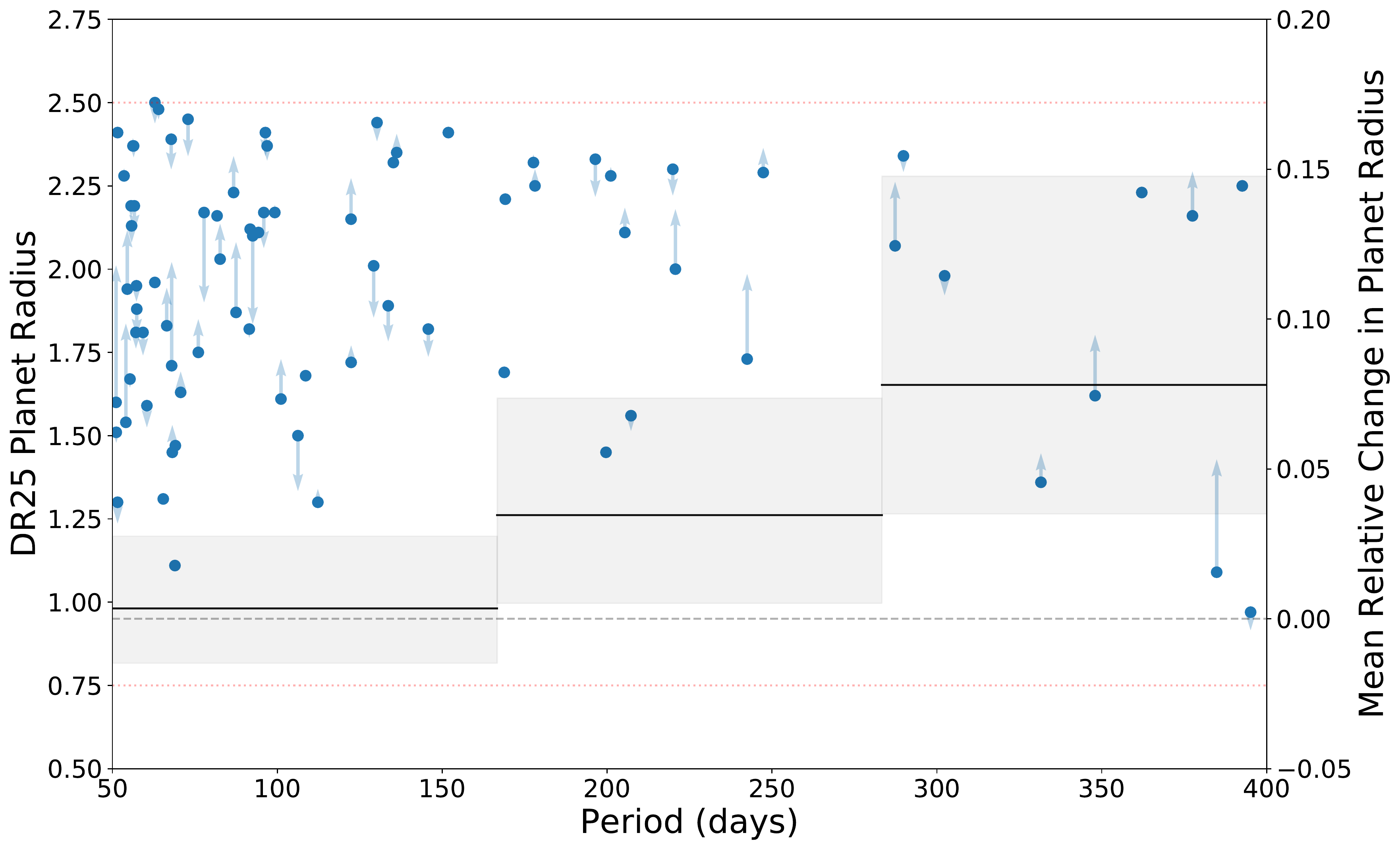} 
  \caption{The planet radii in case 3, using DR25 stellar properties, with the arrows indicating the change in radius when using the \citet{berger20} in case 2 for those PCs with $0.75 \leq \mathrm{radius} \leq 2.5$ $R_{\oplus}$ in both cases.  The solid horizontal lines show the average change in radius in three period bins when changing from case 3 to case 2, averaged over three bins, with values indicated by the right-hand y axis.  The shaded rectangles show the $1 \sigma$ uncertainty.
 } \label{figure:pcCatalogRadChanges}
\end{figure}

There are three ways that a difference in stellar properties can change the occurrence rate:
\begin{itemize}
    \item A change in stellar radius causes a change in planet radius, which will impact the period-radius dependence of the population rate function and may cause the planet to move into or out of the $0.75 \leq \mathrm{radius} \leq 2.5$ $R_{\oplus}$ range.
    \item A change in stellar radius causes the star to be added to or removed from the parent population depending on whether the star becomes smaller or larger than the $1.35 R_{\odot}$ cut.
    \item A change in stellar effective temperature causes the star to be added to or removed from the parent population because it is reclassified as GK or not GK.
\end{itemize}
Figure~\ref{figure:pcCatalogRadChanges} shows the change in planet radius when changing from case 3 (DR25 stellar properties) to case 2, for those planet candidates that are common to both case 2 and case 3.  We see that for periods between 50 and 200 days planet candidates both increase and decrease in size.  For periods greater than 200 days, however, there is clear bias towards larger sizes.  This effect is quantified by computing the average relative change in size in three period bins.  The shortest period bin shows a near-zero average change in size, while the longest-period bin shows an average increase in size of about 8\%, which is about a $2 \sigma$ change. 

\begin{figure}[ht]
  \centering
  \includegraphics[width=0.75\linewidth]{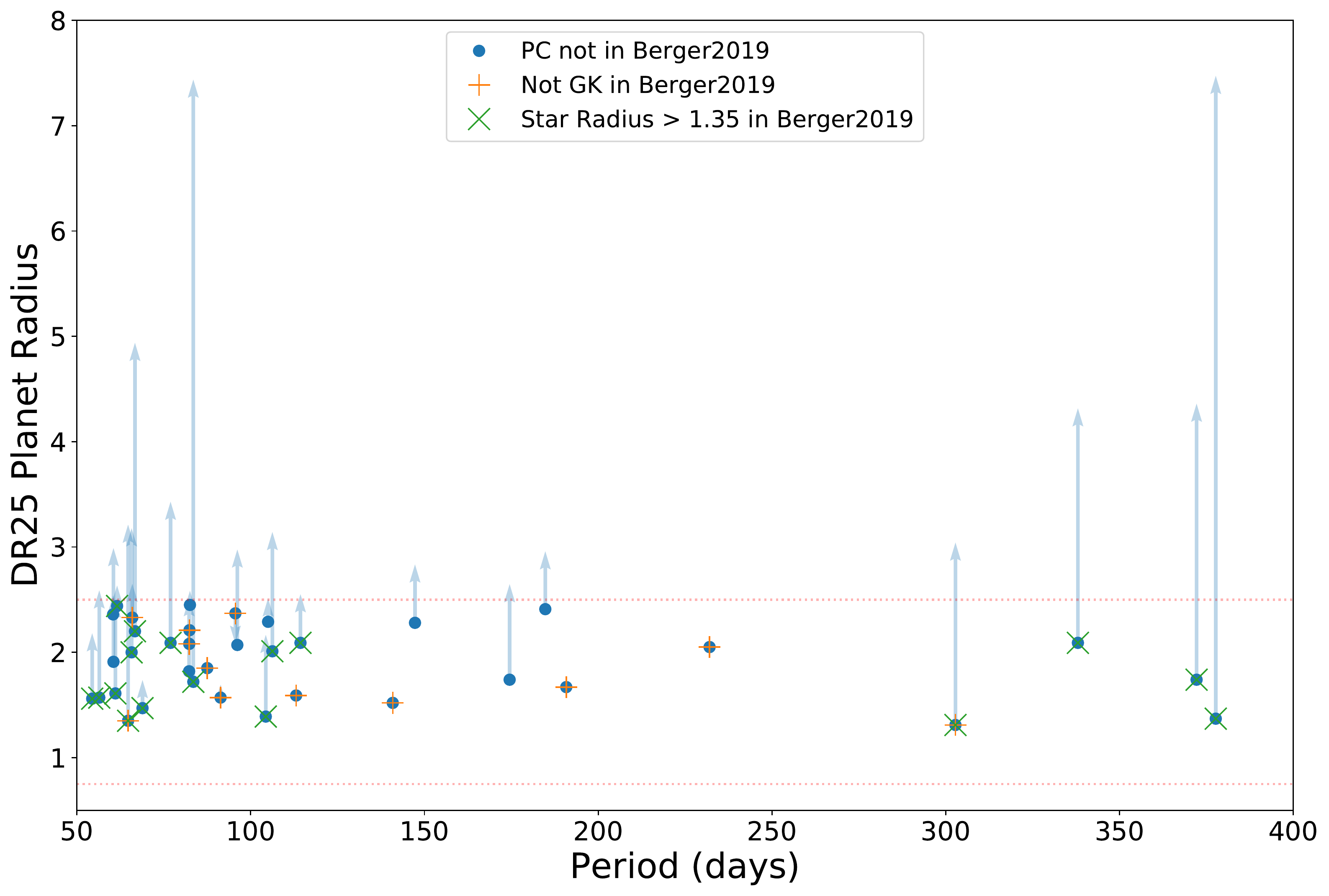} \\
  \includegraphics[width=0.75\linewidth]{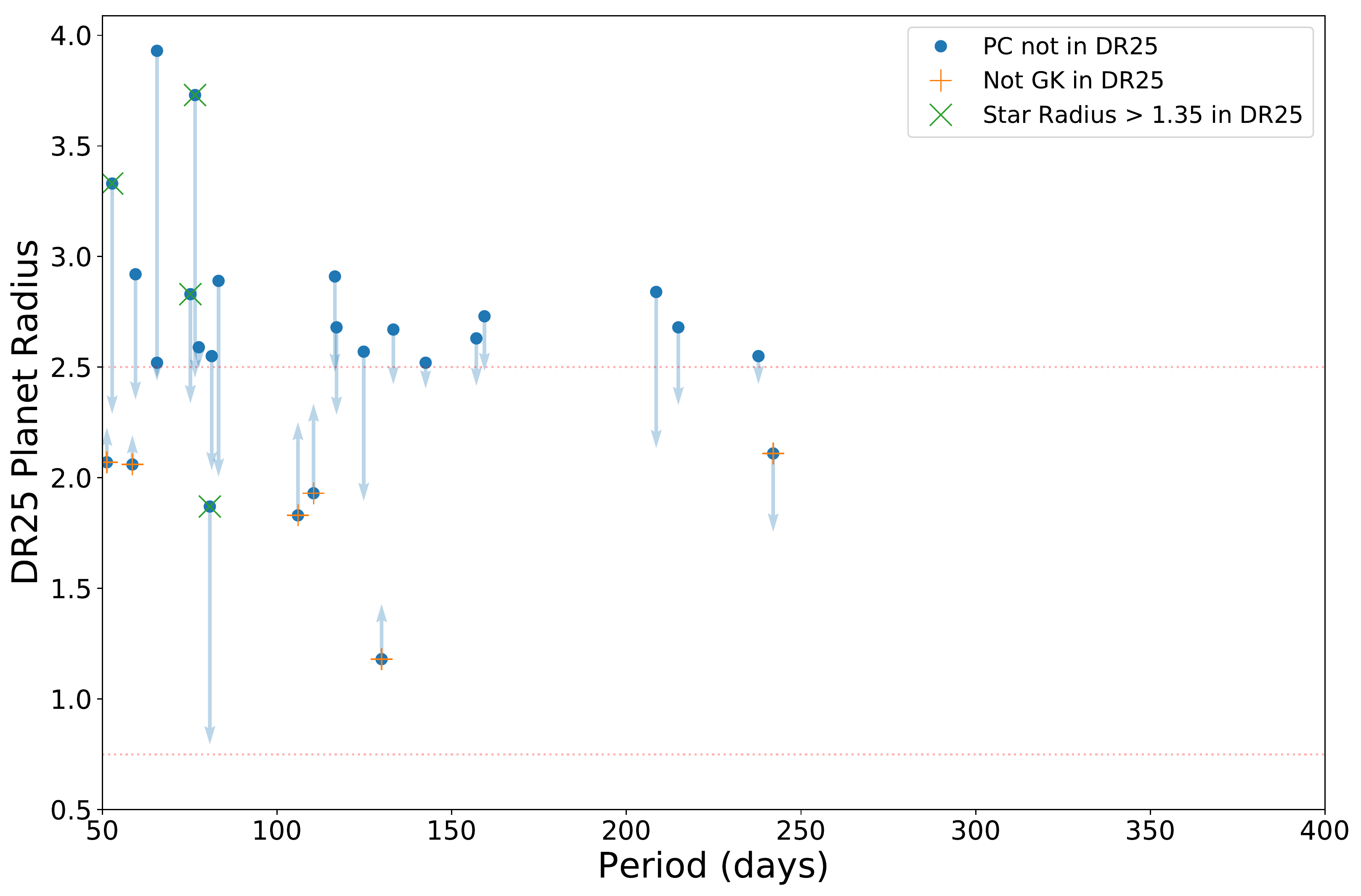}
  \caption{Planet candidates that are not common to case 2 and 3, plotted using the DR25 (case 3) stellar properties, with the arrows indicating the change in radius when using the \citet{berger20} in case 2.  The markers indicate reasons why the PCs present in one case were dropped from the other.  Top: the PCs in case 3 that are not present in case 2.  For most of these PCs, the arrows indicate that their radii using \citet{berger20} stellar properties exceeded $2.5$ $R_{\oplus}$, removing them from the case 3 population.  Other PCs were removed because in case 3 their stellar host radii exceeded $1.35 R_{\odot}$ or were not GK stars.  Bottom: PCs in case 2 that are not present in case 3.  For most of these PCs, they are too large in case 3 using the DR25 stellar properties, and the arrows indicate that these PCs became smaller than $2.5$ $R_{\oplus}$ using the \citet{berger20} stellar properties in case 2.  Other PCs appeared because their stellar hosts were either larger than $1.35 R_{\odot}$ or not GK in case 3 using the DR25 stellar properties, but are smaller and GK in case 2.
 } \label{figure:pcCatalogChanges}
\end{figure}

Figure~\ref{figure:pcCatalogChanges} shows planet candidates that either exited or entered the radius range considered in our occurrence rate in the change from case 3 to case 2.  As in Figure~\ref{figure:pcCatalogRadChanges}, we see that at low period several planet candidates entered our planet radius domain of $0.75 \leq \mathrm{radius} \leq 2.5$ $R_{\oplus}$ while other planet candidates left that domain.  But for longer period, particularly $> 250$ days, planets left our domain by becoming too large while no planets entered our domain.  Thus there is a loss of small exoplanets due to their being larger using \citet{berger20} stellar properties.  In addition, several stars exited or entered our parent population through reclassification due to change in effective temperature.

These figures suggest that the change in planet size when using the \citet{berger20} stellar properties (case 2) is a significant contributing factor in the reduced occurrence rates.  But it would not be correct to conclude that this change in planet size is the ``cause'' of the lower occurrence rate: changes in the parent stellar population also impact occurrence rates through changes in detection completeness and changes in population due to stellar reclassification and which stars pass the stellar size cut.  It is only through computing the full occurrence rate that we can measure the impact of the stellar properties.

\clearpage

\bibliographystyle{apj}
\bibliography{refs}



\end{document}